\documentclass[aps,prl,twocolumn,longbibliography,superscriptaddress]{revtex4-2}

\usepackage{amsfonts}
\usepackage{amsmath}
\usepackage{graphicx}
\usepackage{dsfont}
\usepackage{diagbox}
\usepackage{array}
\usepackage{amssymb}
\usepackage{bbm}
\usepackage{float}
\usepackage{subfigure}
\usepackage{MnSymbol}
\usepackage{url}
\usepackage{times}
\usepackage{manfnt}
\usepackage{booktabs}
\usepackage{xcolor}
\definecolor{darkblue}{HTML}{004D6B}
\definecolor{darkred}{HTML}{8c1515}
\definecolor{darkgreen}{HTML}{006400}
\usepackage{hyperref}
\hypersetup{
    pdftitle={Nishimori's cat},
	colorlinks=true,
	urlcolor=darkred,
	citecolor=darkblue,
	linkcolor=darkred,
	breaklinks
}
\pagecolor{white}
\usepackage{ulem}
\usepackage{physics}
\usepackage{enumitem}
\usepackage{wasysym}

\begin{document}

\title{Nishimori's Cat: Stable Long-Range Entanglement\\from Finite-Depth Unitaries and Weak Measurements}

\author{Guo-Yi Zhu}
\email{gzhu@uni-koeln.de}
\affiliation{Institute for Theoretical Physics, University of Cologne, Z\"ulpicher Straße 77, 50937 Cologne, Germany}

\author{Nathanan Tantivasadakarn}
\affiliation{Walter Burke Institute for Theoretical Physics and Department of Physics, California Institute of Technology, Pasadena, California 91125, USA}
\affiliation{Department of Physics, Harvard University, Cambridge, Massachusetts 02138, USA}

\author{Ashvin Vishwanath}
\affiliation{Department of Physics, Harvard University, Cambridge, Massachusetts 02138, USA}

\author{Simon Trebst}
\affiliation{Institute for Theoretical Physics, University of Cologne, Z\"ulpicher Straße 77, 50937 Cologne, Germany}
\affiliation{Center for Computational Quantum Physics, Flatiron Institute, 162 5th Avenue, New York, New York 10010, USA}

\author{Ruben Verresen}
\affiliation{Department of Physics, Harvard University, Cambridge, Massachusetts 02138, USA}

\date{\today}


\begin{abstract}
In the field of monitored quantum circuits, 
it has remained an open question whether finite-time protocols for preparing long-range entangled states 
lead to phases of matter which are stable to gate imperfections, which can convert projective into weak measurements. 
Here we show that in certain cases, long-range entanglement persists in the presence of weak measurements, and gives rise to novel forms of quantum criticality. 
We demonstrate this explicitly for preparing the two-dimensional Greenberger-Horne-Zeilinger cat state and the three-dimensional toric code as minimal instances. 
%
In contrast to random monitored circuits,
our circuit of gates and measurements is deterministic; the only randomness is in the measurement outcomes.
We show how the randomness in these weak measurements allows us to track the solvable Nishimori line of the random-bond Ising model, rigorously establishing the stability of the glassy long-range entangled states in two and three spatial dimensions. 
Away from this exactly solvable construction, we use hybrid tensor network and Monte Carlo simulations to obtain a nonzero Edwards-Anderson order parameter as an indicator of long-range entanglement in the two-dimensional scenario.
We argue that our protocol admits a natural implementation in existing quantum computing architectures, requiring only a depth-3 circuit on IBM's heavy-hexagon transmon chips.
\end{abstract}

\maketitle


In extended quantum systems, the rich interplay between measurements and quantum correlations point to a plethora of new emergent phenomena. Although measurements are often associated with reducing entanglement, they provide an intriguing loophole enabling fast preparation of long-range entangled (LRE) states, such as macroscopic cat states or topologically ordered states, that are otherwise forbidden. Indeed, while LRE can only be prepared with a unitary quantum circuit whose \textit{depth grows with system size}\cite{Verstraete2006LRB,Aguado08,Koenig09,Chen2010,Zaletel20,Soejima20,ShtengelPollmann,Wei22, Wei2020GHZ, Mooney2021GHZ},
a large class of them can be prepared in \textit{finite time} by simply measuring certain \textit{stabilizers} (finite products of Pauli operators) \cite{Gottesman97}. 
This allows for deterministic state preparation using a finite-depth unitary feedback\cite{Raussendorf2001,Raussendorf2005,Aguado08b,Brennen09,Stace2016,Piroli21,Lucas22measurement}, intimately tied to the idea of quantum error correcting codes\cite{Gottesman2001,Preskill2002}. Moreover, it has recently been shown that measurement-based state preparation protocols also exist for certain non-stabilizer states, including non-Abelian topological order\cite{Verresen2021cat,Nat2021measure,Bravyi2022,Hsieh2022}.

Remarkably, it is not known whether such measurement-induced states form stable phases of matter, which are robust to local perturbations of the preparation protocol. While this question is of clear practical significance, it is also of conceptual interest to explore whether one can extend the familiar notion of stability of phases of matter (primarily developed for solid-state purposes) to the era of quantum simulators and computers\cite{Altman2021_simulators,Gambetta2020}. Here, we explore what happens when the circuit is perturbed prior to measuring. In effect, this turns an originally projective measurement into a \text{weak measurement}, as we will discuss. We ask whether such a generic scenario allows for stable LRE states; and if so, is there a critical point at the boundary of stability?

This motivating question fits naturally into the broader realm of monitored quantum circuits\cite{Potter21review,Fisher2022reviewMIPT}. Recent years have seen immense progress and activity in studying the long-time limit of random unitary gates combined with (projective) measurements. A key result has been that there is an entanglement transition between volume-law and area-law entangled regions as one increases the measurement rate\cite{Li2018,Skinner2019}. Subsequent works also explored how the latter can be in distinct phases of matter\cite{Barkeshli2021measure, Barkeshli2021measuretoriccode, Hsieh2021measure, Buchold22measurespt, Vijay22, Ippoliti22}. While the effects of weak measurements have been partially explored for the case of long-time quantum trajectories\cite{Schomerus2019,Jian20,Altman2020weak, Schomerus2020weakmeasure, Ashida2020, Jian21SYKMIPT, Schiro2021, Biella2021manybodyquantumzeno, Diehl22weakmeasure, Romito2021}, to the best of our knowledge, it has not been explored in the \textit{finite-time} protocols.
This question is especially important in the latter case, since using measurement is then the \textit{only} route towards preparing LRE states. 


\begin{figure*}[t] 
   \centering
   \includegraphics[width=2\columnwidth]{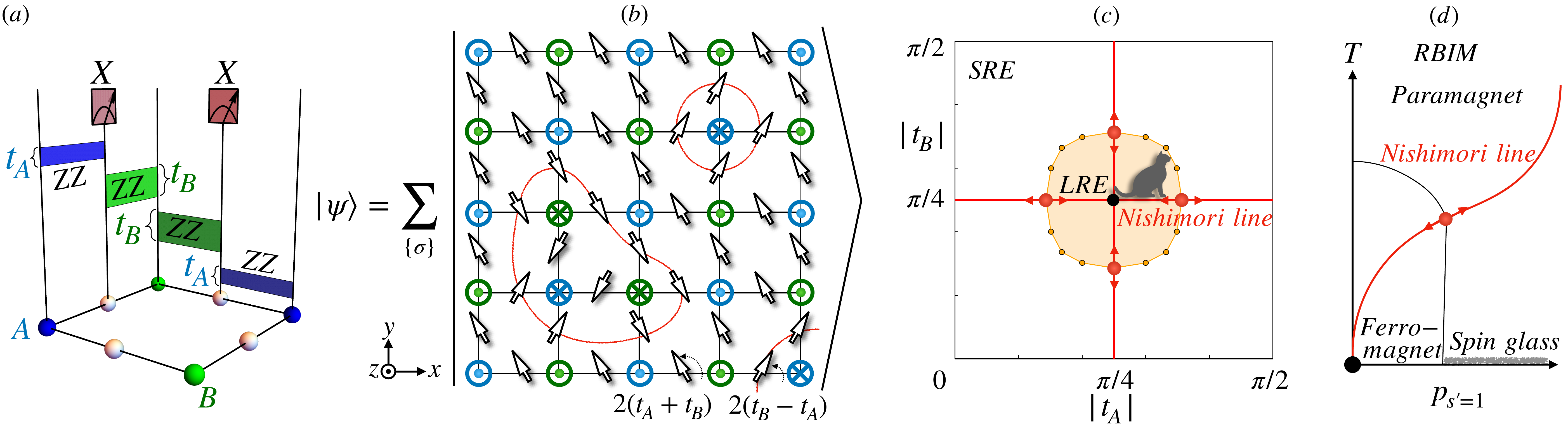} 
   \caption{
   {\bf Circuit and phase diagram for Nishimori's cat from measurements}.
   (a) On the Lieb lattice populated with physical site (blue/green) and ancilla bond (white) spins, 
   	a depth-4 circuit of $e^{-it ZZ}$ gates is applied to nearest-neighbor spins, where the evolution time depends on the site sublattice ($A$-$B$), and the ancilla bond spins are then measured in Pauli-$x$ basis.
   (b) A classical snapshot of the premeasurement wave function. 
   The circles/crosses label spins up($\uparrow$)/down($\downarrow$), and their domain walls are highlighted by red loops. 
  Ising evolution correlates the ancillas to the domain walls.
   (c) Wave function phase diagram and the Nishimori line by tuning evolution times on the A(B) sublattice. The postmeasurement wave function is a LRE disordered cat state inside the yellow region, and a short-ranged entangled (SRE) state outside. 
   A $Z_2$ gauge symmetry emerges along the Nishimori lines (red), which upon gauge symmetrization, can be mapped exactly onto the eponymous line\cite{Nishimori1981}        
   in the  phase diagram of the classical RBIM shown in (d).
   Beyond the Nishimori line, the phase boundaries in (c) are charted out by numerical computations (yellow dots),  
   which have no direct equivalent in the RBIM phase diagram.
   }
   \label{fig:fig1}
\end{figure*}

In this Letter, we establish a stability threshold for various measurement based protocols that induce long range entanglement, with a novel form of quantum criticality at the threshold. For this it is of fundamental importance to recall how one experimentally measures a multi-body stabilizer $\mathcal O$ for an arbitrary state $\ket{\psi}$, such as the two-body Ising interaction for a cat state\cite{Raussendorf2001} or the four-body stabilizers of the toric code\cite{Kitaev2003}. Since most platforms naturally perform \textit{single-site measurements}, one introduces an ancilla qubit and entangles it with $\ket{\psi}$ in such a way that measuring the ancilla effectively measures $\mathcal O$. However, if the entangling operation is not perfect, the net result is to have a (partial collapsing) \textit{weak measurement}\cite{Clerk2010rmp}. This can be seen most clearly from the following identity that transforms real time into imaginary time evolution up to a complex phase factor (derived in the Supplemental Material (SM)\cite{supplement})
\begin{equation}
	\langle\pm_y|e^{-its^z \otimes \mathcal O} |+_x\rangle \propto e^{\pm \frac{\beta}{2} \mathcal O} \textrm{ with } \tanh (\beta/2) = \tan t \,,\label{eq:weak}
\end{equation}
where $\ket{\pm_\alpha}$ are the eigenstates of Pauli matrix $s^\alpha$ on the ancilla qubit.
This is a projective measurement on $\ket{\psi}$ \textit{only} if $t = \frac{\pi}{4}$: then $\beta = \infty$ pins $\mathcal O = \pm 1$ depending on the measurement outcome. 
Eq.~\eqref{eq:weak} gives us two key insights into the correlations resulting from weak measurements (e.g., for times $0 < t < \pi/4$): first, the effective imaginary time-evolution suggests we ought to consider phases which are stable to finite temperature, such as a 2D Ising ferromagnet or 3D discrete gauge theory. Second, the randomness of measurement outcomes introduces effective disorder. A large part of our  analysis is devoted to demonstrating stability against this disorder, which we discuss in detail for the minimal cases of a 2D Greenberger-Horne-Zeilinger (GHZ)-type\cite{GHZ} state, and whose discussion \textit{mutatis mutandis} carries over for the 3D toric code. Crucially, the disorder distribution in our scenario is highly correlated, enabling us to map the entire range between strong and weak measurements $\pi/4 \geq t \geq 0$ {\it exactly} onto the solvable Nishimori line of the random bond Ising model (RBIM)\cite{Nishimori1981}. For instance, for the simple protocol in Eq.~\eqref{eq:weak}, we find a Nishimori critical point at $t_c \approx 0.143\pi$ in 2D. We refer to the stable LRE phase between the GHZ-type fixed point and the Nishimori critical point as \textit{Nishimori's cat}. Our work thereby also establishes a firm connection between monitored circuits and the vast literature on spin glasses. 

\textit{Prior work}.-- 
We further note that finite-time transitions have recently been explored in the context of {teleportation transitions}\cite{Altman2021measure,Chen2022measure}, which again involves projective measurements and where one leaves a \textit{subextensive} region unmeasured, while Ref.~\onlinecite{Garratt22} studied the effect of weak measurement on \textit{removing} quantum correlations of an initially critical state. Finite-depth transitions have also been explored in the context of computational\cite{Browne2008}  
and complexity transitions\cite{Harrow2022}. Finally, we point out an intriguing formal connection to phase transitions in information recovery in surface codes\cite{Preskill2002}, where in the absence/presence of syndrome measurement errors, the problem is also mapped to the 2D RBIM/3D random plaquette Ising gauge model along the Nishimori line.


\textit{Circuit model}.-- To achieve an Ising 
LRE phase, we weakly measure the domain wall operator: $\mathcal{O} = \sigma_i^z \sigma_j^z$, which is weight-three when including additional ancillas (see Eq.~\eqref{eq:weak}). However, one can design a protocol with only two-body evolutions\cite{supplement}.
For this, we consider qubits on the Lieb lattice (Figure \ref{fig:fig1}a), where we denote the target spins on the square lattice as $\sigma_j^{z(x)}$ and the ancillas at the bond centers as $s_{ij}^{z(x)}$.
We entangle these two types of spins by a depth-4 circuit of nearest-neighbor Ising evolutions (Fig.~\ref{fig:fig1}a):
\begin{equation}
\ket{\psi(t_A, t_B)} = e^{-i \sum_{\langle ij\rangle } t_j  \sigma_j^z s_{ij}^z} \ket{+_x}^{\otimes N} \,.
\label{eq:psi}
\end{equation}
Crucially, we have introduced {\it two} evolution times $t_j=t_{A(B)}$ if $j$ belongs to the A or B sublattice (of the original square lattice of site spins), see Fig.~\ref{fig:fig1}a. 
As shown in Fig.~\ref{fig:fig1}b, the pair of gates associated with any given bond effectively rotates the ancilla spin by an angle $2(t_A\pm t_B)$ depending on the alignment of the neighboring spin pair. 
Consequently, measuring the ancilla spin in $x$ direction weakly measures the domain wall of the target spins, which becomes a strong measurement only when both $t_A,t_B \to \frac{\pi}{4}$, in which case $\ket{\psi}$ equals the 2D cluster state\cite{Raussendorf2001}.
More generally, the entire wave function \eqref{eq:psi} can be viewed as a superposition of all allowed $\{\sigma\}$ classical configurations, in which the orientation of ancillas uniquely depends on whether it sits on a domain wall or not (Fig.~\ref{fig:fig1}b). 

The probability of the measurement outcome $s_{ij}^x \to s_{ij} = \pm 1$ is given by Born's rule:
\begin{equation}
\label{eq:ps}
	p_{\{s\}}  \equiv \norm{\bra{\{s\}}\ket{\psi}}^2
	\propto \sum_{\{\sigma\}}e^{-\beta \sum_{ij} (J_{s_{ij}} \sigma_i \sigma_j + h s_{ij})} \,,
\end{equation}
which we recognize as the partition function of the RBIM (with the measurement outcome labeling the random bond configuration), 
where a straightforward computation\cite{supplement} shows that
\[
	 \tanh \frac{\beta}{2} J_+ = \tan t_A \tan t_B,
	 \quad 
	 \tanh \frac{\beta}{2} J_- = -\tan t_A \cot t_B\,,
\] 
and $\beta h = \frac{1}{2}\ln\lvert\tan(t_A+t_B)\tan(t_A-t_B)\rvert$. The subspace $(t_A, t_B=\pi/4)$ of this two-parameter protocol recovers the single-parameter protocol of Eq.~\eqref{eq:weak}. Note that we can interpret the right-hand side of Eq.~\eqref{eq:ps} as a classical partition function $Z_{\{s\}}$, which contains the information of all diagonal correlation functions\cite{Henley04RK, Fradkin04RK, Cirac06} of our postmeasurement quantum state. 

We can thus interpret the ensemble (over all measurement outcomes of the ancillas) as a classical system with disorder $\{s\}$, where frustrated plaquettes $\prod_{l\in \square} s_l = -1$ are said to have an Ising vortex. However, unlike commonly studied disordered models, the disorder distribution in Eq.~\eqref{eq:ps} is highly correlated (making the vortices attractive). In fact, the property that $Z_{\{s\}} \propto p_{\{s\}}$ is akin to the structure Nishimori first uncovered after a gauge transformation \cite{Nishimori1981} for his eponymous line in the RBIM. It implies that certain quantities (like the internal energy) are nonsingular even at the transition. This remarkable fact is naturally explained by our approach, since those quantities can be expressed as {\it linear} functions of the density matrix of the premeasurement wave function, generated by finite-depth unitary circuit. 

To chart out our generic phase diagram in Fig.~\ref{fig:fig1}c, we use the Edwards-Anderson (EA) order parameter as our diagnostic
for the formation of a glassy LRE state\cite{EdwardsAnderson}:
\begin{equation}
q\equiv [\langle \sigma_0\sigma_c\rangle^2] \equiv \sum_{\{s\}} p_{\{s\}} \langle \sigma_0\sigma_c\rangle_{\{s\}}^2 \,,
\end{equation}
where $\sigma_{c(0)}$ is the spin at the central(corner) site of the open square lattice, $[\cdots]$ denotes the measurement (disorder) average, and $\langle \cdots\rangle$ the quantum average of the postmeasurement wave function, equivalent to the classical ensemble average for a given disorder pattern. 
Because of the global Ising symmetry of the protocol, the quantum state is Ising symmetric with $\langle \sigma \rangle = 0$, and a nonzero EA order in thermodynamic limit signifies long-range connected quantum correlation, which serves as lower bound for the quantum mutual information between two sites at a distance\cite{Cirac2008mutualinfo}.
Therefore the ordered phase of this classical description corresponds to the postmeasurement quantum state being a LRE cat state. 


\textit{Nishimori line}.--
Along the line $( t_B=\pi/4)$ (the red horizontal line in Fig.~\ref{fig:fig1}c, although the same discussion also applies to $t_A = \pi/4$), the EA order parameter can be exactly mapped to the magnetization of the Nishimori line in the RBIM, which exhibits a phase transition on crossing the Nishimori multicritical point\cite{Georges1985a,Georges1985b,Binder86rmp, Adler1996, Fisher1997, Read2000, Pujol2001, Chalker2002, Hartmann2004}. 
Importantly, this point is located at a finite $t_A < \pi/4$ in our model, implying stability of the cat state up to a \textit{finite} error threshold at $t_A^c\approx0.143\pi$\cite{Nishimori1981, Harris1988Nishimori, Adler1996, Fisher1997, Read2000, Pujol2001, Chalker2002, Hartmann2004}. 
This can be seen as follows:
First, consider the partition function \eqref{eq:ps} for a given disorder realization. Then as $\beta J_+ = -\beta J_-$ and $\beta h = 0$, our circuit model becomes precisely equivalent to the RBIM with quenched binary bond disorder, where the inverse temperature $\beta \equiv \ln|\tan(t_A+\pi/4)|$ (by setting $J_+=-J_-=1$) is tuned by the unitary evolution time.
Second, consider the disorder ensemble: due to an Ising gauge symmetry in the premeasurement wave function\cite{supplement}, any pair of bond disorder configurations that share the same vortex configuration are gauge equivalent and have the same probability. 
Together, this implies that our possible measurement outcomes $\{s\}$ form a gauge symmetric disorder ensemble generated by {\it gauge symmetrizing} an {\it uncorrelated} bond disorder $\{s'\}$ with probability $p_{s'=1} = 1/(1+e^{2\beta})=(1-\sin(2t_A))/2$, according to 
\begin{equation}
	\sigma_j' = \sigma_j\tau_j,\quad  s_{ij}' = s_{ij} \tau_i\tau_j \,,
\end{equation}
where $\tau_j=\pm 1$ stands for a local $Z_2$ gauge transformation. Then the measurement average can be decomposed to two steps: $[\cdots] = \sum_{\{\tau\}} [\cdots]'$, where $[\cdots]'$ denotes the uncorrelated disorder average as in the RBIM, and $\sum_{\{\tau\}}$ denotes gauge symmetrization.
We thus find that all gauge invariant observables of the Nishimori line in the RBIM (i.e.\ the red line in Fig.~\ref{fig:fig1}d) coincide with those in our model.

The Nishimori line is known to be invariant under a renormalization group flow\cite{Harris1988Nishimori, Harris1989}, which crosses the paramagnetic / ferromagnetic phase boundary at a multicritical point\cite{Nishimori1981, Fisher1997, Read2001}. 
It was mathematically proven that the phase transition happens at {\it finite} critical disorder probability\cite{Nishimori1981, Preskill2002, Preskill2003}. 
Inside the ferromagnetic phase,  $[\langle \sigma_0\sigma_c\rangle]'\neq 0$. 
Nevertheless, our wave function measurement average involves an extra gauge symmetrization, i.e.\ summation over $\tau=\pm 1$, which turns the ferromagnetic phase into a finite-temperature spin glass:
$
	[\langle\sigma_0 \sigma_c\rangle] = 0,
	\ 
	[\langle \sigma_0\sigma_c\rangle^2] \neq 0 \,.
$
That is, while the linear magnetization vanishes, the \textit{nonlinear} EA order parameter keeps track of the magnetization correlation in each gauge sample, because $[\langle\sigma_0\sigma_c\rangle^2] = [\langle\sigma_0\sigma_c\rangle^2]' = [\langle\sigma_0\sigma_c\rangle]'$\cite{Nishimori1981}.
More generally, any odd moment of a $\sigma$ correlation function is odd under gauge transform and thus vanishes under gauge symmetrization. 
Note that this spin glass state should be contrasted to the zero-temperature spin glass in the 2D RBIM (indicated by the gray dashed line in Fig.~\ref{fig:fig1}e).
The robust glassiness of our state against finite temperature originates from the gauge symmetry, analogous to the exactly solvable Mattis spin glass\cite{Mattis1976, Binder86rmp} which gauge symmetrizes the frustration-free Ising ordered phase. 
Nevertheless, away from the limit $t_A\to \pi/4$, our state features a finite density of Ising vortices that is more nontrivial than conventional Mattis spin glasses.


\textit{Beyond the Nishimori line}.--
\begin{figure}[t] 
   \centering
   \includegraphics[width=\columnwidth]{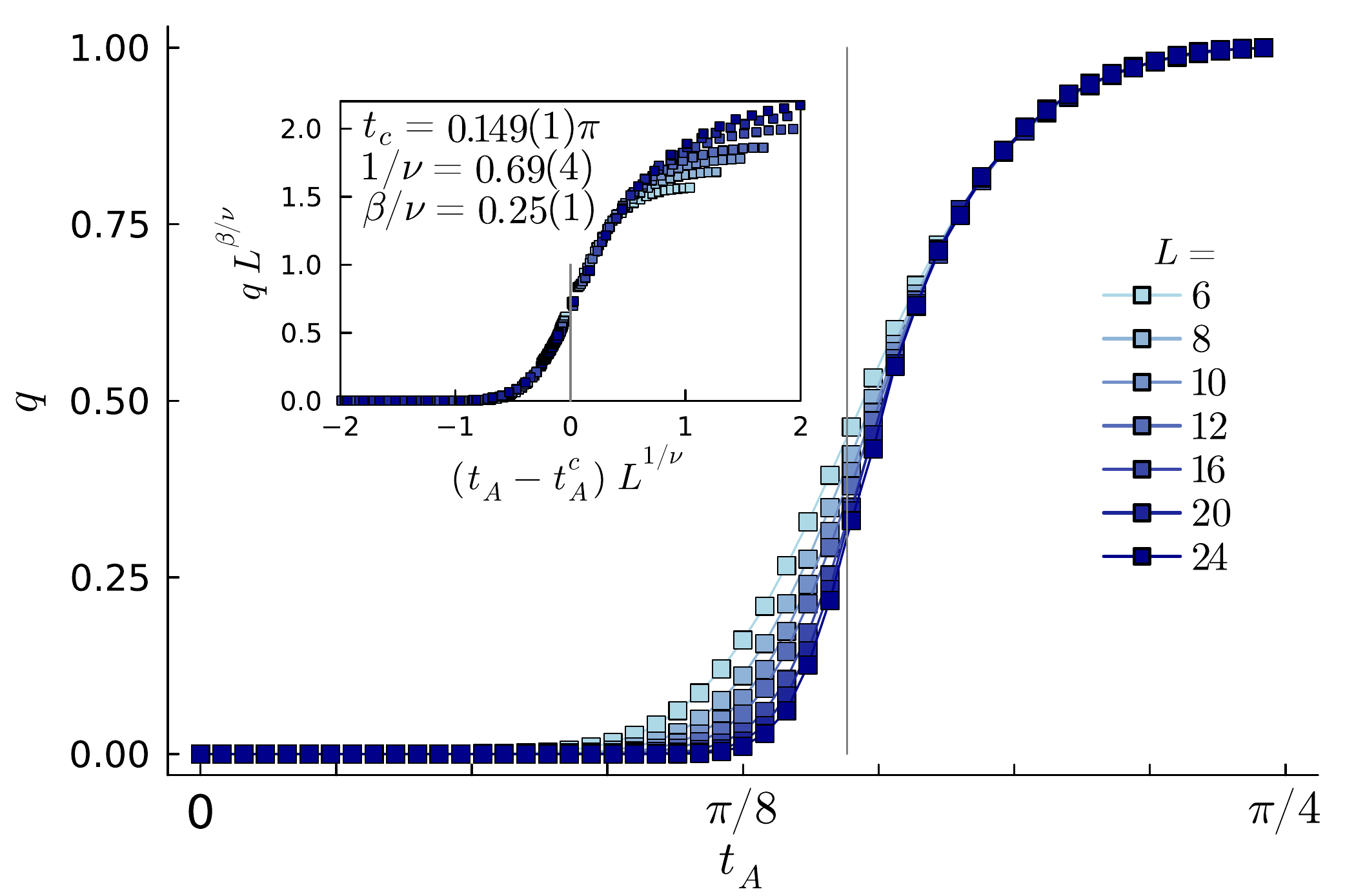} 
   \caption{{\bf Transition from SRE to LRE states after finite evolution time along the Nishimori line} 
    ($t_A, t_B=\pi/4$) in the phase diagram of Fig.~\ref{fig:fig1}(d).
   Shown are results for the EA order parameter from our hybrid Monte-Carlo / tensor-network approach (symbols) 
   for Lieb lattices of varying system sizes with open boundaries.
   The vertical gray line indicates our estimate of the critical point $t_c\approx 0.149\pi$ 
   extracted from a data collapse in a window $0.1\pi\leq t_A\leq 0.2\pi$, 
   fitting a scaling function\cite{FSSpackage} $q=[\langle\sigma_0\sigma_c\rangle^2]\propto L^{-\beta/\nu} f((t_A-t_A^c)L^{1/\nu})$. 
   }
   \label{fig:fig2}
\end{figure}
We expect the phase diagram established on the Nishimori line to be perturbatively robust, because any symmetric perturbation in the circuit away from the Nishimori line can be mapped to a local, Ising-symmetric perturbation in the corresponding classical model. 
For more generic $(t_A, t_B)$, the partition function $Z_{\{s\}}$ of Eq.~\eqref{eq:ps} can still be interpreted as a disordered Ising model, 
albeit one with imbalanced strengths, $J_+$ and $J_-$, of the ferromagnetic and antiferromagnetic bonds signaling the breakdown of the gauge symmetry, i.e.\ we are moving away from the solvable line in the phase diagram of
Fig.~\ref{fig:fig1}c (and out of the plane of the phase diagram in  Fig.~\ref{fig:fig1}d). Indeed, while the ``Nishimori property" $p_{\{s\}} \propto Z_{\{s\}}$ remains, the gauge symmetry was crucial for obtaining exact results along the Nishimori line\cite{Nishimori1981}. This coupling imbalance becomes particularly pronounced when one approaches the diagonal line $t_A=t_B$ in the phase diagram of Fig.~\ref{fig:fig1}c, where the strength of the ferromagnetic bond diverges to infinity. 

For this generic scenario with two timescales $(t_A, t_B)$,  one needs to numerically contract out the {\it entire} tensor network to calculate the disorder probability $p_{\{s\}}$, which is essentially a structured shallow version of the quantum circuit sampling problem~\cite{Lund2017sampling, Harrow2017sampling, Zhang2022sampling}.  
To do so, we develop a hybrid Monte-Carlo/tensor-network approach, which traces out the two degrees of freedoms in different manners:
we sample the ancilla bond spins $\{s\}$ using a standard Metropolis algorithm 
but the weights of the importance sampling are computed by tracing out the site spins $\{\sigma\}$ via a tensor-network algorithm (for details of the algorithm see SM\cite{supplement}).
Despite the considerable cost of such Monte Carlo sweeps, 
this treatment  has the advantage that it effectively avoids the minima of the glassy landscape for the $\{\sigma\}$ spins in the presence of disorder
\footnote{In fact, this hybrid algorithm can be applied to more general two-dimensional measurement problems as long as the wave functions are within the projected entangled pair states manifold of area law entanglement entropy, and the numerical complexity remains classically tractable for shallow depth circuits \cite{Harrow2022}.}.

We use this method to chart out the phase diagram in Fig.~\ref{fig:fig1}c by performing calculations for three scenarios: along the Nishimori line $t_B=\pi/4$ (to validate our approach), along the diagonal line $t_B=t_A$ (with maximal coupling imbalance), as well as for a case in between with $t_B=\pi/5$. 
Along the Nishimori line,
varying the system size and analyzing the finite-size scaling of the EA order parameter as shown in Fig.~\ref{fig:fig2}, 
we can verify the existence of a true {\it critical point} at $t_A^c \approx 0.149 \pi$, 
in reasonable numerical agreement with the location of the multicritical Nishimori point established in large-scale simulations\cite{Nishimori1981, Harris1988Nishimori, Adler1996, Fisher1997, Read2000, Pujol2001, Chalker2002, Hartmann2004} of the RBIM identifying the critical point $t^c_A\approx 0.143\pi$ and $\nu\approx 4/3$.
The numerical results for the diagonal line $t_B=t_A$ and $t_B=\pi/5$ are qualitatively similar and provided in SM\cite{supplement}. 


\textit{Realization in quantum devices}.--
Vying a potential realization of our 2D cat state construction, we note that our protocol employs two basic ingredients that are readily available in current digital quantum computing platforms: 
a two-body Ising evolution and selective measurements for an extensive set of ancilla qubits on every bond. 
A particularly well-suited platform is IBM's quantum computing systems\cite{Gambetta2020, IBMQuantumCloud}, which arrange their superconducting transmon qubits in a heavy-hexagon lattice geometry -- a honeycomb variant of the square Lieb lattice,
which can be realized by a depth-3 circuit and exhibits qualitatively similar physics as discussed above, exhibiting a smaller LRE phase region (for detailed calculations see the SM\cite{supplement}). 
An important question is how to experimentally prove the successful preparation of a LRE state. 
Along the Nishimori line, a large number of distinct ancillas configurations are related by the gauge transformation such that the classical spin snapshot for one ancilla configuration can be transformed to that for another gauge equivalent configuration. 
The current chip sizes (with up to 127 qubits),
allow a brute-force approach by postselecting\cite{Minnich2022ibmMIPT} the same ancillas vortex configuration to measure the EA order, which at worst case costs $O(2^Q)$ number of operations with $Q=18$ being the number of plaquettes. Note the probability of obtaining vortex-free configuration approaches $100\%$ when $t_A\to\pi/4$.


\textit{Glassy topological order.}--
As the 2D Ising protocol measuring domain walls generates Nishimori's cat state with Ising vortex disorder, 
an analogous 3D gauge protocol that weakly measures plaquette fluxes\cite{Liu22coherentnoise} results in a glassy $Z_2$ topological order\cite{Wen2005, Castelnovo2008} with magnetic monopole disorder\cite{Preskill2002}. 
For instance, using Eq.~\eqref{eq:weak} for the four-body plaquette stabilizer $\mathcal O = B_p$ of the toric code on the cubic lattice, results in the correlations of classical 3D $Z_2$ lattice gauge theory (describing the fluctuation of magnetic flux tubes). The latter is well-known to have a finite-temperature transition\cite{Wegner71duality, Kogut79rmp} in the clean case. 
Our correlated disorder distribution is by construction (and like the 2D case) of Nishimori type, which allows us to directly relate the postmeasurement state to the solvable line of the classical 3D random plaquette $Z_2$ gauge model\cite{supplement}. This has an extended deconfined phase with a known transition\cite{Preskill2002, Preskill2003, Matsui2004} which, mapped to our time parametrization, occurs at $t_c\approx 0.192\pi$. For times beyond this critical threshold, we have stable topological order, which can be detected by the perimeter law scaling of the EA analog of the Wilson loop. We note that this protocol only (weakly) measures fluxes; gauge charges remain frozen and absent at all times. See the SM for more details, in particular, how the above solvable path can be achieved using only three-body gates\cite{supplement}.


\textit{Outlook}.--
We have demonstrated that stable LRE phases (2D cat states and 3D topological order) can be realized in fixed-depth unitaries upon relaxing strong to weak measurements. 
The key conceptual finding is that weak measurements can effectively act as a source of thermal fluctuations and correlated disorder 
that conspire to yield precisely Nishimori's critical state. 
The stability of the ordered phase in the classical model implies that the cat state is stable against generic Ising symmetric noise, a detailed study of which is left to future study. 
Unlike deep-depth random unitary circuits that feature fluctuations in the temporal dimension, 
our state exhibits criticality with fluctuations solely in space, reminiscent of projected entangled pair state \textit{wave function deformation criticality}\cite{Henley04RK, Fradkin04RK, Cirac06, Troyer10topocrit, Cirac2012, Zhu19, gmz20fibonacci, gmz20doublesemion, Zhu22fracton} effectively tuned by a deterministic circuit.

Although the focus of the present work was on stable measurement-induced LRE, we note that our mechanism can be used more generally to prepare exotic states, such as {\it deterministically} preparing phase transitions between distinct stable SRE phases in 1D\cite{Wolf06,Smith22,Jones21}, or symmetry-enriched cat states in higher dimensions\cite{pivot}; see the SM for details\cite{supplement}. More generally, it would be interesting to further explore how weak measurements can give rise to new phenomenology in monitored circuits.

We emphasize the implementability of our protocol, 
with regard to the heavy-hexagon geometry 
of the IBM transmon chips, which 
will require only a 
depth-3 circuit to bring Nishimori's cat to life.
Alternatively, Rydberg atom simulators are a highly tunable platform\cite{Browaeys2020,Lukin2021atom256, Browaeys2021} allowing for measuring ancillas\cite{Lukin2022quantumprocessor,Singh22,Zhang21}. The Ising interactions of Rydbergs on sites and bonds of a hexagonal lattice have been argued to generate the requisite unitary evolution\cite{Verresen2021cat}, making this a promising platform for realizing this transition.
While the EA order parameter can in principle be measured in a brute-force manner for current chip sizes, an important open question is whether postselection can be effectively avoided, e.g., by engineering a clever decoder for reading out hidden information\cite{Altman2020qec, Huse2020qec, Noel2022trappedionMIPT, Gullans22decoder, Vasseur2022qec}. 
To implement a minimal instance of glassy topological order via a 3D ``Nishimori code", we anticipate that a two-body Ising evolution on the Raussendorf lattice\cite{Raussendorf2006} is sufficient to give a stable toric code phase in 3D. 

\vspace{2pt}

\textit{Note added.}-- Upon completion of the present manuscript, we became aware of an independent work studying extended long-range entangled phases and transitions from finite-depth unitaries and measurement, which appeared in the same arXiv posting \cite{JYLee}.

\vspace{2pt}


\begin{acknowledgments}
\textit{Acknowledgments.--} 
We thank Ehud Altman, Zhen Bi, Max Block, Michael Buchhold, Matthew Fisher, Sam Garratt, Antoine Georges, Sarang Gopalakrishnan, Wenjie Ji, Roderich Moessner, Vadim Oganesyan, Drew Potter, Miles Stoudenmire, and Sagar Vijay for insightful discussions. 
The Cologne group acknowledges partial funding from the Deutsche Forschungsgemeinschaft (DFG, German Research Foundation) -- project grant 277101999 -- through CRC network SFB/TRR 183 (projects A04, B01).
RV is supported by the Harvard Quantum Initiative Postdoctoral Fellowship in Science and Engineering, and RV and AV by the Simons Collaboration on Ultra-Quantum Matter, which is a grant from the Simons Foundation (651440, AV).
Part of this work was performed by RV and AV at the Aspen Center for Physics, which is supported by National Science Foundation grant PHY-1607611.
NT is supported by the Walter Burke Institute for Theoretical Physics at Caltech.
The numerical simulations were performed on the JUWELS cluster at the Forschungszentrum Juelich. 
The Flatiron Institute is a division of the Simons Foundation.
\end{acknowledgments}

{\it Data availability}.-- The numerical data shown in the figures is available on Zenodo~\cite{zenodo_NishimoriCat}.


\bibliography{measurements}

\begin{thebibliography}{112}%
\makeatletter
\providecommand \@ifxundefined [1]{%
 \@ifx{#1\undefined}
}%
\providecommand \@ifnum [1]{%
 \ifnum #1\expandafter \@firstoftwo
 \else \expandafter \@secondoftwo
 \fi
}%
\providecommand \@ifx [1]{%
 \ifx #1\expandafter \@firstoftwo
 \else \expandafter \@secondoftwo
 \fi
}%
\providecommand \natexlab [1]{#1}%
\providecommand \enquote  [1]{``#1''}%
\providecommand \bibnamefont  [1]{#1}%
\providecommand \bibfnamefont [1]{#1}%
\providecommand \citenamefont [1]{#1}%
\providecommand \href@noop [0]{\@secondoftwo}%
\providecommand \href [0]{\begingroup \@sanitize@url \@href}%
\providecommand \@href[1]{\@@startlink{#1}\@@href}%
\providecommand \@@href[1]{\endgroup#1\@@endlink}%
\providecommand \@sanitize@url [0]{\catcode `\\12\catcode `\$12\catcode
  `\&12\catcode `\#12\catcode `\^12\catcode `\_12\catcode `\%12\relax}%
\providecommand \@@startlink[1]{}%
\providecommand \@@endlink[0]{}%
\providecommand \url  [0]{\begingroup\@sanitize@url \@url }%
\providecommand \@url [1]{\endgroup\@href {#1}{\urlprefix }}%
\providecommand \urlprefix  [0]{URL }%
\providecommand \Eprint [0]{\href }%
\providecommand \doibase [0]{https://doi.org/}%
\providecommand \selectlanguage [0]{\@gobble}%
\providecommand \bibinfo  [0]{\@secondoftwo}%
\providecommand \bibfield  [0]{\@secondoftwo}%
\providecommand \translation [1]{[#1]}%
\providecommand \BibitemOpen [0]{}%
\providecommand \bibitemStop [0]{}%
\providecommand \bibitemNoStop [0]{.\EOS\space}%
\providecommand \EOS [0]{\spacefactor3000\relax}%
\providecommand \BibitemShut  [1]{\csname bibitem#1\endcsname}%
\let\auto@bib@innerbib\@empty
\bibitem [{\citenamefont {Bravyi}\ \emph {et~al.}(2006)\citenamefont {Bravyi},
  \citenamefont {Hastings},\ and\ \citenamefont
  {Verstraete}}]{Verstraete2006LRB}%
  \BibitemOpen
  \bibfield  {author} {\bibinfo {author} {\bibfnamefont {S.}~\bibnamefont
  {Bravyi}}, \bibinfo {author} {\bibfnamefont {M.~B.}\ \bibnamefont
  {Hastings}},\ and\ \bibinfo {author} {\bibfnamefont {F.}~\bibnamefont
  {Verstraete}},\ }\bibfield  {title} {\bibinfo {title} {{Lieb-Robinson Bounds
  and the Generation of Correlations and Topological Quantum Order}},\ }\href
  {https://doi.org/10.1103/PhysRevLett.97.050401} {\bibfield  {journal}
  {\bibinfo  {journal} {Phys. Rev. Lett.}\ }\textbf {\bibinfo {volume} {97}},\
  \bibinfo {pages} {050401} (\bibinfo {year} {2006})}\BibitemShut {NoStop}%
\bibitem [{\citenamefont {Aguado}\ and\ \citenamefont
  {Vidal}(2008)}]{Aguado08}%
  \BibitemOpen
  \bibfield  {author} {\bibinfo {author} {\bibfnamefont {M.}~\bibnamefont
  {Aguado}}\ and\ \bibinfo {author} {\bibfnamefont {G.}~\bibnamefont {Vidal}},\
  }\bibfield  {title} {\bibinfo {title} {{Entanglement Renormalization and
  Topological Order}},\ }\href {https://doi.org/10.1103/PhysRevLett.100.070404}
  {\bibfield  {journal} {\bibinfo  {journal} {Phys. Rev. Lett.}\ }\textbf
  {\bibinfo {volume} {100}},\ \bibinfo {pages} {070404} (\bibinfo {year}
  {2008})}\BibitemShut {NoStop}%
\bibitem [{\citenamefont {K\"onig}\ \emph {et~al.}(2009)\citenamefont
  {K\"onig}, \citenamefont {Reichardt},\ and\ \citenamefont
  {Vidal}}]{Koenig09}%
  \BibitemOpen
  \bibfield  {author} {\bibinfo {author} {\bibfnamefont {R.}~\bibnamefont
  {K\"onig}}, \bibinfo {author} {\bibfnamefont {B.~W.}\ \bibnamefont
  {Reichardt}},\ and\ \bibinfo {author} {\bibfnamefont {G.}~\bibnamefont
  {Vidal}},\ }\bibfield  {title} {\bibinfo {title} {Exact entanglement
  renormalization for string-net models},\ }\href
  {https://doi.org/10.1103/PhysRevB.79.195123} {\bibfield  {journal} {\bibinfo
  {journal} {Phys. Rev. B}\ }\textbf {\bibinfo {volume} {79}},\ \bibinfo
  {pages} {195123} (\bibinfo {year} {2009})}\BibitemShut {NoStop}%
\bibitem [{\citenamefont {Chen}\ \emph {et~al.}(2010)\citenamefont {Chen},
  \citenamefont {Gu},\ and\ \citenamefont {Wen}}]{Chen2010}%
  \BibitemOpen
  \bibfield  {author} {\bibinfo {author} {\bibfnamefont {X.}~\bibnamefont
  {Chen}}, \bibinfo {author} {\bibfnamefont {Z.-C.}\ \bibnamefont {Gu}},\ and\
  \bibinfo {author} {\bibfnamefont {X.-G.}\ \bibnamefont {Wen}},\ }\bibfield
  {title} {\bibinfo {title} {{Local unitary transformation, long-range quantum
  entanglement, wave function renormalization, and topological order}},\ }\href
  {https://doi.org/10.1103/PhysRevB.82.155138} {\bibfield  {journal} {\bibinfo
  {journal} {Phys. Rev. B}\ }\textbf {\bibinfo {volume} {82}},\ \bibinfo
  {pages} {155138} (\bibinfo {year} {2010})}\BibitemShut {NoStop}%
\bibitem [{\citenamefont {Zaletel}\ and\ \citenamefont
  {Pollmann}(2020)}]{Zaletel20}%
  \BibitemOpen
  \bibfield  {author} {\bibinfo {author} {\bibfnamefont {M.~P.}\ \bibnamefont
  {Zaletel}}\ and\ \bibinfo {author} {\bibfnamefont {F.}~\bibnamefont
  {Pollmann}},\ }\bibfield  {title} {\bibinfo {title} {{Isometric Tensor
  Network States in Two Dimensions}},\ }\href
  {https://doi.org/10.1103/PhysRevLett.124.037201} {\bibfield  {journal}
  {\bibinfo  {journal} {Phys. Rev. Lett.}\ }\textbf {\bibinfo {volume} {124}},\
  \bibinfo {pages} {037201} (\bibinfo {year} {2020})}\BibitemShut {NoStop}%
\bibitem [{\citenamefont {Soejima}\ \emph {et~al.}(2020)\citenamefont
  {Soejima}, \citenamefont {Siva}, \citenamefont {Bultinck}, \citenamefont
  {Chatterjee}, \citenamefont {Pollmann},\ and\ \citenamefont
  {Zaletel}}]{Soejima20}%
  \BibitemOpen
  \bibfield  {author} {\bibinfo {author} {\bibfnamefont {T.}~\bibnamefont
  {Soejima}}, \bibinfo {author} {\bibfnamefont {K.}~\bibnamefont {Siva}},
  \bibinfo {author} {\bibfnamefont {N.}~\bibnamefont {Bultinck}}, \bibinfo
  {author} {\bibfnamefont {S.}~\bibnamefont {Chatterjee}}, \bibinfo {author}
  {\bibfnamefont {F.}~\bibnamefont {Pollmann}},\ and\ \bibinfo {author}
  {\bibfnamefont {M.~P.}\ \bibnamefont {Zaletel}},\ }\bibfield  {title}
  {\bibinfo {title} {Isometric tensor network representation of string-net
  liquids},\ }\href {https://doi.org/10.1103/PhysRevB.101.085117} {\bibfield
  {journal} {\bibinfo  {journal} {Phys. Rev. B}\ }\textbf {\bibinfo {volume}
  {101}},\ \bibinfo {pages} {085117} (\bibinfo {year} {2020})}\BibitemShut
  {NoStop}%
\bibitem [{\citenamefont {Liu}\ \emph {et~al.}(2021)\citenamefont {Liu},
  \citenamefont {Shtengel}, \citenamefont {Smith},\ and\ \citenamefont
  {Pollmann}}]{ShtengelPollmann}%
  \BibitemOpen
  \bibfield  {author} {\bibinfo {author} {\bibfnamefont {Y.-J.}\ \bibnamefont
  {Liu}}, \bibinfo {author} {\bibfnamefont {K.}~\bibnamefont {Shtengel}},
  \bibinfo {author} {\bibfnamefont {A.}~\bibnamefont {Smith}},\ and\ \bibinfo
  {author} {\bibfnamefont {F.}~\bibnamefont {Pollmann}},\ }\bibfield  {title}
  {\bibinfo {title} {Methods for simulating string-net states and anyons on a
  digital quantum computer},\ }\href@noop {} {\  (\bibinfo {year} {2021})},\
  \Eprint {https://arxiv.org/abs/2110.02020} {arXiv:2110.02020} \BibitemShut
  {NoStop}%
\bibitem [{\citenamefont {Wei}\ \emph {et~al.}(2022)\citenamefont {Wei},
  \citenamefont {Malz},\ and\ \citenamefont {Cirac}}]{Wei22}%
  \BibitemOpen
  \bibfield  {author} {\bibinfo {author} {\bibfnamefont {Z.-Y.}\ \bibnamefont
  {Wei}}, \bibinfo {author} {\bibfnamefont {D.}~\bibnamefont {Malz}},\ and\
  \bibinfo {author} {\bibfnamefont {J.~I.}\ \bibnamefont {Cirac}},\ }\bibfield
  {title} {\bibinfo {title} {{Sequential Generation of Projected Entangled-Pair
  States}},\ }\href {https://doi.org/10.1103/PhysRevLett.128.010607} {\bibfield
   {journal} {\bibinfo  {journal} {Phys. Rev. Lett.}\ }\textbf {\bibinfo
  {volume} {128}},\ \bibinfo {pages} {010607} (\bibinfo {year}
  {2022})}\BibitemShut {NoStop}%
\bibitem [{\citenamefont {Wei}\ \emph {et~al.}(2020)\citenamefont {Wei},
  \citenamefont {Lauer}, \citenamefont {Srinivasan}, \citenamefont
  {Sundaresan}, \citenamefont {McClure}, \citenamefont {Toyli}, \citenamefont
  {McKay}, \citenamefont {Gambetta},\ and\ \citenamefont
  {Sheldon}}]{Wei2020GHZ}%
  \BibitemOpen
  \bibfield  {author} {\bibinfo {author} {\bibfnamefont {K.~X.}\ \bibnamefont
  {Wei}}, \bibinfo {author} {\bibfnamefont {I.}~\bibnamefont {Lauer}}, \bibinfo
  {author} {\bibfnamefont {S.}~\bibnamefont {Srinivasan}}, \bibinfo {author}
  {\bibfnamefont {N.}~\bibnamefont {Sundaresan}}, \bibinfo {author}
  {\bibfnamefont {D.~T.}\ \bibnamefont {McClure}}, \bibinfo {author}
  {\bibfnamefont {D.}~\bibnamefont {Toyli}}, \bibinfo {author} {\bibfnamefont
  {D.~C.}\ \bibnamefont {McKay}}, \bibinfo {author} {\bibfnamefont {J.~M.}\
  \bibnamefont {Gambetta}},\ and\ \bibinfo {author} {\bibfnamefont
  {S.}~\bibnamefont {Sheldon}},\ }\bibfield  {title} {\bibinfo {title}
  {Verifying multipartite entangled greenberger-horne-zeilinger states via
  multiple quantum coherences},\ }\href
  {https://doi.org/10.1103/PhysRevA.101.032343} {\bibfield  {journal} {\bibinfo
   {journal} {Phys. Rev. A}\ }\textbf {\bibinfo {volume} {101}},\ \bibinfo
  {pages} {032343} (\bibinfo {year} {2020})}\BibitemShut {NoStop}%
\bibitem [{\citenamefont {Mooney}\ \emph {et~al.}(2021)\citenamefont {Mooney},
  \citenamefont {White}, \citenamefont {Hill},\ and\ \citenamefont
  {Hollenberg}}]{Mooney2021GHZ}%
  \BibitemOpen
  \bibfield  {author} {\bibinfo {author} {\bibfnamefont {G.~J.}\ \bibnamefont
  {Mooney}}, \bibinfo {author} {\bibfnamefont {G.~A.~L.}\ \bibnamefont
  {White}}, \bibinfo {author} {\bibfnamefont {C.~D.}\ \bibnamefont {Hill}},\
  and\ \bibinfo {author} {\bibfnamefont {L.~C.~L.}\ \bibnamefont
  {Hollenberg}},\ }\bibfield  {title} {\bibinfo {title} {Generation and
  verification of 27-qubit greenberger-horne-zeilinger states in a
  superconducting quantum computer},\ }\href
  {https://doi.org/10.1088/2399-6528/ac1df7} {\bibfield  {journal} {\bibinfo
  {journal} {Journal of Physics Communications}\ }\textbf {\bibinfo {volume}
  {5}},\ \bibinfo {pages} {095004} (\bibinfo {year} {2021})}\BibitemShut
  {NoStop}%
\bibitem [{\citenamefont {Gottesman}(1997)}]{Gottesman97}%
  \BibitemOpen
  \bibfield  {author} {\bibinfo {author} {\bibfnamefont {D.}~\bibnamefont
  {Gottesman}},\ }\href {https://doi.org/10.48550/ARXIV.QUANT-PH/9705052}
  {\bibinfo {title} {{Stabilizer Codes and Quantum Error Correction}}}
  (\bibinfo {year} {1997}),\ \Eprint {https://arxiv.org/abs/9705052}
  {arXiv.quant-phys:9705052} \BibitemShut {NoStop}%
\bibitem [{\citenamefont {Briegel}\ and\ \citenamefont
  {Raussendorf}(2001)}]{Raussendorf2001}%
  \BibitemOpen
  \bibfield  {author} {\bibinfo {author} {\bibfnamefont {H.~J.}\ \bibnamefont
  {Briegel}}\ and\ \bibinfo {author} {\bibfnamefont {R.}~\bibnamefont
  {Raussendorf}},\ }\bibfield  {title} {\bibinfo {title} {{Persistent
  Entanglement in Arrays of Interacting Particles}},\ }\href
  {https://doi.org/10.1103/PhysRevLett.86.910} {\bibfield  {journal} {\bibinfo
  {journal} {Phys. Rev. Lett.}\ }\textbf {\bibinfo {volume} {86}},\ \bibinfo
  {pages} {910} (\bibinfo {year} {2001})}\BibitemShut {NoStop}%
\bibitem [{\citenamefont {Raussendorf}\ \emph {et~al.}(2005)\citenamefont
  {Raussendorf}, \citenamefont {Bravyi},\ and\ \citenamefont
  {Harrington}}]{Raussendorf2005}%
  \BibitemOpen
  \bibfield  {author} {\bibinfo {author} {\bibfnamefont {R.}~\bibnamefont
  {Raussendorf}}, \bibinfo {author} {\bibfnamefont {S.}~\bibnamefont
  {Bravyi}},\ and\ \bibinfo {author} {\bibfnamefont {J.}~\bibnamefont
  {Harrington}},\ }\bibfield  {title} {\bibinfo {title} {Long-range quantum
  entanglement in noisy cluster states},\ }\href
  {https://doi.org/10.1103/PhysRevA.71.062313} {\bibfield  {journal} {\bibinfo
  {journal} {Phys. Rev. A}\ }\textbf {\bibinfo {volume} {71}},\ \bibinfo
  {pages} {062313} (\bibinfo {year} {2005})}\BibitemShut {NoStop}%
\bibitem [{\citenamefont {Aguado}\ \emph {et~al.}(2008)\citenamefont {Aguado},
  \citenamefont {Brennen}, \citenamefont {Verstraete},\ and\ \citenamefont
  {Cirac}}]{Aguado08b}%
  \BibitemOpen
  \bibfield  {author} {\bibinfo {author} {\bibfnamefont {M.}~\bibnamefont
  {Aguado}}, \bibinfo {author} {\bibfnamefont {G.~K.}\ \bibnamefont {Brennen}},
  \bibinfo {author} {\bibfnamefont {F.}~\bibnamefont {Verstraete}},\ and\
  \bibinfo {author} {\bibfnamefont {J.~I.}\ \bibnamefont {Cirac}},\ }\bibfield
  {title} {\bibinfo {title} {{Creation, Manipulation, and Detection of Abelian
  and Non-Abelian Anyons in Optical Lattices}},\ }\href
  {https://doi.org/10.1103/PhysRevLett.101.260501} {\bibfield  {journal}
  {\bibinfo  {journal} {Phys. Rev. Lett.}\ }\textbf {\bibinfo {volume} {101}},\
  \bibinfo {pages} {260501} (\bibinfo {year} {2008})}\BibitemShut {NoStop}%
\bibitem [{\citenamefont {Brennen}\ \emph {et~al.}(2009)\citenamefont
  {Brennen}, \citenamefont {Aguado},\ and\ \citenamefont {Cirac}}]{Brennen09}%
  \BibitemOpen
  \bibfield  {author} {\bibinfo {author} {\bibfnamefont {G.~K.}\ \bibnamefont
  {Brennen}}, \bibinfo {author} {\bibfnamefont {M.}~\bibnamefont {Aguado}},\
  and\ \bibinfo {author} {\bibfnamefont {J.~I.}\ \bibnamefont {Cirac}},\
  }\bibfield  {title} {\bibinfo {title} {Simulations of quantum double
  models},\ }\href {https://doi.org/10.1088/1367-2630/11/5/053009} {\bibfield
  {journal} {\bibinfo  {journal} {New Journal of Physics}\ }\textbf {\bibinfo
  {volume} {11}},\ \bibinfo {pages} {053009} (\bibinfo {year}
  {2009})}\BibitemShut {NoStop}%
\bibitem [{\citenamefont {Bolt}\ \emph {et~al.}(2016)\citenamefont {Bolt},
  \citenamefont {Duclos-Cianci}, \citenamefont {Poulin},\ and\ \citenamefont
  {Stace}}]{Stace2016}%
  \BibitemOpen
  \bibfield  {author} {\bibinfo {author} {\bibfnamefont {A.}~\bibnamefont
  {Bolt}}, \bibinfo {author} {\bibfnamefont {G.}~\bibnamefont {Duclos-Cianci}},
  \bibinfo {author} {\bibfnamefont {D.}~\bibnamefont {Poulin}},\ and\ \bibinfo
  {author} {\bibfnamefont {T.~M.}\ \bibnamefont {Stace}},\ }\bibfield  {title}
  {\bibinfo {title} {{Foliated Quantum Error-Correcting Codes}},\ }\href
  {https://doi.org/10.1103/PhysRevLett.117.070501} {\bibfield  {journal}
  {\bibinfo  {journal} {Phys. Rev. Lett.}\ }\textbf {\bibinfo {volume} {117}},\
  \bibinfo {pages} {070501} (\bibinfo {year} {2016})}\BibitemShut {NoStop}%
\bibitem [{\citenamefont {Piroli}\ \emph {et~al.}(2021)\citenamefont {Piroli},
  \citenamefont {Styliaris},\ and\ \citenamefont {Cirac}}]{Piroli21}%
  \BibitemOpen
  \bibfield  {author} {\bibinfo {author} {\bibfnamefont {L.}~\bibnamefont
  {Piroli}}, \bibinfo {author} {\bibfnamefont {G.}~\bibnamefont {Styliaris}},\
  and\ \bibinfo {author} {\bibfnamefont {J.~I.}\ \bibnamefont {Cirac}},\
  }\bibfield  {title} {\bibinfo {title} {{Quantum Circuits Assisted by Local
  Operations and Classical Communication: Transformations and Phases of
  Matter}},\ }\href {https://doi.org/10.1103/PhysRevLett.127.220503} {\bibfield
   {journal} {\bibinfo  {journal} {Phys. Rev. Lett.}\ }\textbf {\bibinfo
  {volume} {127}},\ \bibinfo {pages} {220503} (\bibinfo {year}
  {2021})}\BibitemShut {NoStop}%
\bibitem [{\citenamefont {{Friedman}}\ \emph {et~al.}()\citenamefont
  {{Friedman}}, \citenamefont {{Yin}}, \citenamefont {{Hong}},\ and\
  \citenamefont {{Lucas}}}]{Lucas22measurement}%
  \BibitemOpen
  \bibfield  {author} {\bibinfo {author} {\bibfnamefont {A.~J.}\ \bibnamefont
  {{Friedman}}}, \bibinfo {author} {\bibfnamefont {C.}~\bibnamefont {{Yin}}},
  \bibinfo {author} {\bibfnamefont {Y.}~\bibnamefont {{Hong}}},\ and\ \bibinfo
  {author} {\bibfnamefont {A.}~\bibnamefont {{Lucas}}},\ }\bibfield  {title}
  {\bibinfo {title} {{Locality and error correction in quantum dynamics with
  measurement}},\ }\href@noop {} {\ }\Eprint {https://arxiv.org/abs/2206.09929}
  {arXiv:2206.09929} \BibitemShut {NoStop}%
\bibitem [{\citenamefont {Gottesman}\ \emph {et~al.}(2001)\citenamefont
  {Gottesman}, \citenamefont {Kitaev},\ and\ \citenamefont
  {Preskill}}]{Gottesman2001}%
  \BibitemOpen
  \bibfield  {author} {\bibinfo {author} {\bibfnamefont {D.}~\bibnamefont
  {Gottesman}}, \bibinfo {author} {\bibfnamefont {A.}~\bibnamefont {Kitaev}},\
  and\ \bibinfo {author} {\bibfnamefont {J.}~\bibnamefont {Preskill}},\
  }\bibfield  {title} {\bibinfo {title} {{Encoding a qubit in an oscillator}},\
  }\href {https://doi.org/10.1103/PhysRevA.64.012310} {\bibfield  {journal}
  {\bibinfo  {journal} {Phys. Rev. A}\ }\textbf {\bibinfo {volume} {64}},\
  \bibinfo {pages} {012310} (\bibinfo {year} {2001})}\BibitemShut {NoStop}%
\bibitem [{\citenamefont {Dennis}\ \emph {et~al.}(2002)\citenamefont {Dennis},
  \citenamefont {Kitaev}, \citenamefont {Landahl},\ and\ \citenamefont
  {Preskill}}]{Preskill2002}%
  \BibitemOpen
  \bibfield  {author} {\bibinfo {author} {\bibfnamefont {E.}~\bibnamefont
  {Dennis}}, \bibinfo {author} {\bibfnamefont {A.}~\bibnamefont {Kitaev}},
  \bibinfo {author} {\bibfnamefont {A.}~\bibnamefont {Landahl}},\ and\ \bibinfo
  {author} {\bibfnamefont {J.}~\bibnamefont {Preskill}},\ }\bibfield  {title}
  {\bibinfo {title} {Topological quantum memory},\ }\href
  {https://doi.org/10.1063/1.1499754} {\bibfield  {journal} {\bibinfo
  {journal} {Journal of Mathematical Physics}\ }\textbf {\bibinfo {volume}
  {43}},\ \bibinfo {pages} {4452} (\bibinfo {year} {2002})}\BibitemShut
  {NoStop}%
\bibitem [{\citenamefont {Verresen}\ \emph {et~al.}(2021)\citenamefont
  {Verresen}, \citenamefont {Tantivasadakarn},\ and\ \citenamefont
  {Vishwanath}}]{Verresen2021cat}%
  \BibitemOpen
  \bibfield  {author} {\bibinfo {author} {\bibfnamefont {R.}~\bibnamefont
  {Verresen}}, \bibinfo {author} {\bibfnamefont {N.}~\bibnamefont
  {Tantivasadakarn}},\ and\ \bibinfo {author} {\bibfnamefont {A.}~\bibnamefont
  {Vishwanath}},\ }\bibfield  {title} {\bibinfo {title} {{Efficiently preparing
  Schr{\"o}dinger’s cat, fractons and non-Abelian topological order in
  quantum devices}},\ }\href@noop {} {\  (\bibinfo {year} {2021})},\ \Eprint
  {https://arxiv.org/abs/2112.03061} {arXiv:2112.03061} \BibitemShut {NoStop}%
\bibitem [{\citenamefont {Tantivasadakarn}\ \emph
  {et~al.}(2021{\natexlab{a}})\citenamefont {Tantivasadakarn}, \citenamefont
  {Thorngren}, \citenamefont {Vishwanath},\ and\ \citenamefont
  {Verresen}}]{Nat2021measure}%
  \BibitemOpen
  \bibfield  {author} {\bibinfo {author} {\bibfnamefont {N.}~\bibnamefont
  {Tantivasadakarn}}, \bibinfo {author} {\bibfnamefont {R.}~\bibnamefont
  {Thorngren}}, \bibinfo {author} {\bibfnamefont {A.}~\bibnamefont
  {Vishwanath}},\ and\ \bibinfo {author} {\bibfnamefont {R.}~\bibnamefont
  {Verresen}},\ }\bibfield  {title} {\bibinfo {title} {Long-range entanglement
  from measuring symmetry-protected topological phases},\ }\href@noop {} {\
  (\bibinfo {year} {2021}{\natexlab{a}})},\ \Eprint
  {https://arxiv.org/abs/2112.01519} {arXiv:2112.01519} \BibitemShut {NoStop}%
\bibitem [{\citenamefont {Bravyi}\ \emph {et~al.}(2022)\citenamefont {Bravyi},
  \citenamefont {Kim}, \citenamefont {Kliesch},\ and\ \citenamefont
  {Koenig}}]{Bravyi2022}%
  \BibitemOpen
  \bibfield  {author} {\bibinfo {author} {\bibfnamefont {S.}~\bibnamefont
  {Bravyi}}, \bibinfo {author} {\bibfnamefont {I.}~\bibnamefont {Kim}},
  \bibinfo {author} {\bibfnamefont {A.}~\bibnamefont {Kliesch}},\ and\ \bibinfo
  {author} {\bibfnamefont {R.}~\bibnamefont {Koenig}},\ }\bibfield  {title}
  {\bibinfo {title} {Adaptive constant-depth circuits for manipulating
  non-abelian anyons},\ }\href@noop {} {\  (\bibinfo {year} {2022})},\ \Eprint
  {https://arxiv.org/abs/2205.01933} {arXiv:2205.01933} \BibitemShut {NoStop}%
\bibitem [{\citenamefont {Lu}\ \emph {et~al.}(2022)\citenamefont {Lu},
  \citenamefont {Lessa}, \citenamefont {Kim},\ and\ \citenamefont
  {Hsieh}}]{Hsieh2022}%
  \BibitemOpen
  \bibfield  {author} {\bibinfo {author} {\bibfnamefont {T.-C.}\ \bibnamefont
  {Lu}}, \bibinfo {author} {\bibfnamefont {L.~A.}\ \bibnamefont {Lessa}},
  \bibinfo {author} {\bibfnamefont {I.~H.}\ \bibnamefont {Kim}},\ and\ \bibinfo
  {author} {\bibfnamefont {T.~H.}\ \bibnamefont {Hsieh}},\ }\bibfield  {title}
  {\bibinfo {title} {Measurement as a shortcut to long-range entangled quantum
  matter},\ }\href@noop {} {\  (\bibinfo {year} {2022})},\ \Eprint
  {https://arxiv.org/abs/2206.13527} {arXiv:2206.13527} \BibitemShut {NoStop}%
\bibitem [{\citenamefont {Altman}\ \emph {et~al.}(2021)\citenamefont {Altman},
  \citenamefont {Brown}, \citenamefont {Carleo}, \citenamefont {Carr},
  \citenamefont {Demler}, \citenamefont {Chin}, \citenamefont {DeMarco},
  \citenamefont {Economou}, \citenamefont {Eriksson}, \citenamefont {Fu},
  \citenamefont {Greiner}, \citenamefont {Hazzard}, \citenamefont {Hulet},
  \citenamefont {Koll\'ar}, \citenamefont {Lev}, \citenamefont {Lukin},
  \citenamefont {Ma}, \citenamefont {Mi}, \citenamefont {Misra}, \citenamefont
  {Monroe}, \citenamefont {Murch}, \citenamefont {Nazario}, \citenamefont {Ni},
  \citenamefont {Potter}, \citenamefont {Roushan}, \citenamefont {Saffman},
  \citenamefont {Schleier-Smith}, \citenamefont {Siddiqi}, \citenamefont
  {Simmonds}, \citenamefont {Singh}, \citenamefont {Spielman}, \citenamefont
  {Temme}, \citenamefont {Weiss}, \citenamefont {Vu\ifmmode \check{c}\else
  \v{c}\fi{}kovi\ifmmode~\acute{c}\else \'{c}\fi{}}, \citenamefont
  {Vuleti\ifmmode~\acute{c}\else \'{c}\fi{}}, \citenamefont {Ye},\ and\
  \citenamefont {Zwierlein}}]{Altman2021_simulators}%
  \BibitemOpen
  \bibfield  {author} {\bibinfo {author} {\bibfnamefont {E.}~\bibnamefont
  {Altman}}, \bibinfo {author} {\bibfnamefont {K.~R.}\ \bibnamefont {Brown}},
  \bibinfo {author} {\bibfnamefont {G.}~\bibnamefont {Carleo}}, \bibinfo
  {author} {\bibfnamefont {L.~D.}\ \bibnamefont {Carr}}, \bibinfo {author}
  {\bibfnamefont {E.}~\bibnamefont {Demler}}, \bibinfo {author} {\bibfnamefont
  {C.}~\bibnamefont {Chin}}, \bibinfo {author} {\bibfnamefont {B.}~\bibnamefont
  {DeMarco}}, \bibinfo {author} {\bibfnamefont {S.~E.}\ \bibnamefont
  {Economou}}, \bibinfo {author} {\bibfnamefont {M.~A.}\ \bibnamefont
  {Eriksson}}, \bibinfo {author} {\bibfnamefont {K.-M.~C.}\ \bibnamefont {Fu}},
  \bibinfo {author} {\bibfnamefont {M.}~\bibnamefont {Greiner}}, \bibinfo
  {author} {\bibfnamefont {K.~R.}\ \bibnamefont {Hazzard}}, \bibinfo {author}
  {\bibfnamefont {R.~G.}\ \bibnamefont {Hulet}}, \bibinfo {author}
  {\bibfnamefont {A.~J.}\ \bibnamefont {Koll\'ar}}, \bibinfo {author}
  {\bibfnamefont {B.~L.}\ \bibnamefont {Lev}}, \bibinfo {author} {\bibfnamefont
  {M.~D.}\ \bibnamefont {Lukin}}, \bibinfo {author} {\bibfnamefont
  {R.}~\bibnamefont {Ma}}, \bibinfo {author} {\bibfnamefont {X.}~\bibnamefont
  {Mi}}, \bibinfo {author} {\bibfnamefont {S.}~\bibnamefont {Misra}}, \bibinfo
  {author} {\bibfnamefont {C.}~\bibnamefont {Monroe}}, \bibinfo {author}
  {\bibfnamefont {K.}~\bibnamefont {Murch}}, \bibinfo {author} {\bibfnamefont
  {Z.}~\bibnamefont {Nazario}}, \bibinfo {author} {\bibfnamefont {K.-K.}\
  \bibnamefont {Ni}}, \bibinfo {author} {\bibfnamefont {A.~C.}\ \bibnamefont
  {Potter}}, \bibinfo {author} {\bibfnamefont {P.}~\bibnamefont {Roushan}},
  \bibinfo {author} {\bibfnamefont {M.}~\bibnamefont {Saffman}}, \bibinfo
  {author} {\bibfnamefont {M.}~\bibnamefont {Schleier-Smith}}, \bibinfo
  {author} {\bibfnamefont {I.}~\bibnamefont {Siddiqi}}, \bibinfo {author}
  {\bibfnamefont {R.}~\bibnamefont {Simmonds}}, \bibinfo {author}
  {\bibfnamefont {M.}~\bibnamefont {Singh}}, \bibinfo {author} {\bibfnamefont
  {I.}~\bibnamefont {Spielman}}, \bibinfo {author} {\bibfnamefont
  {K.}~\bibnamefont {Temme}}, \bibinfo {author} {\bibfnamefont {D.~S.}\
  \bibnamefont {Weiss}}, \bibinfo {author} {\bibfnamefont {J.}~\bibnamefont
  {Vu\ifmmode \check{c}\else \v{c}\fi{}kovi\ifmmode~\acute{c}\else
  \'{c}\fi{}}}, \bibinfo {author} {\bibfnamefont {V.}~\bibnamefont
  {Vuleti\ifmmode~\acute{c}\else \'{c}\fi{}}}, \bibinfo {author} {\bibfnamefont
  {J.}~\bibnamefont {Ye}},\ and\ \bibinfo {author} {\bibfnamefont
  {M.}~\bibnamefont {Zwierlein}},\ }\bibfield  {title} {\bibinfo {title}
  {Quantum simulators: Architectures and opportunities},\ }\href
  {https://doi.org/10.1103/PRXQuantum.2.017003} {\bibfield  {journal} {\bibinfo
   {journal} {PRX Quantum}\ }\textbf {\bibinfo {volume} {2}},\ \bibinfo {pages}
  {017003} (\bibinfo {year} {2021})}\BibitemShut {NoStop}%
\bibitem [{\citenamefont {{Córcoles}}\ \emph {et~al.}(2020)\citenamefont
  {{Córcoles}}, \citenamefont {{Kandala}}, \citenamefont {{Javadi-Abhari}},
  \citenamefont {{McClure}}, \citenamefont {{Cross}}, \citenamefont {{Temme}},
  \citenamefont {{Nation}}, \citenamefont {{Steffen}},\ and\ \citenamefont
  {{Gambetta}}}]{Gambetta2020}%
  \BibitemOpen
  \bibfield  {author} {\bibinfo {author} {\bibfnamefont {A.~D.}\ \bibnamefont
  {{Córcoles}}}, \bibinfo {author} {\bibfnamefont {A.}~\bibnamefont
  {{Kandala}}}, \bibinfo {author} {\bibfnamefont {A.}~\bibnamefont
  {{Javadi-Abhari}}}, \bibinfo {author} {\bibfnamefont {D.~T.}\ \bibnamefont
  {{McClure}}}, \bibinfo {author} {\bibfnamefont {A.~W.}\ \bibnamefont
  {{Cross}}}, \bibinfo {author} {\bibfnamefont {K.}~\bibnamefont {{Temme}}},
  \bibinfo {author} {\bibfnamefont {P.~D.}\ \bibnamefont {{Nation}}}, \bibinfo
  {author} {\bibfnamefont {M.}~\bibnamefont {{Steffen}}},\ and\ \bibinfo
  {author} {\bibfnamefont {J.~M.}\ \bibnamefont {{Gambetta}}},\ }\bibfield
  {title} {\bibinfo {title} {{Challenges and Opportunities of Near-Term Quantum
  Computing Systems}},\ }\href {https://doi.org/10.1109/JPROC.2019.2954005}
  {\bibfield  {journal} {\bibinfo  {journal} {Proceedings of the IEEE}\
  }\textbf {\bibinfo {volume} {108}},\ \bibinfo {pages} {1338} (\bibinfo {year}
  {2020})}\BibitemShut {NoStop}%
\bibitem [{\citenamefont {Potter}\ and\ \citenamefont
  {Vasseur}(2021)}]{Potter21review}%
  \BibitemOpen
  \bibfield  {author} {\bibinfo {author} {\bibfnamefont {A.~C.}\ \bibnamefont
  {Potter}}\ and\ \bibinfo {author} {\bibfnamefont {R.}~\bibnamefont
  {Vasseur}},\ }\href {https://doi.org/10.48550/ARXIV.2111.08018} {\bibinfo
  {title} {Entanglement dynamics in hybrid quantum circuits}} (\bibinfo {year}
  {2021})\BibitemShut {NoStop}%
\bibitem [{\citenamefont {Fisher}\ \emph {et~al.}(2022)\citenamefont {Fisher},
  \citenamefont {Khemani}, \citenamefont {Nahum},\ and\ \citenamefont
  {Vijay}}]{Fisher2022reviewMIPT}%
  \BibitemOpen
  \bibfield  {author} {\bibinfo {author} {\bibfnamefont {M.~P.~A.}\
  \bibnamefont {Fisher}}, \bibinfo {author} {\bibfnamefont {V.}~\bibnamefont
  {Khemani}}, \bibinfo {author} {\bibfnamefont {A.}~\bibnamefont {Nahum}},\
  and\ \bibinfo {author} {\bibfnamefont {S.}~\bibnamefont {Vijay}},\ }\bibfield
   {title} {\bibinfo {title} {{Random Quantum Circuits}},\ }\href@noop {} {\
  (\bibinfo {year} {2022})},\ \Eprint {https://arxiv.org/abs/2207.14280}
  {arXiv:2207.14280} \BibitemShut {NoStop}%
\bibitem [{\citenamefont {Li}\ \emph {et~al.}(2018)\citenamefont {Li},
  \citenamefont {Chen},\ and\ \citenamefont {Fisher}}]{Li2018}%
  \BibitemOpen
  \bibfield  {author} {\bibinfo {author} {\bibfnamefont {Y.}~\bibnamefont
  {Li}}, \bibinfo {author} {\bibfnamefont {X.}~\bibnamefont {Chen}},\ and\
  \bibinfo {author} {\bibfnamefont {M.~P.~A.}\ \bibnamefont {Fisher}},\
  }\bibfield  {title} {\bibinfo {title} {{Quantum Zeno effect and the many-body
  entanglement transition}},\ }\href
  {https://doi.org/10.1103/PhysRevB.98.205136} {\bibfield  {journal} {\bibinfo
  {journal} {Phys. Rev. B}\ }\textbf {\bibinfo {volume} {98}},\ \bibinfo
  {pages} {205136} (\bibinfo {year} {2018})}\BibitemShut {NoStop}%
\bibitem [{\citenamefont {Skinner}\ \emph {et~al.}(2019)\citenamefont
  {Skinner}, \citenamefont {Ruhman},\ and\ \citenamefont
  {Nahum}}]{Skinner2019}%
  \BibitemOpen
  \bibfield  {author} {\bibinfo {author} {\bibfnamefont {B.}~\bibnamefont
  {Skinner}}, \bibinfo {author} {\bibfnamefont {J.}~\bibnamefont {Ruhman}},\
  and\ \bibinfo {author} {\bibfnamefont {A.}~\bibnamefont {Nahum}},\ }\bibfield
   {title} {\bibinfo {title} {{Measurement-Induced Phase Transitions in the
  Dynamics of Entanglement}},\ }\href
  {https://doi.org/10.1103/PhysRevX.9.031009} {\bibfield  {journal} {\bibinfo
  {journal} {Phys. Rev. X}\ }\textbf {\bibinfo {volume} {9}},\ \bibinfo {pages}
  {031009} (\bibinfo {year} {2019})}\BibitemShut {NoStop}%
\bibitem [{\citenamefont {{Lavasani}}\ \emph {et~al.}(2021)\citenamefont
  {{Lavasani}}, \citenamefont {{Alavirad}},\ and\ \citenamefont
  {{Barkeshli}}}]{Barkeshli2021measure}%
  \BibitemOpen
  \bibfield  {author} {\bibinfo {author} {\bibfnamefont {A.}~\bibnamefont
  {{Lavasani}}}, \bibinfo {author} {\bibfnamefont {Y.}~\bibnamefont
  {{Alavirad}}},\ and\ \bibinfo {author} {\bibfnamefont {M.}~\bibnamefont
  {{Barkeshli}}},\ }\bibfield  {title} {\bibinfo {title} {{Measurement-induced
  topological entanglement transitions in symmetric random quantum circuits}},\
  }\href {https://doi.org/10.1038/s41567-020-01112-z} {\bibfield  {journal}
  {\bibinfo  {journal} {Nature Physics}\ }\textbf {\bibinfo {volume} {17}},\
  \bibinfo {pages} {342} (\bibinfo {year} {2021})}\BibitemShut {NoStop}%
\bibitem [{\citenamefont {Lavasani}\ \emph {et~al.}(2021)\citenamefont
  {Lavasani}, \citenamefont {Alavirad},\ and\ \citenamefont
  {Barkeshli}}]{Barkeshli2021measuretoriccode}%
  \BibitemOpen
  \bibfield  {author} {\bibinfo {author} {\bibfnamefont {A.}~\bibnamefont
  {Lavasani}}, \bibinfo {author} {\bibfnamefont {Y.}~\bibnamefont {Alavirad}},\
  and\ \bibinfo {author} {\bibfnamefont {M.}~\bibnamefont {Barkeshli}},\
  }\bibfield  {title} {\bibinfo {title} {Topological order and criticality in
  $(2+1)\mathrm{D}$ monitored random quantum circuits},\ }\href
  {https://doi.org/10.1103/PhysRevLett.127.235701} {\bibfield  {journal}
  {\bibinfo  {journal} {Phys. Rev. Lett.}\ }\textbf {\bibinfo {volume} {127}},\
  \bibinfo {pages} {235701} (\bibinfo {year} {2021})}\BibitemShut {NoStop}%
\bibitem [{\citenamefont {Sang}\ and\ \citenamefont
  {Hsieh}(2021)}]{Hsieh2021measure}%
  \BibitemOpen
  \bibfield  {author} {\bibinfo {author} {\bibfnamefont {S.}~\bibnamefont
  {Sang}}\ and\ \bibinfo {author} {\bibfnamefont {T.~H.}\ \bibnamefont
  {Hsieh}},\ }\bibfield  {title} {\bibinfo {title} {{Measurement-protected
  quantum phases}},\ }\href {https://doi.org/10.1103/PhysRevResearch.3.023200}
  {\bibfield  {journal} {\bibinfo  {journal} {Phys. Rev. Research}\ }\textbf
  {\bibinfo {volume} {3}},\ \bibinfo {pages} {023200} (\bibinfo {year}
  {2021})}\BibitemShut {NoStop}%
\bibitem [{\citenamefont {{Klocke}}\ and\ \citenamefont
  {{Buchhold}}(2022)}]{Buchold22measurespt}%
  \BibitemOpen
  \bibfield  {author} {\bibinfo {author} {\bibfnamefont {K.}~\bibnamefont
  {{Klocke}}}\ and\ \bibinfo {author} {\bibfnamefont {M.}~\bibnamefont
  {{Buchhold}}},\ }\bibfield  {title} {\bibinfo {title} {{Topological order and
  entanglement dynamics in the measurement-only XZZX quantum code}},\
  }\href@noop {} {\  (\bibinfo {year} {2022})},\ \Eprint
  {https://arxiv.org/abs/2204.08489} {arXiv:2204.08489} \BibitemShut {NoStop}%
\bibitem [{\citenamefont {Lavasani}\ \emph {et~al.}(2022)\citenamefont
  {Lavasani}, \citenamefont {Luo},\ and\ \citenamefont {Vijay}}]{Vijay22}%
  \BibitemOpen
  \bibfield  {author} {\bibinfo {author} {\bibfnamefont {A.}~\bibnamefont
  {Lavasani}}, \bibinfo {author} {\bibfnamefont {Z.-X.}\ \bibnamefont {Luo}},\
  and\ \bibinfo {author} {\bibfnamefont {S.}~\bibnamefont {Vijay}},\ }\bibfield
   {title} {\bibinfo {title} {{Monitored Quantum Dynamics and the Kitaev Spin
  Liquid}},\ }\href@noop {} {\  (\bibinfo {year} {2022})},\ \Eprint
  {https://arxiv.org/abs/2207.02877} {arXiv:2207.02877} \BibitemShut {NoStop}%
\bibitem [{\citenamefont {Sriram}\ \emph {et~al.}(2022)\citenamefont {Sriram},
  \citenamefont {Rakovszky}, \citenamefont {Khemani},\ and\ \citenamefont
  {Ippoliti}}]{Ippoliti22}%
  \BibitemOpen
  \bibfield  {author} {\bibinfo {author} {\bibfnamefont {A.}~\bibnamefont
  {Sriram}}, \bibinfo {author} {\bibfnamefont {T.}~\bibnamefont {Rakovszky}},
  \bibinfo {author} {\bibfnamefont {V.}~\bibnamefont {Khemani}},\ and\ \bibinfo
  {author} {\bibfnamefont {M.}~\bibnamefont {Ippoliti}},\ }\bibfield  {title}
  {\bibinfo {title} {{Topology, criticality, and dynamically generated qubits
  in a stochastic measurement-only Kitaev model}},\ }\href@noop {} {\
  (\bibinfo {year} {2022})},\ \Eprint {https://arxiv.org/abs/2207.07096}
  {arXiv:2207.07096} \BibitemShut {NoStop}%
\bibitem [{\citenamefont {Szyniszewski}\ \emph {et~al.}(2019)\citenamefont
  {Szyniszewski}, \citenamefont {Romito},\ and\ \citenamefont
  {Schomerus}}]{Schomerus2019}%
  \BibitemOpen
  \bibfield  {author} {\bibinfo {author} {\bibfnamefont {M.}~\bibnamefont
  {Szyniszewski}}, \bibinfo {author} {\bibfnamefont {A.}~\bibnamefont
  {Romito}},\ and\ \bibinfo {author} {\bibfnamefont {H.}~\bibnamefont
  {Schomerus}},\ }\bibfield  {title} {\bibinfo {title} {Entanglement transition
  from variable-strength weak measurements},\ }\href
  {https://doi.org/10.1103/PhysRevB.100.064204} {\bibfield  {journal} {\bibinfo
   {journal} {Phys. Rev. B}\ }\textbf {\bibinfo {volume} {100}},\ \bibinfo
  {pages} {064204} (\bibinfo {year} {2019})}\BibitemShut {NoStop}%
\bibitem [{\citenamefont {Jian}\ \emph {et~al.}(2020)\citenamefont {Jian},
  \citenamefont {You}, \citenamefont {Vasseur},\ and\ \citenamefont
  {Ludwig}}]{Jian20}%
  \BibitemOpen
  \bibfield  {author} {\bibinfo {author} {\bibfnamefont {C.-M.}\ \bibnamefont
  {Jian}}, \bibinfo {author} {\bibfnamefont {Y.-Z.}\ \bibnamefont {You}},
  \bibinfo {author} {\bibfnamefont {R.}~\bibnamefont {Vasseur}},\ and\ \bibinfo
  {author} {\bibfnamefont {A.~W.~W.}\ \bibnamefont {Ludwig}},\ }\bibfield
  {title} {\bibinfo {title} {Measurement-induced criticality in random quantum
  circuits},\ }\href {https://doi.org/10.1103/PhysRevB.101.104302} {\bibfield
  {journal} {\bibinfo  {journal} {Phys. Rev. B}\ }\textbf {\bibinfo {volume}
  {101}},\ \bibinfo {pages} {104302} (\bibinfo {year} {2020})}\BibitemShut
  {NoStop}%
\bibitem [{\citenamefont {Bao}\ \emph {et~al.}(2020)\citenamefont {Bao},
  \citenamefont {Choi},\ and\ \citenamefont {Altman}}]{Altman2020weak}%
  \BibitemOpen
  \bibfield  {author} {\bibinfo {author} {\bibfnamefont {Y.}~\bibnamefont
  {Bao}}, \bibinfo {author} {\bibfnamefont {S.}~\bibnamefont {Choi}},\ and\
  \bibinfo {author} {\bibfnamefont {E.}~\bibnamefont {Altman}},\ }\bibfield
  {title} {\bibinfo {title} {Theory of the phase transition in random unitary
  circuits with measurements},\ }\href
  {https://doi.org/10.1103/PhysRevB.101.104301} {\bibfield  {journal} {\bibinfo
   {journal} {Phys. Rev. B}\ }\textbf {\bibinfo {volume} {101}},\ \bibinfo
  {pages} {104301} (\bibinfo {year} {2020})}\BibitemShut {NoStop}%
\bibitem [{\citenamefont {Szyniszewski}\ \emph {et~al.}(2020)\citenamefont
  {Szyniszewski}, \citenamefont {Romito},\ and\ \citenamefont
  {Schomerus}}]{Schomerus2020weakmeasure}%
  \BibitemOpen
  \bibfield  {author} {\bibinfo {author} {\bibfnamefont {M.}~\bibnamefont
  {Szyniszewski}}, \bibinfo {author} {\bibfnamefont {A.}~\bibnamefont
  {Romito}},\ and\ \bibinfo {author} {\bibfnamefont {H.}~\bibnamefont
  {Schomerus}},\ }\bibfield  {title} {\bibinfo {title} {{Universality of
  Entanglement Transitions from Stroboscopic to Continuous Measurements}},\
  }\href {https://doi.org/10.1103/PhysRevLett.125.210602} {\bibfield  {journal}
  {\bibinfo  {journal} {Phys. Rev. Lett.}\ }\textbf {\bibinfo {volume} {125}},\
  \bibinfo {pages} {210602} (\bibinfo {year} {2020})}\BibitemShut {NoStop}%
\bibitem [{\citenamefont {Fuji}\ and\ \citenamefont
  {Ashida}(2020)}]{Ashida2020}%
  \BibitemOpen
  \bibfield  {author} {\bibinfo {author} {\bibfnamefont {Y.}~\bibnamefont
  {Fuji}}\ and\ \bibinfo {author} {\bibfnamefont {Y.}~\bibnamefont {Ashida}},\
  }\bibfield  {title} {\bibinfo {title} {Measurement-induced quantum
  criticality under continuous monitoring},\ }\href
  {https://doi.org/10.1103/PhysRevB.102.054302} {\bibfield  {journal} {\bibinfo
   {journal} {Phys. Rev. B}\ }\textbf {\bibinfo {volume} {102}},\ \bibinfo
  {pages} {054302} (\bibinfo {year} {2020})}\BibitemShut {NoStop}%
\bibitem [{\citenamefont {Jian}\ \emph {et~al.}(2021)\citenamefont {Jian},
  \citenamefont {Liu}, \citenamefont {Chen}, \citenamefont {Swingle},\ and\
  \citenamefont {Zhang}}]{Jian21SYKMIPT}%
  \BibitemOpen
  \bibfield  {author} {\bibinfo {author} {\bibfnamefont {S.-K.}\ \bibnamefont
  {Jian}}, \bibinfo {author} {\bibfnamefont {C.}~\bibnamefont {Liu}}, \bibinfo
  {author} {\bibfnamefont {X.}~\bibnamefont {Chen}}, \bibinfo {author}
  {\bibfnamefont {B.}~\bibnamefont {Swingle}},\ and\ \bibinfo {author}
  {\bibfnamefont {P.}~\bibnamefont {Zhang}},\ }\bibfield  {title} {\bibinfo
  {title} {{Measurement-Induced Phase Transition in the Monitored
  Sachdev-Ye-Kitaev Model}},\ }\href
  {https://doi.org/10.1103/PhysRevLett.127.140601} {\bibfield  {journal}
  {\bibinfo  {journal} {Phys. Rev. Lett.}\ }\textbf {\bibinfo {volume} {127}},\
  \bibinfo {pages} {140601} (\bibinfo {year} {2021})}\BibitemShut {NoStop}%
\bibitem [{\citenamefont {Turkeshi}\ \emph {et~al.}(2021)\citenamefont
  {Turkeshi}, \citenamefont {Biella}, \citenamefont {Fazio}, \citenamefont
  {Dalmonte},\ and\ \citenamefont {Schir\'o}}]{Schiro2021}%
  \BibitemOpen
  \bibfield  {author} {\bibinfo {author} {\bibfnamefont {X.}~\bibnamefont
  {Turkeshi}}, \bibinfo {author} {\bibfnamefont {A.}~\bibnamefont {Biella}},
  \bibinfo {author} {\bibfnamefont {R.}~\bibnamefont {Fazio}}, \bibinfo
  {author} {\bibfnamefont {M.}~\bibnamefont {Dalmonte}},\ and\ \bibinfo
  {author} {\bibfnamefont {M.}~\bibnamefont {Schir\'o}},\ }\bibfield  {title}
  {\bibinfo {title} {{Measurement-induced entanglement transitions in the
  quantum Ising chain: From infinite to zero clicks}},\ }\href
  {https://doi.org/10.1103/PhysRevB.103.224210} {\bibfield  {journal} {\bibinfo
   {journal} {Phys. Rev. B}\ }\textbf {\bibinfo {volume} {103}},\ \bibinfo
  {pages} {224210} (\bibinfo {year} {2021})}\BibitemShut {NoStop}%
\bibitem [{\citenamefont {Biella}\ and\ \citenamefont
  {Schir{\'{o}}}(2021)}]{Biella2021manybodyquantumzeno}%
  \BibitemOpen
  \bibfield  {author} {\bibinfo {author} {\bibfnamefont {A.}~\bibnamefont
  {Biella}}\ and\ \bibinfo {author} {\bibfnamefont {M.}~\bibnamefont
  {Schir{\'{o}}}},\ }\bibfield  {title} {\bibinfo {title} {Many-{B}ody
  {Q}uantum {Z}eno {E}ffect and {M}easurement-{I}nduced {S}ubradiance
  {T}ransition},\ }\href {https://doi.org/10.22331/q-2021-08-19-528} {\bibfield
   {journal} {\bibinfo  {journal} {{Quantum}}\ }\textbf {\bibinfo {volume}
  {5}},\ \bibinfo {pages} {528} (\bibinfo {year} {2021})}\BibitemShut {NoStop}%
\bibitem [{\citenamefont {M\"uller}\ \emph {et~al.}(2022)\citenamefont
  {M\"uller}, \citenamefont {Diehl},\ and\ \citenamefont
  {Buchhold}}]{Diehl22weakmeasure}%
  \BibitemOpen
  \bibfield  {author} {\bibinfo {author} {\bibfnamefont {T.}~\bibnamefont
  {M\"uller}}, \bibinfo {author} {\bibfnamefont {S.}~\bibnamefont {Diehl}},\
  and\ \bibinfo {author} {\bibfnamefont {M.}~\bibnamefont {Buchhold}},\
  }\bibfield  {title} {\bibinfo {title} {{Measurement-Induced Dark State Phase
  Transitions in Long-Ranged Fermion Systems}},\ }\href
  {https://doi.org/10.1103/PhysRevLett.128.010605} {\bibfield  {journal}
  {\bibinfo  {journal} {Phys. Rev. Lett.}\ }\textbf {\bibinfo {volume} {128}},\
  \bibinfo {pages} {010605} (\bibinfo {year} {2022})}\BibitemShut {NoStop}%
\bibitem [{\citenamefont {{Kells}}\ \emph {et~al.}(2021)\citenamefont
  {{Kells}}, \citenamefont {{Meidan}},\ and\ \citenamefont
  {{Romito}}}]{Romito2021}%
  \BibitemOpen
  \bibfield  {author} {\bibinfo {author} {\bibfnamefont {G.}~\bibnamefont
  {{Kells}}}, \bibinfo {author} {\bibfnamefont {D.}~\bibnamefont {{Meidan}}},\
  and\ \bibinfo {author} {\bibfnamefont {A.}~\bibnamefont {{Romito}}},\
  }\bibfield  {title} {\bibinfo {title} {{Topological transitions with
  continuously monitored free fermions}},\ }\href@noop {} {\  (\bibinfo {year}
  {2021})},\ \Eprint {https://arxiv.org/abs/2112.09787} {arXiv:2112.09787}
  \BibitemShut {NoStop}%
\bibitem [{\citenamefont {Nishimori}(1981)}]{Nishimori1981}%
  \BibitemOpen
  \bibfield  {author} {\bibinfo {author} {\bibfnamefont {H.}~\bibnamefont
  {Nishimori}},\ }\bibfield  {title} {\bibinfo {title} {{Internal Energy,
  Specific Heat and Correlation Function of the Bond-Random Ising Model}},\
  }\href {https://doi.org/10.1143/PTP.66.1169} {\bibfield  {journal} {\bibinfo
  {journal} {Progress of Theoretical Physics}\ }\textbf {\bibinfo {volume}
  {66}},\ \bibinfo {pages} {1169} (\bibinfo {year} {1981})}\BibitemShut
  {NoStop}%
\bibitem [{\citenamefont {Kitaev}(2003)}]{Kitaev2003}%
  \BibitemOpen
  \bibfield  {author} {\bibinfo {author} {\bibfnamefont {A.~Y.}\ \bibnamefont
  {Kitaev}},\ }\bibfield  {title} {\bibinfo {title} {Fault-tolerant quantum
  computation by anyons},\ }\href
  {https://doi.org/http://dx.doi.org/10.1016/S0003-4916(02)00018-0} {\bibfield
  {journal} {\bibinfo  {journal} {Annals of Physics}\ }\textbf {\bibinfo
  {volume} {303}},\ \bibinfo {pages} {2 } (\bibinfo {year} {2003})}\BibitemShut
  {NoStop}%
\bibitem [{\citenamefont {Clerk}\ \emph {et~al.}(2010)\citenamefont {Clerk},
  \citenamefont {Devoret}, \citenamefont {Girvin}, \citenamefont {Marquardt},\
  and\ \citenamefont {Schoelkopf}}]{Clerk2010rmp}%
  \BibitemOpen
  \bibfield  {author} {\bibinfo {author} {\bibfnamefont {A.~A.}\ \bibnamefont
  {Clerk}}, \bibinfo {author} {\bibfnamefont {M.~H.}\ \bibnamefont {Devoret}},
  \bibinfo {author} {\bibfnamefont {S.~M.}\ \bibnamefont {Girvin}}, \bibinfo
  {author} {\bibfnamefont {F.}~\bibnamefont {Marquardt}},\ and\ \bibinfo
  {author} {\bibfnamefont {R.~J.}\ \bibnamefont {Schoelkopf}},\ }\bibfield
  {title} {\bibinfo {title} {Introduction to quantum noise, measurement, and
  amplification},\ }\href {https://doi.org/10.1103/RevModPhys.82.1155}
  {\bibfield  {journal} {\bibinfo  {journal} {Rev. Mod. Phys.}\ }\textbf
  {\bibinfo {volume} {82}},\ \bibinfo {pages} {1155} (\bibinfo {year}
  {2010})}\BibitemShut {NoStop}%
\bibitem [{sup()}]{supplement}%
  \BibitemOpen
  \href@noop {} {\bibinfo {title} {{See the Supplemental Material for details
  of our circuit model and its relation to Nishimori physics, a tensor-network
  representation of the monitored quantum state, numerical data on and off the
  Nishimori line along with details on the hybrid tensor network / Monte Carlo
  algorithm, a discussion of the alternative heavy-hexagon lattice geometry,
  and the route to glassy topological order in a 3D Nishimori
  code.}}}\BibitemShut {Stop}%
\bibitem [{\citenamefont {Greenberger}\ \emph {et~al.}(1989)\citenamefont
  {Greenberger}, \citenamefont {Horne},\ and\ \citenamefont {Zeilinger}}]{GHZ}%
  \BibitemOpen
  \bibfield  {author} {\bibinfo {author} {\bibfnamefont {D.~M.}\ \bibnamefont
  {Greenberger}}, \bibinfo {author} {\bibfnamefont {M.~A.}\ \bibnamefont
  {Horne}},\ and\ \bibinfo {author} {\bibfnamefont {A.}~\bibnamefont
  {Zeilinger}},\ }\bibfield  {title} {\bibinfo {title} {{Going Beyond Bell's
  Theorem}},\ }in\ \href {https://doi.org/10.48550/arXiv.0712.0921} {\emph
  {\bibinfo {booktitle} {{Bell's Theorem, Quantum Theory, and Conceptions of
  the Universe}}}}\ (\bibinfo  {publisher} {Kluwer},\ \bibinfo {year} {1989})\
  pp.\ \bibinfo {pages} {69--72},\ \Eprint {https://arxiv.org/abs/0712.0921}
  {arXiv:0712.0921} \BibitemShut {NoStop}%
\bibitem [{\citenamefont {Bao}\ \emph {et~al.}(2021)\citenamefont {Bao},
  \citenamefont {Block},\ and\ \citenamefont {Altman}}]{Altman2021measure}%
  \BibitemOpen
  \bibfield  {author} {\bibinfo {author} {\bibfnamefont {Y.}~\bibnamefont
  {Bao}}, \bibinfo {author} {\bibfnamefont {M.}~\bibnamefont {Block}},\ and\
  \bibinfo {author} {\bibfnamefont {E.}~\bibnamefont {Altman}},\ }\bibfield
  {title} {\bibinfo {title} {{Finite time teleportation phase transition in
  random quantum circuits}},\ }\href@noop {} {\  (\bibinfo {year} {2021})},\
  \Eprint {https://arxiv.org/abs/2110.06963} {arXiv:2110.06963} \BibitemShut
  {NoStop}%
\bibitem [{\citenamefont {Liu}\ \emph {et~al.}(2022)\citenamefont {Liu},
  \citenamefont {Zhou},\ and\ \citenamefont {Chen}}]{Chen2022measure}%
  \BibitemOpen
  \bibfield  {author} {\bibinfo {author} {\bibfnamefont {H.}~\bibnamefont
  {Liu}}, \bibinfo {author} {\bibfnamefont {T.}~\bibnamefont {Zhou}},\ and\
  \bibinfo {author} {\bibfnamefont {X.}~\bibnamefont {Chen}},\ }\bibfield
  {title} {\bibinfo {title} {Measurement induced entanglement transition in two
  dimensional shallow circuit},\ }\href@noop {} {\  (\bibinfo {year} {2022})},\
  \Eprint {https://arxiv.org/abs/2203.07510} {arXiv:2203.07510} \BibitemShut
  {NoStop}%
\bibitem [{\citenamefont {Garratt}\ \emph {et~al.}(2022)\citenamefont
  {Garratt}, \citenamefont {Weinstein},\ and\ \citenamefont
  {Altman}}]{Garratt22}%
  \BibitemOpen
  \bibfield  {author} {\bibinfo {author} {\bibfnamefont {S.~J.}\ \bibnamefont
  {Garratt}}, \bibinfo {author} {\bibfnamefont {Z.}~\bibnamefont {Weinstein}},\
  and\ \bibinfo {author} {\bibfnamefont {E.}~\bibnamefont {Altman}},\
  }\bibfield  {title} {\bibinfo {title} {{Measurements conspire nonlocally to
  restructure critical quantum states}},\ }\href@noop {} {\  (\bibinfo {year}
  {2022})},\ \Eprint {https://arxiv.org/abs/2207.09476} {arXiv:2207.09476}
  \BibitemShut {NoStop}%
\bibitem [{\citenamefont {Browne}\ \emph {et~al.}(2008)\citenamefont {Browne},
  \citenamefont {Elliott}, \citenamefont {Flammia}, \citenamefont {Merkel},
  \citenamefont {Miyake},\ and\ \citenamefont {Short}}]{Browne2008}%
  \BibitemOpen
  \bibfield  {author} {\bibinfo {author} {\bibfnamefont {D.~E.}\ \bibnamefont
  {Browne}}, \bibinfo {author} {\bibfnamefont {M.~B.}\ \bibnamefont {Elliott}},
  \bibinfo {author} {\bibfnamefont {S.~T.}\ \bibnamefont {Flammia}}, \bibinfo
  {author} {\bibfnamefont {S.~T.}\ \bibnamefont {Merkel}}, \bibinfo {author}
  {\bibfnamefont {A.}~\bibnamefont {Miyake}},\ and\ \bibinfo {author}
  {\bibfnamefont {A.~J.}\ \bibnamefont {Short}},\ }\bibfield  {title} {\bibinfo
  {title} {Phase transition of computational power in the resource states for
  one-way quantum computation},\ }\href
  {https://doi.org/10.1088/1367-2630/10/2/023010} {\bibfield  {journal}
  {\bibinfo  {journal} {New Journal of Physics}\ }\textbf {\bibinfo {volume}
  {10}},\ \bibinfo {pages} {023010} (\bibinfo {year} {2008})}\BibitemShut
  {NoStop}%
\bibitem [{\citenamefont {Napp}\ \emph {et~al.}(2022)\citenamefont {Napp},
  \citenamefont {La~Placa}, \citenamefont {Dalzell}, \citenamefont
  {Brand\~ao},\ and\ \citenamefont {Harrow}}]{Harrow2022}%
  \BibitemOpen
  \bibfield  {author} {\bibinfo {author} {\bibfnamefont {J.~C.}\ \bibnamefont
  {Napp}}, \bibinfo {author} {\bibfnamefont {R.~L.}\ \bibnamefont {La~Placa}},
  \bibinfo {author} {\bibfnamefont {A.~M.}\ \bibnamefont {Dalzell}}, \bibinfo
  {author} {\bibfnamefont {F.~G. S.~L.}\ \bibnamefont {Brand\~ao}},\ and\
  \bibinfo {author} {\bibfnamefont {A.~W.}\ \bibnamefont {Harrow}},\ }\bibfield
   {title} {\bibinfo {title} {{Efficient Classical Simulation of Random Shallow
  2D Quantum Circuits}},\ }\href {https://doi.org/10.1103/PhysRevX.12.021021}
  {\bibfield  {journal} {\bibinfo  {journal} {Phys. Rev. X}\ }\textbf {\bibinfo
  {volume} {12}},\ \bibinfo {pages} {021021} (\bibinfo {year}
  {2022})}\BibitemShut {NoStop}%
\bibitem [{\citenamefont {Henley}(2004)}]{Henley04RK}%
  \BibitemOpen
  \bibfield  {author} {\bibinfo {author} {\bibfnamefont {C.~L.}\ \bibnamefont
  {Henley}},\ }\bibfield  {title} {\bibinfo {title} {{From classical to quantum
  dynamics at Rokhsar{\textendash}Kivelson points}},\ }\href
  {https://doi.org/10.1088/0953-8984/16/11/045} {\bibfield  {journal} {\bibinfo
   {journal} {Journal of Physics: Condensed Matter}\ }\textbf {\bibinfo
  {volume} {16}},\ \bibinfo {pages} {S891} (\bibinfo {year}
  {2004})}\BibitemShut {NoStop}%
\bibitem [{\citenamefont {Ardonne}\ \emph {et~al.}(2004)\citenamefont
  {Ardonne}, \citenamefont {Fendley},\ and\ \citenamefont
  {Fradkin}}]{Fradkin04RK}%
  \BibitemOpen
  \bibfield  {author} {\bibinfo {author} {\bibfnamefont {E.}~\bibnamefont
  {Ardonne}}, \bibinfo {author} {\bibfnamefont {P.}~\bibnamefont {Fendley}},\
  and\ \bibinfo {author} {\bibfnamefont {E.}~\bibnamefont {Fradkin}},\
  }\bibfield  {title} {\bibinfo {title} {{Topological order and conformal
  quantum critical points}},\ }\href
  {https://doi.org/10.1016/j.aop.2004.01.004} {\bibfield  {journal} {\bibinfo
  {journal} {Annals of Physics}\ }\textbf {\bibinfo {volume} {310}},\ \bibinfo
  {pages} {493} (\bibinfo {year} {2004})}\BibitemShut {NoStop}%
\bibitem [{\citenamefont {Verstraete}\ \emph {et~al.}(2006)\citenamefont
  {Verstraete}, \citenamefont {Wolf}, \citenamefont {Perez-Garcia},\ and\
  \citenamefont {Cirac}}]{Cirac06}%
  \BibitemOpen
  \bibfield  {author} {\bibinfo {author} {\bibfnamefont {F.}~\bibnamefont
  {Verstraete}}, \bibinfo {author} {\bibfnamefont {M.~M.}\ \bibnamefont
  {Wolf}}, \bibinfo {author} {\bibfnamefont {D.}~\bibnamefont {Perez-Garcia}},\
  and\ \bibinfo {author} {\bibfnamefont {J.~I.}\ \bibnamefont {Cirac}},\
  }\bibfield  {title} {\bibinfo {title} {{Criticality, the Area Law, and the
  Computational Power of Projected Entangled Pair States}},\ }\href
  {https://doi.org/10.1103/PhysRevLett.96.220601} {\bibfield  {journal}
  {\bibinfo  {journal} {Phys. Rev. Lett.}\ }\textbf {\bibinfo {volume} {96}},\
  \bibinfo {pages} {220601} (\bibinfo {year} {2006})}\BibitemShut {NoStop}%
\bibitem [{\citenamefont {Edwards}\ and\ \citenamefont
  {Anderson}(1975)}]{EdwardsAnderson}%
  \BibitemOpen
  \bibfield  {author} {\bibinfo {author} {\bibfnamefont {S.~F.}\ \bibnamefont
  {Edwards}}\ and\ \bibinfo {author} {\bibfnamefont {P.~W.}\ \bibnamefont
  {Anderson}},\ }\bibfield  {title} {\bibinfo {title} {Theory of spin
  glasses},\ }\href {https://doi.org/10.1088/0305-4608/5/5/017} {\bibfield
  {journal} {\bibinfo  {journal} {Journal of Physics F: Metal Physics}\
  }\textbf {\bibinfo {volume} {5}},\ \bibinfo {pages} {965} (\bibinfo {year}
  {1975})}\BibitemShut {NoStop}%
\bibitem [{\citenamefont {Wolf}\ \emph {et~al.}(2008)\citenamefont {Wolf},
  \citenamefont {Verstraete}, \citenamefont {Hastings},\ and\ \citenamefont
  {Cirac}}]{Cirac2008mutualinfo}%
  \BibitemOpen
  \bibfield  {author} {\bibinfo {author} {\bibfnamefont {M.~M.}\ \bibnamefont
  {Wolf}}, \bibinfo {author} {\bibfnamefont {F.}~\bibnamefont {Verstraete}},
  \bibinfo {author} {\bibfnamefont {M.~B.}\ \bibnamefont {Hastings}},\ and\
  \bibinfo {author} {\bibfnamefont {J.~I.}\ \bibnamefont {Cirac}},\ }\bibfield
  {title} {\bibinfo {title} {Area laws in quantum systems: Mutual information
  and correlations},\ }\href {https://doi.org/10.1103/PhysRevLett.100.070502}
  {\bibfield  {journal} {\bibinfo  {journal} {Phys. Rev. Lett.}\ }\textbf
  {\bibinfo {volume} {100}},\ \bibinfo {pages} {070502} (\bibinfo {year}
  {2008})}\BibitemShut {NoStop}%
\bibitem [{\citenamefont {Georges}\ \emph
  {et~al.}(1985{\natexlab{a}})\citenamefont {Georges}, \citenamefont {Hansel},\
  and\ \citenamefont {Le~Doussal}}]{Georges1985a}%
  \BibitemOpen
  \bibfield  {author} {\bibinfo {author} {\bibfnamefont {A.}~\bibnamefont
  {Georges}}, \bibinfo {author} {\bibfnamefont {D.}~\bibnamefont {Hansel}},\
  and\ \bibinfo {author} {\bibfnamefont {P.}~\bibnamefont {Le~Doussal}},\
  }\bibfield  {title} {\bibinfo {title} {{Exact properties of spin glasses. -
  I. 2D supersymmetry and Nishimori's result}},\ }\href
  {https://doi.org/10.1051/jphys:019850046080130900} {\bibfield  {journal}
  {\bibinfo  {journal} {J. Phys. France}\ }\textbf {\bibinfo {volume} {46}},\
  \bibinfo {pages} {1309} (\bibinfo {year} {1985}{\natexlab{a}})}\BibitemShut
  {NoStop}%
\bibitem [{\citenamefont {Georges}\ \emph
  {et~al.}(1985{\natexlab{b}})\citenamefont {Georges}, \citenamefont {Hansel},
  \citenamefont {Le~Doussal},\ and\ \citenamefont {Bouchaud}}]{Georges1985b}%
  \BibitemOpen
  \bibfield  {author} {\bibinfo {author} {\bibfnamefont {A.}~\bibnamefont
  {Georges}}, \bibinfo {author} {\bibfnamefont {D.}~\bibnamefont {Hansel}},
  \bibinfo {author} {\bibfnamefont {P.}~\bibnamefont {Le~Doussal}},\ and\
  \bibinfo {author} {\bibfnamefont {J.-P.}\ \bibnamefont {Bouchaud}},\
  }\bibfield  {title} {\bibinfo {title} {{Exact properties of spin glasses. II.
  Nishimori's line : new results and physical implications}},\ }\href
  {https://doi.org/10.1051/jphys:0198500460110182700} {\bibfield  {journal}
  {\bibinfo  {journal} {J. Phys. France}\ }\textbf {\bibinfo {volume} {46}},\
  \bibinfo {pages} {1827} (\bibinfo {year} {1985}{\natexlab{b}})}\BibitemShut
  {NoStop}%
\bibitem [{\citenamefont {Binder}\ and\ \citenamefont
  {Young}(1986)}]{Binder86rmp}%
  \BibitemOpen
  \bibfield  {author} {\bibinfo {author} {\bibfnamefont {K.}~\bibnamefont
  {Binder}}\ and\ \bibinfo {author} {\bibfnamefont {A.~P.}\ \bibnamefont
  {Young}},\ }\bibfield  {title} {\bibinfo {title} {{Spin glasses: Experimental
  facts, theoretical concepts, and open questions}},\ }\href
  {https://doi.org/10.1103/RevModPhys.58.801} {\bibfield  {journal} {\bibinfo
  {journal} {Rev. Mod. Phys.}\ }\textbf {\bibinfo {volume} {58}},\ \bibinfo
  {pages} {801} (\bibinfo {year} {1986})}\BibitemShut {NoStop}%
\bibitem [{\citenamefont {Singh}\ and\ \citenamefont
  {Adler}(1996)}]{Adler1996}%
  \BibitemOpen
  \bibfield  {author} {\bibinfo {author} {\bibfnamefont {R.~R.~P.}\
  \bibnamefont {Singh}}\ and\ \bibinfo {author} {\bibfnamefont
  {J.}~\bibnamefont {Adler}},\ }\bibfield  {title} {\bibinfo {title}
  {{High-temperature expansion study of the Nishimori multicritical point in
  two and four dimensions}},\ }\href {https://doi.org/10.1103/PhysRevB.54.364}
  {\bibfield  {journal} {\bibinfo  {journal} {Phys. Rev. B}\ }\textbf {\bibinfo
  {volume} {54}},\ \bibinfo {pages} {364} (\bibinfo {year} {1996})}\BibitemShut
  {NoStop}%
\bibitem [{\citenamefont {Cho}\ and\ \citenamefont
  {Fisher}(1997)}]{Fisher1997}%
  \BibitemOpen
  \bibfield  {author} {\bibinfo {author} {\bibfnamefont {S.}~\bibnamefont
  {Cho}}\ and\ \bibinfo {author} {\bibfnamefont {M.~P.~A.}\ \bibnamefont
  {Fisher}},\ }\bibfield  {title} {\bibinfo {title} {{Criticality in the
  two-dimensional random-bond Ising model}},\ }\href
  {https://doi.org/10.1103/PhysRevB.55.1025} {\bibfield  {journal} {\bibinfo
  {journal} {Phys. Rev. B}\ }\textbf {\bibinfo {volume} {55}},\ \bibinfo
  {pages} {1025} (\bibinfo {year} {1997})}\BibitemShut {NoStop}%
\bibitem [{\citenamefont {Read}\ and\ \citenamefont {Ludwig}(2000)}]{Read2000}%
  \BibitemOpen
  \bibfield  {author} {\bibinfo {author} {\bibfnamefont {N.}~\bibnamefont
  {Read}}\ and\ \bibinfo {author} {\bibfnamefont {A.~W.~W.}\ \bibnamefont
  {Ludwig}},\ }\bibfield  {title} {\bibinfo {title} {{Absence of a metallic
  phase in random-bond Ising models in two dimensions: Applications to
  disordered superconductors and paired quantum Hall states}},\ }\href
  {https://doi.org/10.1103/PhysRevB.63.024404} {\bibfield  {journal} {\bibinfo
  {journal} {Phys. Rev. B}\ }\textbf {\bibinfo {volume} {63}},\ \bibinfo
  {pages} {024404} (\bibinfo {year} {2000})}\BibitemShut {NoStop}%
\bibitem [{\citenamefont {Honecker}\ \emph {et~al.}(2001)\citenamefont
  {Honecker}, \citenamefont {Picco},\ and\ \citenamefont {Pujol}}]{Pujol2001}%
  \BibitemOpen
  \bibfield  {author} {\bibinfo {author} {\bibfnamefont {A.}~\bibnamefont
  {Honecker}}, \bibinfo {author} {\bibfnamefont {M.}~\bibnamefont {Picco}},\
  and\ \bibinfo {author} {\bibfnamefont {P.}~\bibnamefont {Pujol}},\ }\bibfield
   {title} {\bibinfo {title} {{Universality Class of the Nishimori Point in the
  2D $\ifmmode\pm\else\textpm\fi{}\mathit{J}$ Random-Bond Ising Model}},\
  }\href {https://doi.org/10.1103/PhysRevLett.87.047201} {\bibfield  {journal}
  {\bibinfo  {journal} {Phys. Rev. Lett.}\ }\textbf {\bibinfo {volume} {87}},\
  \bibinfo {pages} {047201} (\bibinfo {year} {2001})}\BibitemShut {NoStop}%
\bibitem [{\citenamefont {Merz}\ and\ \citenamefont
  {Chalker}(2002)}]{Chalker2002}%
  \BibitemOpen
  \bibfield  {author} {\bibinfo {author} {\bibfnamefont {F.}~\bibnamefont
  {Merz}}\ and\ \bibinfo {author} {\bibfnamefont {J.~T.}\ \bibnamefont
  {Chalker}},\ }\bibfield  {title} {\bibinfo {title} {{Two-dimensional
  random-bond Ising model, free fermions, and the network model}},\ }\href
  {https://doi.org/10.1103/PhysRevB.65.054425} {\bibfield  {journal} {\bibinfo
  {journal} {Phys. Rev. B}\ }\textbf {\bibinfo {volume} {65}},\ \bibinfo
  {pages} {054425} (\bibinfo {year} {2002})}\BibitemShut {NoStop}%
\bibitem [{\citenamefont {Amoruso}\ and\ \citenamefont
  {Hartmann}(2004)}]{Hartmann2004}%
  \BibitemOpen
  \bibfield  {author} {\bibinfo {author} {\bibfnamefont {C.}~\bibnamefont
  {Amoruso}}\ and\ \bibinfo {author} {\bibfnamefont {A.~K.}\ \bibnamefont
  {Hartmann}},\ }\bibfield  {title} {\bibinfo {title} {{Domain-wall energies
  and magnetization of the two-dimensional random-bond Ising model}},\ }\href
  {https://doi.org/10.1103/PhysRevB.70.134425} {\bibfield  {journal} {\bibinfo
  {journal} {Phys. Rev. B}\ }\textbf {\bibinfo {volume} {70}},\ \bibinfo
  {pages} {134425} (\bibinfo {year} {2004})}\BibitemShut {NoStop}%
\bibitem [{\citenamefont {Le~Doussal}\ and\ \citenamefont
  {Harris}(1988)}]{Harris1988Nishimori}%
  \BibitemOpen
  \bibfield  {author} {\bibinfo {author} {\bibfnamefont {P.}~\bibnamefont
  {Le~Doussal}}\ and\ \bibinfo {author} {\bibfnamefont {A.~B.}\ \bibnamefont
  {Harris}},\ }\bibfield  {title} {\bibinfo {title} {{Location of the Ising
  Spin-Glass Multicritical Point on Nishimori's Line}},\ }\href
  {https://doi.org/10.1103/PhysRevLett.61.625} {\bibfield  {journal} {\bibinfo
  {journal} {Phys. Rev. Lett.}\ }\textbf {\bibinfo {volume} {61}},\ \bibinfo
  {pages} {625} (\bibinfo {year} {1988})}\BibitemShut {NoStop}%
\bibitem [{\citenamefont {Le~Doussal}\ and\ \citenamefont
  {Harris}(1989)}]{Harris1989}%
  \BibitemOpen
  \bibfield  {author} {\bibinfo {author} {\bibfnamefont {P.}~\bibnamefont
  {Le~Doussal}}\ and\ \bibinfo {author} {\bibfnamefont {A.~B.}\ \bibnamefont
  {Harris}},\ }\bibfield  {title} {\bibinfo {title} {{\ensuremath{\epsilon}
  expansion for the Nishimori multicritical point of spin glasses}},\ }\href
  {https://doi.org/10.1103/PhysRevB.40.9249} {\bibfield  {journal} {\bibinfo
  {journal} {Phys. Rev. B}\ }\textbf {\bibinfo {volume} {40}},\ \bibinfo
  {pages} {9249} (\bibinfo {year} {1989})}\BibitemShut {NoStop}%
\bibitem [{\citenamefont {Gruzberg}\ \emph {et~al.}(2001)\citenamefont
  {Gruzberg}, \citenamefont {Read},\ and\ \citenamefont {Ludwig}}]{Read2001}%
  \BibitemOpen
  \bibfield  {author} {\bibinfo {author} {\bibfnamefont {I.~A.}\ \bibnamefont
  {Gruzberg}}, \bibinfo {author} {\bibfnamefont {N.}~\bibnamefont {Read}},\
  and\ \bibinfo {author} {\bibfnamefont {A.~W.~W.}\ \bibnamefont {Ludwig}},\
  }\bibfield  {title} {\bibinfo {title} {{Random-bond Ising model in two
  dimensions: The Nishimori line and supersymmetry}},\ }\href
  {https://doi.org/10.1103/PhysRevB.63.104422} {\bibfield  {journal} {\bibinfo
  {journal} {Phys. Rev. B}\ }\textbf {\bibinfo {volume} {63}},\ \bibinfo
  {pages} {104422} (\bibinfo {year} {2001})}\BibitemShut {NoStop}%
\bibitem [{\citenamefont {Wang}\ \emph {et~al.}(2003)\citenamefont {Wang},
  \citenamefont {Harrington},\ and\ \citenamefont {Preskill}}]{Preskill2003}%
  \BibitemOpen
  \bibfield  {author} {\bibinfo {author} {\bibfnamefont {C.}~\bibnamefont
  {Wang}}, \bibinfo {author} {\bibfnamefont {J.}~\bibnamefont {Harrington}},\
  and\ \bibinfo {author} {\bibfnamefont {J.}~\bibnamefont {Preskill}},\
  }\bibfield  {title} {\bibinfo {title} {{Confinement-Higgs transition in a
  disordered gauge theory and the accuracy threshold for quantum memory}},\
  }\href {https://doi.org/https://doi.org/10.1016/S0003-4916(02)00019-2}
  {\bibfield  {journal} {\bibinfo  {journal} {Annals of Physics}\ }\textbf
  {\bibinfo {volume} {303}},\ \bibinfo {pages} {31} (\bibinfo {year}
  {2003})}\BibitemShut {NoStop}%
\bibitem [{\citenamefont {Mattis}(1976)}]{Mattis1976}%
  \BibitemOpen
  \bibfield  {author} {\bibinfo {author} {\bibfnamefont {D.}~\bibnamefont
  {Mattis}},\ }\bibfield  {title} {\bibinfo {title} {Solvable spin systems with
  random interactions},\ }\href
  {https://doi.org/https://doi.org/10.1016/0375-9601(76)90396-0} {\bibfield
  {journal} {\bibinfo  {journal} {Physics Letters A}\ }\textbf {\bibinfo
  {volume} {56}},\ \bibinfo {pages} {421} (\bibinfo {year} {1976})}\BibitemShut
  {NoStop}%
\bibitem [{\citenamefont {{Melchert}}(2009)}]{FSSpackage}%
  \BibitemOpen
  \bibfield  {author} {\bibinfo {author} {\bibfnamefont {O.}~\bibnamefont
  {{Melchert}}},\ }\bibfield  {title} {\bibinfo {title} {{autoScale.py - A
  program for automatic finite-size scaling analyses: A user's guide}},\
  }\href@noop {} {\  (\bibinfo {year} {2009})},\ \Eprint
  {https://arxiv.org/abs/0910.5403} {arXiv:0910.5403} \BibitemShut {NoStop}%
\bibitem [{\citenamefont {Lund}\ \emph {et~al.}(2017)\citenamefont {Lund},
  \citenamefont {Bremner},\ and\ \citenamefont {Ralph}}]{Lund2017sampling}%
  \BibitemOpen
  \bibfield  {author} {\bibinfo {author} {\bibfnamefont {A.~P.}\ \bibnamefont
  {Lund}}, \bibinfo {author} {\bibfnamefont {M.~J.}\ \bibnamefont {Bremner}},\
  and\ \bibinfo {author} {\bibfnamefont {T.~C.}\ \bibnamefont {Ralph}},\
  }\bibfield  {title} {\bibinfo {title} {Quantum sampling problems,
  bosonsampling and quantum supremacy},\ }\href@noop {} {\bibfield  {journal}
  {\bibinfo  {journal} {npj Quantum Information}\ }\textbf {\bibinfo {volume}
  {3}},\ \bibinfo {pages} {1} (\bibinfo {year} {2017})}\BibitemShut {NoStop}%
\bibitem [{\citenamefont {Harrow}\ and\ \citenamefont
  {Montanaro}(2017)}]{Harrow2017sampling}%
  \BibitemOpen
  \bibfield  {author} {\bibinfo {author} {\bibfnamefont {A.~W.}\ \bibnamefont
  {Harrow}}\ and\ \bibinfo {author} {\bibfnamefont {A.}~\bibnamefont
  {Montanaro}},\ }\bibfield  {title} {\bibinfo {title} {Quantum computational
  supremacy},\ }\href@noop {} {\bibfield  {journal} {\bibinfo  {journal}
  {Nature}\ }\textbf {\bibinfo {volume} {549}},\ \bibinfo {pages} {203}
  (\bibinfo {year} {2017})}\BibitemShut {NoStop}%
\bibitem [{\citenamefont {Pan}\ \emph {et~al.}(2022)\citenamefont {Pan},
  \citenamefont {Chen},\ and\ \citenamefont {Zhang}}]{Zhang2022sampling}%
  \BibitemOpen
  \bibfield  {author} {\bibinfo {author} {\bibfnamefont {F.}~\bibnamefont
  {Pan}}, \bibinfo {author} {\bibfnamefont {K.}~\bibnamefont {Chen}},\ and\
  \bibinfo {author} {\bibfnamefont {P.}~\bibnamefont {Zhang}},\ }\bibfield
  {title} {\bibinfo {title} {Solving the sampling problem of the sycamore
  quantum circuits},\ }\href {https://doi.org/10.1103/PhysRevLett.129.090502}
  {\bibfield  {journal} {\bibinfo  {journal} {Phys. Rev. Lett.}\ }\textbf
  {\bibinfo {volume} {129}},\ \bibinfo {pages} {090502} (\bibinfo {year}
  {2022})}\BibitemShut {NoStop}%
\bibitem [{Note1()}]{Note1}%
  \BibitemOpen
  \bibinfo {note} {In fact, this hybrid algorithm can be applied to more
  general two-dimensional measurement problems as long as the wave functions
  are within the projected entangled pair states manifold of area law
  entanglement entropy, and the numerical complexity remains classically
  tractable for shallow depth circuits \cite {Harrow2022}.}\BibitemShut {Stop}%
\bibitem [{IBM()}]{IBMQuantumCloud}%
  \BibitemOpen
  \href {https://www.ibm.com/quantum-computing/systems/} {\bibinfo {title}
  {https://www.ibm.com/quantum-computing/}}\BibitemShut {NoStop}%
\bibitem [{\citenamefont {Koh}\ \emph {et~al.}(2022)\citenamefont {Koh},
  \citenamefont {Sun}, \citenamefont {Motta},\ and\ \citenamefont
  {Minnich}}]{Minnich2022ibmMIPT}%
  \BibitemOpen
  \bibfield  {author} {\bibinfo {author} {\bibfnamefont {J.~M.}\ \bibnamefont
  {Koh}}, \bibinfo {author} {\bibfnamefont {S.-N.}\ \bibnamefont {Sun}},
  \bibinfo {author} {\bibfnamefont {M.}~\bibnamefont {Motta}},\ and\ \bibinfo
  {author} {\bibfnamefont {A.~J.}\ \bibnamefont {Minnich}},\ }\bibfield
  {title} {\bibinfo {title} {{Experimental Realization of a Measurement-Induced
  Entanglement Phase Transition on a Superconducting Quantum Processor}},\
  }\href@noop {} {\  (\bibinfo {year} {2022})},\ \Eprint
  {https://arxiv.org/abs/2203.04338} {arXiv:2203.04338} \BibitemShut {NoStop}%
\bibitem [{\citenamefont {Yang}\ and\ \citenamefont
  {Liu}(2022)}]{Liu22coherentnoise}%
  \BibitemOpen
  \bibfield  {author} {\bibinfo {author} {\bibfnamefont {Q.}~\bibnamefont
  {Yang}}\ and\ \bibinfo {author} {\bibfnamefont {D.~E.}\ \bibnamefont {Liu}},\
  }\bibfield  {title} {\bibinfo {title} {Effect of quantum error correction on
  detection-induced coherent errors},\ }\href
  {https://doi.org/10.1103/PhysRevA.105.022434} {\bibfield  {journal} {\bibinfo
   {journal} {Phys. Rev. A}\ }\textbf {\bibinfo {volume} {105}},\ \bibinfo
  {pages} {022434} (\bibinfo {year} {2022})}\BibitemShut {NoStop}%
\bibitem [{\citenamefont {Hamma}\ \emph {et~al.}(2005)\citenamefont {Hamma},
  \citenamefont {Zanardi},\ and\ \citenamefont {Wen}}]{Wen2005}%
  \BibitemOpen
  \bibfield  {author} {\bibinfo {author} {\bibfnamefont {A.}~\bibnamefont
  {Hamma}}, \bibinfo {author} {\bibfnamefont {P.}~\bibnamefont {Zanardi}},\
  and\ \bibinfo {author} {\bibfnamefont {X.-G.}\ \bibnamefont {Wen}},\
  }\bibfield  {title} {\bibinfo {title} {String and membrane condensation on
  three-dimensional lattices},\ }\href
  {https://doi.org/10.1103/PhysRevB.72.035307} {\bibfield  {journal} {\bibinfo
  {journal} {Phys. Rev. B}\ }\textbf {\bibinfo {volume} {72}},\ \bibinfo
  {pages} {035307} (\bibinfo {year} {2005})}\BibitemShut {NoStop}%
\bibitem [{\citenamefont {Castelnovo}\ and\ \citenamefont
  {Chamon}(2008)}]{Castelnovo2008}%
  \BibitemOpen
  \bibfield  {author} {\bibinfo {author} {\bibfnamefont {C.}~\bibnamefont
  {Castelnovo}}\ and\ \bibinfo {author} {\bibfnamefont {C.}~\bibnamefont
  {Chamon}},\ }\bibfield  {title} {\bibinfo {title} {Topological order in a
  three-dimensional toric code at finite temperature},\ }\href
  {https://doi.org/10.1103/PhysRevB.78.155120} {\bibfield  {journal} {\bibinfo
  {journal} {Phys. Rev. B}\ }\textbf {\bibinfo {volume} {78}},\ \bibinfo
  {pages} {155120} (\bibinfo {year} {2008})}\BibitemShut {NoStop}%
\bibitem [{\citenamefont {Wegner}(1971)}]{Wegner71duality}%
  \BibitemOpen
  \bibfield  {author} {\bibinfo {author} {\bibfnamefont {F.~J.}\ \bibnamefont
  {Wegner}},\ }\bibfield  {title} {\bibinfo {title} {{Duality in Generalized
  Ising Models and Phase Transitions without Local Order Parameters}},\ }\href
  {https://doi.org/10.1063/1.1665530} {\bibfield  {journal} {\bibinfo
  {journal} {Journal of Mathematical Physics}\ }\textbf {\bibinfo {volume}
  {12}},\ \bibinfo {pages} {2259} (\bibinfo {year} {1971})}\BibitemShut
  {NoStop}%
\bibitem [{\citenamefont {Kogut}(1979)}]{Kogut79rmp}%
  \BibitemOpen
  \bibfield  {author} {\bibinfo {author} {\bibfnamefont {J.~B.}\ \bibnamefont
  {Kogut}},\ }\bibfield  {title} {\bibinfo {title} {An introduction to lattice
  gauge theory and spin systems},\ }\href
  {https://doi.org/10.1103/RevModPhys.51.659} {\bibfield  {journal} {\bibinfo
  {journal} {Rev. Mod. Phys.}\ }\textbf {\bibinfo {volume} {51}},\ \bibinfo
  {pages} {659} (\bibinfo {year} {1979})}\BibitemShut {NoStop}%
\bibitem [{\citenamefont {Ohno}\ \emph {et~al.}(2004)\citenamefont {Ohno},
  \citenamefont {Arakawa}, \citenamefont {Ichinose},\ and\ \citenamefont
  {Matsui}}]{Matsui2004}%
  \BibitemOpen
  \bibfield  {author} {\bibinfo {author} {\bibfnamefont {T.}~\bibnamefont
  {Ohno}}, \bibinfo {author} {\bibfnamefont {G.}~\bibnamefont {Arakawa}},
  \bibinfo {author} {\bibfnamefont {I.}~\bibnamefont {Ichinose}},\ and\
  \bibinfo {author} {\bibfnamefont {T.}~\bibnamefont {Matsui}},\ }\bibfield
  {title} {\bibinfo {title} {{Phase structure of the random-plaquette Z2 gauge
  model: accuracy threshold for a toric quantum memory}},\ }\href
  {https://doi.org/https://doi.org/10.1016/j.nuclphysb.2004.07.003} {\bibfield
  {journal} {\bibinfo  {journal} {Nuclear Physics B}\ }\textbf {\bibinfo
  {volume} {697}},\ \bibinfo {pages} {462} (\bibinfo {year}
  {2004})}\BibitemShut {NoStop}%
\bibitem [{\citenamefont {{Castelnovo}}\ \emph {et~al.}(2010)\citenamefont
  {{Castelnovo}}, \citenamefont {{Trebst}},\ and\ \citenamefont
  {{Troyer}}}]{Troyer10topocrit}%
  \BibitemOpen
  \bibfield  {author} {\bibinfo {author} {\bibfnamefont {C.}~\bibnamefont
  {{Castelnovo}}}, \bibinfo {author} {\bibfnamefont {S.}~\bibnamefont
  {{Trebst}}},\ and\ \bibinfo {author} {\bibfnamefont {M.}~\bibnamefont
  {{Troyer}}},\ }\bibfield  {title} {\bibinfo {title} {{Topological Order and
  Quantum Criticality}},\ }in\ \href {https://doi.org/10.1201/b10273-14} {\emph
  {\bibinfo {booktitle} {Understanding Quantum Phase Transitions}}}\ (\bibinfo
  {publisher} {Taylor \& Francis},\ \bibinfo {year} {2010})\ pp.\ \bibinfo
  {pages} {169--192},\ \Eprint {https://arxiv.org/abs/0912.3272}
  {arXiv:0912.3272} \BibitemShut {NoStop}%
\bibitem [{\citenamefont {Schuch}\ \emph {et~al.}(2013)\citenamefont {Schuch},
  \citenamefont {Poilblanc}, \citenamefont {Cirac},\ and\ \citenamefont
  {P\'erez-Garc\'{\i}a}}]{Cirac2012}%
  \BibitemOpen
  \bibfield  {author} {\bibinfo {author} {\bibfnamefont {N.}~\bibnamefont
  {Schuch}}, \bibinfo {author} {\bibfnamefont {D.}~\bibnamefont {Poilblanc}},
  \bibinfo {author} {\bibfnamefont {J.~I.}\ \bibnamefont {Cirac}},\ and\
  \bibinfo {author} {\bibfnamefont {D.}~\bibnamefont {P\'erez-Garc\'{\i}a}},\
  }\bibfield  {title} {\bibinfo {title} {{Topological Order in the Projected
  Entangled-Pair States Formalism: Transfer Operator and Boundary
  Hamiltonians}},\ }\href {https://doi.org/10.1103/PhysRevLett.111.090501}
  {\bibfield  {journal} {\bibinfo  {journal} {Phys. Rev. Lett.}\ }\textbf
  {\bibinfo {volume} {111}},\ \bibinfo {pages} {090501} (\bibinfo {year}
  {2013})}\BibitemShut {NoStop}%
\bibitem [{\citenamefont {Zhu}\ and\ \citenamefont {Zhang}(2019)}]{Zhu19}%
  \BibitemOpen
  \bibfield  {author} {\bibinfo {author} {\bibfnamefont {G.-Y.}\ \bibnamefont
  {Zhu}}\ and\ \bibinfo {author} {\bibfnamefont {G.-M.}\ \bibnamefont
  {Zhang}},\ }\bibfield  {title} {\bibinfo {title} {{Gapless Coulomb State
  Emerging from a Self-Dual Topological Tensor-Network State}},\ }\href
  {https://doi.org/10.1103/PhysRevLett.122.176401} {\bibfield  {journal}
  {\bibinfo  {journal} {Phys. Rev. Lett.}\ }\textbf {\bibinfo {volume} {122}},\
  \bibinfo {pages} {176401} (\bibinfo {year} {2019})}\BibitemShut {NoStop}%
\bibitem [{\citenamefont {Xu}\ \emph {et~al.}(2020)\citenamefont {Xu},
  \citenamefont {Zhang},\ and\ \citenamefont {Zhang}}]{gmz20fibonacci}%
  \BibitemOpen
  \bibfield  {author} {\bibinfo {author} {\bibfnamefont {W.-T.}\ \bibnamefont
  {Xu}}, \bibinfo {author} {\bibfnamefont {Q.}~\bibnamefont {Zhang}},\ and\
  \bibinfo {author} {\bibfnamefont {G.-M.}\ \bibnamefont {Zhang}},\ }\bibfield
  {title} {\bibinfo {title} {{Tensor Network Approach to Phase Transitions of a
  Non-Abelian Topological Phase}},\ }\href
  {https://doi.org/10.1103/PhysRevLett.124.130603} {\bibfield  {journal}
  {\bibinfo  {journal} {Phys. Rev. Lett.}\ }\textbf {\bibinfo {volume} {124}},\
  \bibinfo {pages} {130603} (\bibinfo {year} {2020})}\BibitemShut {NoStop}%
\bibitem [{\citenamefont {{Zhang}}\ \emph {et~al.}(2020)\citenamefont
  {{Zhang}}, \citenamefont {{Xu}}, \citenamefont {{Wang}},\ and\ \citenamefont
  {{Zhang}}}]{gmz20doublesemion}%
  \BibitemOpen
  \bibfield  {author} {\bibinfo {author} {\bibfnamefont {Q.}~\bibnamefont
  {{Zhang}}}, \bibinfo {author} {\bibfnamefont {W.-T.}\ \bibnamefont {{Xu}}},
  \bibinfo {author} {\bibfnamefont {Z.-Q.}\ \bibnamefont {{Wang}}},\ and\
  \bibinfo {author} {\bibfnamefont {G.-M.}\ \bibnamefont {{Zhang}}},\
  }\bibfield  {title} {\bibinfo {title} {{Non-Hermitian effects of the
  intrinsic signs in topologically ordered wavefunctions}},\ }\href
  {https://doi.org/10.1038/s42005-020-00479-y} {\bibfield  {journal} {\bibinfo
  {journal} {Communications Physics}\ }\textbf {\bibinfo {volume} {3}},\
  \bibinfo {eid} {209} (\bibinfo {year} {2020})}\BibitemShut {NoStop}%
\bibitem [{\citenamefont {Zhu}\ \emph {et~al.}(2022)\citenamefont {Zhu},
  \citenamefont {Chen}, \citenamefont {Ye},\ and\ \citenamefont
  {Trebst}}]{Zhu22fracton}%
  \BibitemOpen
  \bibfield  {author} {\bibinfo {author} {\bibfnamefont {G.-Y.}\ \bibnamefont
  {Zhu}}, \bibinfo {author} {\bibfnamefont {J.-Y.}\ \bibnamefont {Chen}},
  \bibinfo {author} {\bibfnamefont {P.}~\bibnamefont {Ye}},\ and\ \bibinfo
  {author} {\bibfnamefont {S.}~\bibnamefont {Trebst}},\ }\bibfield  {title}
  {\bibinfo {title} {Topological fracton quantum phase transitions by tuning
  exact tensor network states},\ }\href@noop {} {\  (\bibinfo {year} {2022})},\
  \Eprint {https://arxiv.org/abs/2203.00015} {arXiv:2203.00015} \BibitemShut
  {NoStop}%
\bibitem [{\citenamefont {Wolf}\ \emph {et~al.}(2006)\citenamefont {Wolf},
  \citenamefont {Ortiz}, \citenamefont {Verstraete},\ and\ \citenamefont
  {Cirac}}]{Wolf06}%
  \BibitemOpen
  \bibfield  {author} {\bibinfo {author} {\bibfnamefont {M.~M.}\ \bibnamefont
  {Wolf}}, \bibinfo {author} {\bibfnamefont {G.}~\bibnamefont {Ortiz}},
  \bibinfo {author} {\bibfnamefont {F.}~\bibnamefont {Verstraete}},\ and\
  \bibinfo {author} {\bibfnamefont {J.~I.}\ \bibnamefont {Cirac}},\ }\bibfield
  {title} {\bibinfo {title} {{Quantum Phase Transitions in Matrix Product
  Systems}},\ }\href {https://doi.org/10.1103/PhysRevLett.97.110403} {\bibfield
   {journal} {\bibinfo  {journal} {Phys. Rev. Lett.}\ }\textbf {\bibinfo
  {volume} {97}},\ \bibinfo {pages} {110403} (\bibinfo {year}
  {2006})}\BibitemShut {NoStop}%
\bibitem [{\citenamefont {Smith}\ \emph {et~al.}(2022)\citenamefont {Smith},
  \citenamefont {Jobst}, \citenamefont {Green},\ and\ \citenamefont
  {Pollmann}}]{Smith22}%
  \BibitemOpen
  \bibfield  {author} {\bibinfo {author} {\bibfnamefont {A.}~\bibnamefont
  {Smith}}, \bibinfo {author} {\bibfnamefont {B.}~\bibnamefont {Jobst}},
  \bibinfo {author} {\bibfnamefont {A.~G.}\ \bibnamefont {Green}},\ and\
  \bibinfo {author} {\bibfnamefont {F.}~\bibnamefont {Pollmann}},\ }\bibfield
  {title} {\bibinfo {title} {Crossing a topological phase transition with a
  quantum computer},\ }\href
  {https://doi.org/10.1103/PhysRevResearch.4.L022020} {\bibfield  {journal}
  {\bibinfo  {journal} {Phys. Rev. Research}\ }\textbf {\bibinfo {volume}
  {4}},\ \bibinfo {pages} {L022020} (\bibinfo {year} {2022})}\BibitemShut
  {NoStop}%
\bibitem [{\citenamefont {Jones}\ \emph {et~al.}(2021)\citenamefont {Jones},
  \citenamefont {Bibo}, \citenamefont {Jobst}, \citenamefont {Pollmann},
  \citenamefont {Smith},\ and\ \citenamefont {Verresen}}]{Jones21}%
  \BibitemOpen
  \bibfield  {author} {\bibinfo {author} {\bibfnamefont {N.~G.}\ \bibnamefont
  {Jones}}, \bibinfo {author} {\bibfnamefont {J.}~\bibnamefont {Bibo}},
  \bibinfo {author} {\bibfnamefont {B.}~\bibnamefont {Jobst}}, \bibinfo
  {author} {\bibfnamefont {F.}~\bibnamefont {Pollmann}}, \bibinfo {author}
  {\bibfnamefont {A.}~\bibnamefont {Smith}},\ and\ \bibinfo {author}
  {\bibfnamefont {R.}~\bibnamefont {Verresen}},\ }\bibfield  {title} {\bibinfo
  {title} {Skeleton of matrix-product-state-solvable models connecting
  topological phases of matter},\ }\href
  {https://doi.org/10.1103/PhysRevResearch.3.033265} {\bibfield  {journal}
  {\bibinfo  {journal} {Phys. Rev. Research}\ }\textbf {\bibinfo {volume}
  {3}},\ \bibinfo {pages} {033265} (\bibinfo {year} {2021})}\BibitemShut
  {NoStop}%
\bibitem [{\citenamefont {Tantivasadakarn}\ \emph
  {et~al.}(2021{\natexlab{b}})\citenamefont {Tantivasadakarn}, \citenamefont
  {Thorngren}, \citenamefont {Vishwanath},\ and\ \citenamefont
  {Verresen}}]{pivot}%
  \BibitemOpen
  \bibfield  {author} {\bibinfo {author} {\bibfnamefont {N.}~\bibnamefont
  {Tantivasadakarn}}, \bibinfo {author} {\bibfnamefont {R.}~\bibnamefont
  {Thorngren}}, \bibinfo {author} {\bibfnamefont {A.}~\bibnamefont
  {Vishwanath}},\ and\ \bibinfo {author} {\bibfnamefont {R.}~\bibnamefont
  {Verresen}},\ }\bibfield  {title} {\bibinfo {title} {{Pivot Hamiltonians as
  generators of symmetry and entanglement}},\ }\href@noop {} {\  (\bibinfo
  {year} {2021}{\natexlab{b}})},\ \Eprint {https://arxiv.org/abs/2110.07599}
  {arXiv:2110.07599} \BibitemShut {NoStop}%
\bibitem [{\citenamefont {Browaeys}\ and\ \citenamefont
  {Lahaye}(2020)}]{Browaeys2020}%
  \BibitemOpen
  \bibfield  {author} {\bibinfo {author} {\bibfnamefont {A.}~\bibnamefont
  {Browaeys}}\ and\ \bibinfo {author} {\bibfnamefont {T.}~\bibnamefont
  {Lahaye}},\ }\bibfield  {title} {\bibinfo {title} {{Many-body physics with
  individually controlled Rydberg atoms}},\ }\href
  {https://doi.org/10.1038/s41567-019-0733-z} {\bibfield  {journal} {\bibinfo
  {journal} {Nature Physics}\ }\textbf {\bibinfo {volume} {16}},\ \bibinfo
  {pages} {132} (\bibinfo {year} {2020})}\BibitemShut {NoStop}%
\bibitem [{\citenamefont {{Ebadi}}\ \emph {et~al.}(2021)\citenamefont
  {{Ebadi}}, \citenamefont {{Wang}}, \citenamefont {{Levine}}, \citenamefont
  {{Keesling}}, \citenamefont {{Semeghini}}, \citenamefont {{Omran}},
  \citenamefont {{Bluvstein}}, \citenamefont {{Samajdar}}, \citenamefont
  {{Pichler}}, \citenamefont {{Ho}}, \citenamefont {{Choi}}, \citenamefont
  {{Sachdev}}, \citenamefont {{Greiner}}, \citenamefont {{Vuleti{\'c}}},\ and\
  \citenamefont {{Lukin}}}]{Lukin2021atom256}%
  \BibitemOpen
  \bibfield  {author} {\bibinfo {author} {\bibfnamefont {S.}~\bibnamefont
  {{Ebadi}}}, \bibinfo {author} {\bibfnamefont {T.~T.}\ \bibnamefont {{Wang}}},
  \bibinfo {author} {\bibfnamefont {H.}~\bibnamefont {{Levine}}}, \bibinfo
  {author} {\bibfnamefont {A.}~\bibnamefont {{Keesling}}}, \bibinfo {author}
  {\bibfnamefont {G.}~\bibnamefont {{Semeghini}}}, \bibinfo {author}
  {\bibfnamefont {A.}~\bibnamefont {{Omran}}}, \bibinfo {author} {\bibfnamefont
  {D.}~\bibnamefont {{Bluvstein}}}, \bibinfo {author} {\bibfnamefont
  {R.}~\bibnamefont {{Samajdar}}}, \bibinfo {author} {\bibfnamefont
  {H.}~\bibnamefont {{Pichler}}}, \bibinfo {author} {\bibfnamefont {W.~W.}\
  \bibnamefont {{Ho}}}, \bibinfo {author} {\bibfnamefont {S.}~\bibnamefont
  {{Choi}}}, \bibinfo {author} {\bibfnamefont {S.}~\bibnamefont {{Sachdev}}},
  \bibinfo {author} {\bibfnamefont {M.}~\bibnamefont {{Greiner}}}, \bibinfo
  {author} {\bibfnamefont {V.}~\bibnamefont {{Vuleti{\'c}}}},\ and\ \bibinfo
  {author} {\bibfnamefont {M.~D.}\ \bibnamefont {{Lukin}}},\ }\bibfield
  {title} {\bibinfo {title} {{Quantum phases of matter on a 256-atom
  programmable quantum simulator}},\ }\href
  {https://doi.org/10.1038/s41586-021-03582-4} {\bibfield  {journal} {\bibinfo
  {journal} {Nature}\ }\textbf {\bibinfo {volume} {595}},\ \bibinfo {pages}
  {227} (\bibinfo {year} {2021})}\BibitemShut {NoStop}%
\bibitem [{\citenamefont {{Scholl}}\ \emph {et~al.}(2021)\citenamefont
  {{Scholl}}, \citenamefont {{Schuler}}, \citenamefont {{Williams}},
  \citenamefont {{Eberharter}}, \citenamefont {{Barredo}}, \citenamefont
  {{Schymik}}, \citenamefont {{Lienhard}}, \citenamefont {{Henry}},
  \citenamefont {{Lang}}, \citenamefont {{Lahaye}}, \citenamefont
  {{L{\"a}uchli}},\ and\ \citenamefont {{Browaeys}}}]{Browaeys2021}%
  \BibitemOpen
  \bibfield  {author} {\bibinfo {author} {\bibfnamefont {P.}~\bibnamefont
  {{Scholl}}}, \bibinfo {author} {\bibfnamefont {M.}~\bibnamefont {{Schuler}}},
  \bibinfo {author} {\bibfnamefont {H.~J.}\ \bibnamefont {{Williams}}},
  \bibinfo {author} {\bibfnamefont {A.~A.}\ \bibnamefont {{Eberharter}}},
  \bibinfo {author} {\bibfnamefont {D.}~\bibnamefont {{Barredo}}}, \bibinfo
  {author} {\bibfnamefont {K.-N.}\ \bibnamefont {{Schymik}}}, \bibinfo {author}
  {\bibfnamefont {V.}~\bibnamefont {{Lienhard}}}, \bibinfo {author}
  {\bibfnamefont {L.-P.}\ \bibnamefont {{Henry}}}, \bibinfo {author}
  {\bibfnamefont {T.~C.}\ \bibnamefont {{Lang}}}, \bibinfo {author}
  {\bibfnamefont {T.}~\bibnamefont {{Lahaye}}}, \bibinfo {author}
  {\bibfnamefont {A.~M.}\ \bibnamefont {{L{\"a}uchli}}},\ and\ \bibinfo
  {author} {\bibfnamefont {A.}~\bibnamefont {{Browaeys}}},\ }\bibfield  {title}
  {\bibinfo {title} {{Quantum simulation of 2D antiferromagnets with hundreds
  of Rydberg atoms}},\ }\href {https://doi.org/10.1038/s41586-021-03585-1}
  {\bibfield  {journal} {\bibinfo  {journal} {Nature}\ }\textbf {\bibinfo
  {volume} {595}},\ \bibinfo {pages} {233} (\bibinfo {year}
  {2021})}\BibitemShut {NoStop}%
\bibitem [{\citenamefont {{Bluvstein}}\ \emph {et~al.}(2022)\citenamefont
  {{Bluvstein}}, \citenamefont {{Levine}}, \citenamefont {{Semeghini}},
  \citenamefont {{Wang}}, \citenamefont {{Ebadi}}, \citenamefont
  {{Kalinowski}}, \citenamefont {{Keesling}}, \citenamefont {{Maskara}},
  \citenamefont {{Pichler}}, \citenamefont {{Greiner}}, \citenamefont
  {{Vuleti{\'c}}},\ and\ \citenamefont {{Lukin}}}]{Lukin2022quantumprocessor}%
  \BibitemOpen
  \bibfield  {author} {\bibinfo {author} {\bibfnamefont {D.}~\bibnamefont
  {{Bluvstein}}}, \bibinfo {author} {\bibfnamefont {H.}~\bibnamefont
  {{Levine}}}, \bibinfo {author} {\bibfnamefont {G.}~\bibnamefont
  {{Semeghini}}}, \bibinfo {author} {\bibfnamefont {T.~T.}\ \bibnamefont
  {{Wang}}}, \bibinfo {author} {\bibfnamefont {S.}~\bibnamefont {{Ebadi}}},
  \bibinfo {author} {\bibfnamefont {M.}~\bibnamefont {{Kalinowski}}}, \bibinfo
  {author} {\bibfnamefont {A.}~\bibnamefont {{Keesling}}}, \bibinfo {author}
  {\bibfnamefont {N.}~\bibnamefont {{Maskara}}}, \bibinfo {author}
  {\bibfnamefont {H.}~\bibnamefont {{Pichler}}}, \bibinfo {author}
  {\bibfnamefont {M.}~\bibnamefont {{Greiner}}}, \bibinfo {author}
  {\bibfnamefont {V.}~\bibnamefont {{Vuleti{\'c}}}},\ and\ \bibinfo {author}
  {\bibfnamefont {M.~D.}\ \bibnamefont {{Lukin}}},\ }\bibfield  {title}
  {\bibinfo {title} {{A quantum processor based on coherent transport of
  entangled atom arrays}},\ }\href {https://doi.org/10.1038/s41586-022-04592-6}
  {\bibfield  {journal} {\bibinfo  {journal} {\nat}\ }\textbf {\bibinfo
  {volume} {604}},\ \bibinfo {pages} {451} (\bibinfo {year}
  {2022})}\BibitemShut {NoStop}%
\bibitem [{\citenamefont {Singh}\ \emph {et~al.}(2022)\citenamefont {Singh},
  \citenamefont {Anand}, \citenamefont {Pocklington}, \citenamefont {Kemp},\
  and\ \citenamefont {Bernien}}]{Singh22}%
  \BibitemOpen
  \bibfield  {author} {\bibinfo {author} {\bibfnamefont {K.}~\bibnamefont
  {Singh}}, \bibinfo {author} {\bibfnamefont {S.}~\bibnamefont {Anand}},
  \bibinfo {author} {\bibfnamefont {A.}~\bibnamefont {Pocklington}}, \bibinfo
  {author} {\bibfnamefont {J.~T.}\ \bibnamefont {Kemp}},\ and\ \bibinfo
  {author} {\bibfnamefont {H.}~\bibnamefont {Bernien}},\ }\bibfield  {title}
  {\bibinfo {title} {Dual-element, two-dimensional atom array with
  continuous-mode operation},\ }\href
  {https://doi.org/10.1103/PhysRevX.12.011040} {\bibfield  {journal} {\bibinfo
  {journal} {Phys. Rev. X}\ }\textbf {\bibinfo {volume} {12}},\ \bibinfo
  {pages} {011040} (\bibinfo {year} {2022})}\BibitemShut {NoStop}%
\bibitem [{\citenamefont {Zhang}\ \emph {et~al.}(2021)\citenamefont {Zhang},
  \citenamefont {Picard}, \citenamefont {Cairncross}, \citenamefont {Wang},
  \citenamefont {Yu}, \citenamefont {Fang},\ and\ \citenamefont
  {Ni}}]{Zhang21}%
  \BibitemOpen
  \bibfield  {author} {\bibinfo {author} {\bibfnamefont {J.~T.}\ \bibnamefont
  {Zhang}}, \bibinfo {author} {\bibfnamefont {L.~R.~B.}\ \bibnamefont
  {Picard}}, \bibinfo {author} {\bibfnamefont {W.~B.}\ \bibnamefont
  {Cairncross}}, \bibinfo {author} {\bibfnamefont {K.}~\bibnamefont {Wang}},
  \bibinfo {author} {\bibfnamefont {Y.}~\bibnamefont {Yu}}, \bibinfo {author}
  {\bibfnamefont {F.}~\bibnamefont {Fang}},\ and\ \bibinfo {author}
  {\bibfnamefont {K.-K.}\ \bibnamefont {Ni}},\ }\bibfield  {title} {\bibinfo
  {title} {An optical tweezer array of ground-state polar molecules},\
  }\href@noop {} {\  (\bibinfo {year} {2021})},\ \Eprint
  {https://arxiv.org/abs/2112.00991} {arXiv:2112.00991} \BibitemShut {NoStop}%
\bibitem [{\citenamefont {Choi}\ \emph {et~al.}(2020)\citenamefont {Choi},
  \citenamefont {Bao}, \citenamefont {Qi},\ and\ \citenamefont
  {Altman}}]{Altman2020qec}%
  \BibitemOpen
  \bibfield  {author} {\bibinfo {author} {\bibfnamefont {S.}~\bibnamefont
  {Choi}}, \bibinfo {author} {\bibfnamefont {Y.}~\bibnamefont {Bao}}, \bibinfo
  {author} {\bibfnamefont {X.-L.}\ \bibnamefont {Qi}},\ and\ \bibinfo {author}
  {\bibfnamefont {E.}~\bibnamefont {Altman}},\ }\bibfield  {title} {\bibinfo
  {title} {{Quantum Error Correction in Scrambling Dynamics and
  Measurement-Induced Phase Transition}},\ }\href
  {https://doi.org/10.1103/PhysRevLett.125.030505} {\bibfield  {journal}
  {\bibinfo  {journal} {Phys. Rev. Lett.}\ }\textbf {\bibinfo {volume} {125}},\
  \bibinfo {pages} {030505} (\bibinfo {year} {2020})}\BibitemShut {NoStop}%
\bibitem [{\citenamefont {Gullans}\ and\ \citenamefont
  {Huse}(2020)}]{Huse2020qec}%
  \BibitemOpen
  \bibfield  {author} {\bibinfo {author} {\bibfnamefont {M.~J.}\ \bibnamefont
  {Gullans}}\ and\ \bibinfo {author} {\bibfnamefont {D.~A.}\ \bibnamefont
  {Huse}},\ }\bibfield  {title} {\bibinfo {title} {{Dynamical Purification
  Phase Transition Induced by Quantum Measurements}},\ }\href
  {https://doi.org/10.1103/PhysRevX.10.041020} {\bibfield  {journal} {\bibinfo
  {journal} {Phys. Rev. X}\ }\textbf {\bibinfo {volume} {10}},\ \bibinfo
  {pages} {041020} (\bibinfo {year} {2020})}\BibitemShut {NoStop}%
\bibitem [{\citenamefont {Noel}\ \emph {et~al.}(2022)\citenamefont {Noel},
  \citenamefont {Niroula}, \citenamefont {Zhu}, \citenamefont {Risinger},
  \citenamefont {Egan}, \citenamefont {Biswas}, \citenamefont {Cetina},
  \citenamefont {Gorshkov}, \citenamefont {Gullans}, \citenamefont {Huse},\
  and\ \citenamefont {Monroe}}]{Noel2022trappedionMIPT}%
  \BibitemOpen
  \bibfield  {author} {\bibinfo {author} {\bibfnamefont {C.}~\bibnamefont
  {Noel}}, \bibinfo {author} {\bibfnamefont {P.}~\bibnamefont {Niroula}},
  \bibinfo {author} {\bibfnamefont {D.}~\bibnamefont {Zhu}}, \bibinfo {author}
  {\bibfnamefont {A.}~\bibnamefont {Risinger}}, \bibinfo {author}
  {\bibfnamefont {L.}~\bibnamefont {Egan}}, \bibinfo {author} {\bibfnamefont
  {D.}~\bibnamefont {Biswas}}, \bibinfo {author} {\bibfnamefont
  {M.}~\bibnamefont {Cetina}}, \bibinfo {author} {\bibfnamefont {A.~V.}\
  \bibnamefont {Gorshkov}}, \bibinfo {author} {\bibfnamefont {M.~J.}\
  \bibnamefont {Gullans}}, \bibinfo {author} {\bibfnamefont {D.~A.}\
  \bibnamefont {Huse}},\ and\ \bibinfo {author} {\bibfnamefont
  {C.}~\bibnamefont {Monroe}},\ }\bibfield  {title} {\bibinfo {title}
  {Measurement-induced quantum phases realized in a trapped-ion quantum
  computer},\ }\href {https://doi.org/10.1038/s41567-022-01619-7} {\bibfield
  {journal} {\bibinfo  {journal} {Nature Physics}\ }\textbf {\bibinfo {volume}
  {18}},\ \bibinfo {pages} {760} (\bibinfo {year} {2022})}\BibitemShut
  {NoStop}%
\bibitem [{\citenamefont {Dehghani}\ \emph {et~al.}(2022)\citenamefont
  {Dehghani}, \citenamefont {Lavasani}, \citenamefont {Hafezi},\ and\
  \citenamefont {Gullans}}]{Gullans22decoder}%
  \BibitemOpen
  \bibfield  {author} {\bibinfo {author} {\bibfnamefont {H.}~\bibnamefont
  {Dehghani}}, \bibinfo {author} {\bibfnamefont {A.}~\bibnamefont {Lavasani}},
  \bibinfo {author} {\bibfnamefont {M.}~\bibnamefont {Hafezi}},\ and\ \bibinfo
  {author} {\bibfnamefont {M.~J.}\ \bibnamefont {Gullans}},\ }\bibfield
  {title} {\bibinfo {title} {{Neural-Network Decoders for Measurement Induced
  Phase Transitions}},\ }\href@noop {} {\  (\bibinfo {year} {2022})},\ \Eprint
  {https://arxiv.org/abs/2204.10904} {arXiv:2204.10904} \BibitemShut {NoStop}%
\bibitem [{\citenamefont {Barratt}\ \emph {et~al.}(2022)\citenamefont
  {Barratt}, \citenamefont {Agarwal}, \citenamefont {Potter}, \citenamefont
  {Gopalakrishnan},\ and\ \citenamefont {Vasseur}}]{Vasseur2022qec}%
  \BibitemOpen
  \bibfield  {author} {\bibinfo {author} {\bibfnamefont {F.}~\bibnamefont
  {Barratt}}, \bibinfo {author} {\bibfnamefont {U.}~\bibnamefont {Agarwal}},
  \bibinfo {author} {\bibfnamefont {A.~C.}\ \bibnamefont {Potter}}, \bibinfo
  {author} {\bibfnamefont {S.}~\bibnamefont {Gopalakrishnan}},\ and\ \bibinfo
  {author} {\bibfnamefont {R.}~\bibnamefont {Vasseur}},\ }\bibfield  {title}
  {\bibinfo {title} {Transitions in the learnability of global charges from
  local measurements},\ }\href@noop {} {\  (\bibinfo {year} {2022})},\ \Eprint
  {https://arxiv.org/abs/2206.12429} {arXiv:2206.12429} \BibitemShut {NoStop}%
\bibitem [{\citenamefont {Raussendorf}\ \emph {et~al.}(2006)\citenamefont
  {Raussendorf}, \citenamefont {Harrington},\ and\ \citenamefont
  {Goyal}}]{Raussendorf2006}%
  \BibitemOpen
  \bibfield  {author} {\bibinfo {author} {\bibfnamefont {R.}~\bibnamefont
  {Raussendorf}}, \bibinfo {author} {\bibfnamefont {J.}~\bibnamefont
  {Harrington}},\ and\ \bibinfo {author} {\bibfnamefont {K.}~\bibnamefont
  {Goyal}},\ }\bibfield  {title} {\bibinfo {title} {{A fault-tolerant one-way
  quantum computer}},\ }\href
  {https://doi.org/https://doi.org/10.1016/j.aop.2006.01.012} {\bibfield
  {journal} {\bibinfo  {journal} {Annals of Physics}\ }\textbf {\bibinfo
  {volume} {321}},\ \bibinfo {pages} {2242} (\bibinfo {year}
  {2006})}\BibitemShut {NoStop}%
\bibitem [{\citenamefont {Lee}\ \emph {et~al.}(2022)\citenamefont {Lee},
  \citenamefont {Ji}, \citenamefont {Bi},\ and\ \citenamefont
  {Fisher}}]{JYLee}%
  \BibitemOpen
  \bibfield  {author} {\bibinfo {author} {\bibfnamefont {J.~Y.}\ \bibnamefont
  {Lee}}, \bibinfo {author} {\bibfnamefont {W.}~\bibnamefont {Ji}}, \bibinfo
  {author} {\bibfnamefont {Z.}~\bibnamefont {Bi}},\ and\ \bibinfo {author}
  {\bibfnamefont {M.~P.~A.}\ \bibnamefont {Fisher}},\ }\bibfield  {title}
  {\bibinfo {title} {{Measurement-Prepared Quantum Criticality: from Ising
  model to gauge theory, and beyond}},\ }\href@noop {} {\  (\bibinfo {year}
  {2022})},\ \Eprint {https://arxiv.org/abs/2208.11699} {arXiv:2208.11699}
  \BibitemShut {NoStop}%
\bibitem [{\citenamefont {{Zhu}}\ \emph {et~al.}(2023)\citenamefont {{Zhu}},
  \citenamefont {{Tantivasadakarn}}, \citenamefont {{Vishwanath}},
  \citenamefont {{Trebst}},\ and\ \citenamefont
  {{Verresen}}}]{zenodo_NishimoriCat}%
  \BibitemOpen
  \bibfield  {author} {\bibinfo {author} {\bibfnamefont {G.-Y.}\ \bibnamefont
  {{Zhu}}}, \bibinfo {author} {\bibfnamefont {N.}~\bibnamefont
  {{Tantivasadakarn}}}, \bibinfo {author} {\bibfnamefont {A.}~\bibnamefont
  {{Vishwanath}}}, \bibinfo {author} {\bibfnamefont {S.}~\bibnamefont
  {{Trebst}}},\ and\ \bibinfo {author} {\bibfnamefont {R.}~\bibnamefont
  {{Verresen}}},\ }\bibfield  {title} {\bibinfo {title} {{Data for
  ``Nishimori's cat: stable long-range entanglement from finite-depth unitaries
  and weak measurements"}}\ }\href {https://doi.org/10.5281/zenodo.10025335}
  {10.5281/zenodo.10025335} (\bibinfo {year} {2023})\BibitemShut {NoStop}%
\end{thebibliography}%


\clearpage
\appendix

\begin{center}
{\textbf{SUPPLEMENTAL MATERIAL}}
\end{center}

In this supplemental material, we first show details of our weak measurement protocol, and then discuss the resultant correlated disorder, for which we show how the gauge symmetry emerging in a restricted perturbation line allows analytical treatment. 
For more general perturbation we discuss the tensor-network representation, and propose a hybrid tensor-network / Monte Carlo algorithm to perform the numerical calculations. Relatedly, we also provide numerical results for the alternative heavy-hexagon lattice geometry. 
In the end, we discuss the route to the 3D glassy topological order, as well as the 1D SRE phase transition.

\section{Appendix A: Multibody measurement via ancilla}
\label{appendixA}

In order to realize interesting long-range-entangled (LRE) states under finite-depth circuit, we would like to measure certain stabilizers defined as multibody Pauli string operators. 
To achieve this by more realistic single-site measurement, we can introduce ancilla spins, which are entangled with the target spins in such a way that measuring the ancillas indirectly measures the target stabilizers. 
The parameter that controls the entanglement between target spins and ancillas effectively controls the strength of measurement.  

First let us show the way to measure a single arbitrary Pauli string operator $\mathcal{O}$ (satisfying $\mathcal{O}^\dag = \mathcal{O}$, $\mathcal{O}^2 = 1$) acting on the physical spins. By preparing the ancillas in the $x$-basis, we time evolve it with the physical spins by $s^z \otimes \mathcal O$, after which we projectively measure the ancilla in $y$-basis. As a result, we get an effective non-unitary operator as follows: 
\begin{equation}
\begin{split}
M_s
&=\langle s^y\equiv s|e^{-its^z \otimes \mathcal O} |+_x\rangle\\
&=
\begin{cases}
\frac{1-i}{2} \left( \cos t +  \mathcal{O}\sin t \right), & s=+1\\
\frac{1+i}{2} \left( \cos t - \mathcal{O}\sin t  \right), & s=-1\\
\end{cases}\\
 &\propto e^{\frac{1}{2}\beta s \mathcal{O}}\, ,
\end{split}
\label{Ref:eqn:SI1}
\end{equation}
up to a phase factor, where the real evolution time is effectively transformed into an imaginary time $\beta \equiv 2\tanh^{-1} ( \tan t )$. The phase factor is fixed for each measurement outcome, and thus can be gauged away from the postmeasurement wave function without any physical consequence. The physical meaning of the postmeasurement non-unitary operator is that of a weak measurement (implemented via an imaginary time evolution), where the effective temperature tunes the strength of the measurement. At zero temperature ($\beta=\infty$), it becomes a projector for a strong measurement. 

Second, to measure a set of (mutually commuting) $\{\mathcal{O}_j\}$ stabilizers, one can simply introduce a set of ancilla spins $\{s_j\}$ and repeatedly apply the weak measurement operator \eqref{Ref:eqn:SI1} for every $j$ term to reach the manybody non-unitary operator $\exp(-\frac{1}{2} \beta \sum_j s_j \mathcal{O}_j)$,  which stabilizes the stabilizer state at the limit $\beta\to\infty$. Note that this can be achieved by a {\it finite-depth} circuit, where the depth is bounded by the maximal number of stabilizers that share common spin, instead of diverging with the system size. 

Thirdly, let us focus on stabilizers $\{\mathcal{O}_j\}$ being purely Pauli Z strings, and apply the weak measurement for them upon a spin product state $\ket{+}$ in $x$-basis, which have one to one correspondence with certain classical model in the same spatial dimension analogous to the Rokhsar-Kivelson state\cite{Henley04RK, Fradkin04RK, Cirac06}. 
Namely, we turn on the fixed-depth circuit implementing $\exp(-i t \sum_j s_j^z\otimes \mathcal{O}_j)$, and then measure all the ancillas in $y$ basis, which results in a postmeasurement state for the target spins as follows:
\begin{equation}
e^{-\frac{1}{2} \beta \sum_j s_j \mathcal{O}_j} \ket{+}.
\end{equation}
All diagonal correlations in the postmeasurement state are described by the classical model capturing the fluctuation of the stabilizers\cite{Henley04RK, Fradkin04RK, Cirac06}:
\begin{equation}
\norm{e^{-\frac{1}{2} \beta \sum_j s_j \mathcal{O}_j} \ket{+}}^2 = \sum_{\{\sigma\}}e^{-\beta \sum_j s_j\mathcal{O}_j}.
\end{equation}

Nevertheless, the multi-body unitary gate evolution may render difficulty in experiment. It is thus desirable to break such entanglers into smaller pieces while reaching the same result. For example, it can always be decomposed into two Pauli string operators $\mathcal{O}_A$ and $ \mathcal{O}_B$ such that $\mathcal{O} = \mathcal{O}_A \mathcal{O}_B$.
Then by separately entangling them with the ancilla and then measuring the ancilla, we get the effective non-unitary remnant operator parametrized by two evolution times (where for convenience we now instead measure in the $x$-basis, which also closely matches the cluster state set-up):
\begin{equation}
\begin{split}
M_s
=&\bra{ s^x\equiv s} e^{-i s^z \otimes (t_A \mathcal{O}_A+t_B\mathcal{O}_B)} |+\rangle\\
=&
\begin{cases}
\cos t_A\cos t_B - \sin t_A \sin t_B \mathcal{O} , & s=+1\\
-i\mathcal{O}_B \left(\cos t_A\sin t_B + \sin t_A \cos t_B \mathcal{O}\right) , & s=-1\\
\end{cases}\\
 \propto& e^{-\frac{1}{2} \beta (J_s \mathcal{O} + h s)}\, .
\end{split}
\label{eq:twotermprotocol}
\end{equation}
Note that since $|i\mathcal O_B|^2 =1$, it does not affect the resulting classical partition function:
\begin{equation}
	\label{eq:partition_function}
	p_{\{s\}}  \equiv \norm{\bra{\{s\}}\ket{\psi}}^2
	=\bra{+} M^\dag M \ket{+}
	 \propto  \sum_{\{\sigma\}}e^{-\beta \sum_{j} (J_{s_j} \mathcal{O}_j + h s_j)} \,,
\end{equation}
and the resultant effective couplings are determined by the measurement outcomes as follows:
\begin{equation}
\begin{split}
&
 \tanh \frac{\beta}{2} J_+ = \tan t_A \tan t_B,\quad \tanh \frac{\beta}{2} J_- = -\tan t_A \cot t_B\, ,\\
&
\beta h = \frac{1}{2}\ln\left|\tan(t_A+t_B)\tan(t_A-t_B)\right| \,.
\end{split}
\end{equation}
This two-parameter protocol~\eqref{eq:twotermprotocol} contains a subspace $t_B=\pi/4$ that recovers the one-parameter protocol~\eqref{Ref:eqn:SI1}, up to a measurement outcome dependent local basis transformation. 
Even though the circuit is deterministic, the measurement inevitably introduces inherent randomness underlying the fundamental quantum mechanics. We can treat the random measurement outcome of ancillas $\{s=\pm 1\}$  as a disorder sample, which compose the disorder ensemble. The probability is dictated by the Born's rule $p_{\{s\}} = \norm{\bra{\{s\}}\ket{\psi}}^2$. 

Below we apply this engineering principle to  two minimal representative examples: 
\begin{itemize}
\item{Ising protocol:} in a bipartite lattice in any dimension, we place ancillas on the bond centers, and choose $\mathcal{O}_{A(B)} = \sigma_{A(B)}^z$ for the site-A(B) sublattice, thus reaching an effective model with nearest-neighbor Ising interaction $\mathcal{O} = \sigma_A^z\sigma_B^z$. 
Note that the Ising symmetry in the effective classical model originates from the initial state $\ket{+}$ and the measurement observable $\mathcal{O}$ being Ising symmetric. In view of the circuit, the Ising symmetry in the postmeasurement state $\prod_j \sigma_j^x \propto \prod_j \sigma_j^x\prod_l s_l^x$ can be guaranteed if the all the gates preserve the global Ising symmetry including both the spins and ancillas. 
\item{Ising gauge protocol:} in a hyper-cubic lattice, we place ancillas to the plaquette centers, and choose $\mathcal{O}_{A} = \sigma_l^z \sigma_u^z$, while $\mathcal{O}_{B} = \sigma_r^z \sigma_d^z$, where $l,u,r,d$ labels the physical spins on the left, up, right, and down edge of the plaquette. In this way, we reach an effective model with plaquette Ising interaction $\mathcal{O} = \sigma_l^z\sigma_u^z\sigma_r^z\sigma_d^z$. 
Note that the initial state $\ket{+}$ and the measurement observable $\mathcal{O}$ guarantees the 1-form symmetry resulting in charge-free condition, and the 2-form symmetry present in strong measurement limit is perturbatively stable.
\end{itemize}
The recipe can be applied to arbitrary dimensions, but in the following we will mainly focus on the 2D Ising protocol, while the 3D Ising gauge protocol and the 1D protocol will be discussed in the end.


\section{Appendix B: Analytic discussions for 2D Ising protocol}
In this section we focus on the 2D Ising protocol and analytically discuss the correlated disorder ensemble as well as the gauge symmetric Nishimori line. 

\subsection{Disorder correlation}
To characterize the disorder ensemble for the 2-body Ising protocol, a central quantity is the string correlation:
\begin{equation}
\begin{split}
&[\prod_{l\in \mathcal{S}_{ij}}s_{l}] =  \bra{\psi} \prod_{l\in \mathcal{S}_{ij}}s_{l}^x \ket{\psi} \\
 =& 
\left( \cos(2t_A)\cos(2t_B)\right)^{|\mathcal{S}_{ij}|} 
+ \delta_{i,j} \left( -\sin(2t_A)\sin(2t_B) \right)^{|\mathcal{S}_{ij}|}\,,\\
\end{split}
\label{eq:wilsonloop}
\end{equation}
where $[\cdots]$ denotes measurement average over the measurement samples $\{s\}$; $\mathcal{S}_{ij}$ denotes a string configuration \textit{along the links} of the lattice terminating at site $i$ and $j$; $|\mathcal{S}_{ij}|$ counts its length; and the $\delta_{ij}$ in the second contribution accounts for the correlation of the closed loop when $i$ is the same site as $j$. 
It gives the measurement average of a single local ancilla:
\begin{equation}
[s] = \bra{\psi} s^x \ket{\psi} = \cos(2t_A)\cos(2t_B) \,.
\end{equation} 
Thus the connected correlation 
\begin{equation}
\begin{split}
&[\prod_{l\in \mathcal{S}_{ij}}s_{l}] - \prod_{l\in \mathcal{S}_{ij}}[s_{l}] =  \delta_{i,j} \left( -\sin(2t_A)\sin(2t_B) \right)^{|\mathcal{S}_{ij}|}\,,\\
\end{split}
\end{equation}
is nonzero if and only if the ancillas form a {\it closed loop}. 
For example, the four ancillas surrounding a plaquette:
\begin{equation}
\begin{split}
&[\prod_{l\in \square} s_l] = \left(\cos(2t_A)\cos(2t_B)\right)^4 + \left(\sin(2t_A)\sin(2t_B)\right)^4 \,.
\end{split}
\label{eq:ssss}
\end{equation} 
The correlation vanishes in the limit $t_A=0$ or $t_B=0$, and becomes strongest when $t_A=t_B=\pi/4$. 
When an open string is decorated by site spins $\sigma$ at the end points, this defines a string operator
\begin{equation}
\begin{split}
 \bra{\psi} \sigma_i^z \left(\prod_{l\in \mathcal{S}_{ij}}s_{l}^x \right)  \sigma_j^z \ket{\psi} 
 &= 
\left( -\sin(2t_A)\sin(2t_B) \right)^{|\mathcal{S}_{ij}|} \,,
\end{split}
\label{eq:wilsonline}
\end{equation}
which can be interpreted as the gauge invariant Wilson line when gauge symmetry is present, i.e.\ $t_{B(A)}=\pi/4$. This Wilson line is always exponential decaying whenever away from the limit $t_A=t_B=\pi/4$. 

The non-local correlation derived in this section tells us that the disorder ensemble is highly correlated in general. Nevertheless, we will provide two treatments
\begin{itemize}
\item If the measurement is weakened along a restricted line such that gauge symmetry emerges, the disorder ensemble can be {\it uncorrelated} upon {\it gauge fixing}. Then the problem can be treated with standard disorder methods. 
\item For general perturbation without an apparent gauge symmetry, we can represent the probability function as a tensor-network. Then we propose a hybrid Monte Carlo and Tensor-network numerical method. 
\end{itemize}


\subsection{Gauge symmetric Nishimori line}
In this section we focus on the line of varying $t_A$ while fixing $t_{B}=\pi/4$ (or vice versa), and show the emergent $Z_2$ gauge symmetry as well as its consequences, and draw the connection to the classical Nishimori line in random bond Ising model upon gauge fixing. 

The disorder probability function along this line is simplified to
\begin{equation}
	p_{\{s\}} = Z_{\{s\}} \propto  \sum_{\{\sigma\}}e^{-\beta \sum_{ij} s_{ij} \sigma_i \sigma_j } \,,
\end{equation}
which also serves as the partition function. 
The classical energy is invariant under a gauge symmetry
\begin{equation}
\sigma_j' = \sigma_j\tau_j,
\quad
s_{ij}' = s_{ij}\tau_i\tau_j\, ,
\end{equation}
where $\{\tau_j=\pm 1\}$ specifies the Ising gauge configuration. 
Note that the configuration $\{s\}$ can be faithfully represented as a string configuration along the dual lattice, by interpreting $s=1$ as the vacuum while $s=-1$ as a string segment. The open string terminates at a pair of frustrated plaquettes where $\prod_{l\in \square} s_l=-1$, which defines the Ising vortex. The vortex distribution is invariant under the gauge transformation. The vortex pairs generally have effectively attractive interaction because the open string defect generally costs energy that scales linearly with its length in the Ising model in the ordered phase. 

The gauge symmetry allows us to view the dummy variable in the probability function as a gauge variable, and to avoid confusion let us relabel this dummy variable in the probability function as $\{\tau\}$, while that in the partition function as $\{\sigma\}$. Their respective roles are most transparently seen in the measurement averaged free energy by putting together the probability function and partition function:
\begin{equation}
\begin{split}
F 
=&\sum_{\{s\}} p_{\{s\}} \ln Z_{\{s\}}\\
=&\sum_{\{s\}}
\left(\sum_{\{\tau\}}e^{-\beta \sum_{ij} s_{ij} \tau_i \tau_j }\right)
\ln \left(
\sum_{\{\sigma\}}e^{-\beta \sum_{ij} s_{ij} \sigma_i \sigma_j }
\right)\\
=&\sum_{\{\tau\}}\sum_{\{s'\}}
e^{-\beta \sum_{ij} s'_{ij} }
\ln \left(
\sum_{\{\sigma'\}}e^{-\beta \sum_{ij} s'_{ij} \sigma'_i \sigma'_j }
\right).
\end{split}
\end{equation}
Note the connection to Nishimori's operation in Ref.\cite{Nishimori1981}: Nishimori started from the RBIM with uncorrelated disorder and performed gauge symmetrization to obtain the form $p_{\{s\}} \propto Z_{\{s\}}$, going from the bottom line to the first line in the equation above. In contrast, starting from our model we are fixing the gauge to establish its connection to the RBIM. 

In this way, the measurement average for the $n^\text{th}$ moment of an arbitrary correlation function can be brought to the gauge symmetrized form:
\begin{align}
[\langle\sigma_i\cdots \sigma_j\rangle^n]
&=
\sum_{\{s\}} p_{\{s\}} \langle \sigma_i\cdots \sigma_j\rangle_{\{s\}}^n\nonumber\\
&=
\sum_{\{\tau\}} \left(\tau_i\cdots\tau_j\right)^n
\left( 
\sum_{\{s'\}} e^{-\beta \sum_{ij} s_{ij}'} \langle \sigma_i\cdots\sigma_j\rangle_{\{s_{ij}'\}}^n 
\right)\nonumber\\
&\equiv
\sum_{\{\tau\}} \left(\tau_i\cdots\tau_j\right)^n
[\langle \sigma_i\cdots\sigma_j\rangle^n]'
\,.
\label{eq:uncorrdisorder}
\end{align}
where $[\cdots]'$ denotes the average over an {\it uncorrelated} bond disorder $\{s'\}$ with probability $p_{s'=1} = 1/(1+e^{2\beta})$ in a fixed gauge configuration, and $\sum_{\{\tau\}}$ denotes the gauge symmetrization i.e. superposition over all allowed gauge configurations. 
The probability function fulfils the Nishimori condition\cite{Nishimori1981}, which can be understood as equating the temperature of the spin with an effective ``temperature" of the disorder ensemble. 
As a result of gauge symmetrization, the measurement average of the magnetization remains always zero:
\begin{equation}
[\langle \sigma\rangle] =0.
\end{equation}
But the EA order parameter equals the magnetization in the RBIM along the Nishimori line
\begin{equation}
 [\langle\sigma\rangle^2] = [\langle\sigma\rangle]' \,,
\end{equation}
which can be proved as follows
\begin{equation*}
\begin{split}
[\langle \sigma_j\rangle]' 
\propto& \sum_{\{\tau\}}\sum_{\{s'\}} e^{-\beta \sum_{ij} s_{ij}'} \langle \sigma_j\rangle_{\{s'\}}\\
= & \sum_{\{s\}} \left(\sum_{\{\tau\}}\tau_j e^{-\beta \sum_{ij} s_{ij}\tau_i\tau_j}\right) \langle \sigma_j\rangle_{\{s\}}\\
= & \sum_{\{s\}} p_{\{s\}} \langle \sigma_j\rangle^2_{\{s\}}\\
= &[\langle \sigma_j\rangle^2] \,.
\end{split}
\end{equation*}
Here the gauge symmetry is exploited to lift the moment of the correlation function at the Nishimori's temperature. 
It is straightforward to prove a similar equality between even moments and odd moments of correlation functions\cite{Nishimori1981}
\begin{equation}
 [\langle\sigma_i\cdots\sigma_j\rangle^{2n}] =  [\langle\sigma_i\cdots\sigma_j\rangle^{2n-1}]' \,.
\end{equation}
Combined with the gauge invariance of even moments:
$
 [\langle\sigma_i\cdots\sigma_j\rangle^{2n}] =  [\langle\sigma_i\cdots\sigma_j\rangle^{2n}]' 
$,
one can derive $[\langle\sigma\rangle^2]' = [\langle \sigma\rangle]'$\cite{Nishimori1981}, which was a key result being used to argue that the Nishimori line of RBIM cannot pass through the conventional spin glass phase with vanishing magnetization but nonzero EA order. We emphasize that it does not contradict our glassy state satisfying $[\langle \sigma\rangle^2] \neq 0,\ [\langle \sigma\rangle]=0$, which originates from the gauge symmetry. 
Moreover, away from the Nishimori line in the phase diagram of RBIM, one can still perform similar gauge trick to prove a rigorous correlation inequality between the Nishimori temperature and any other temperature, for a given probability\cite{Nishimori1981}. This constrains the topology of the classical phase diagram and tells us the existence of a stable ferromagnetic phase along the Nishimori line.


\begin{figure*}[t!] 
   \centering
   \includegraphics[width=\linewidth]{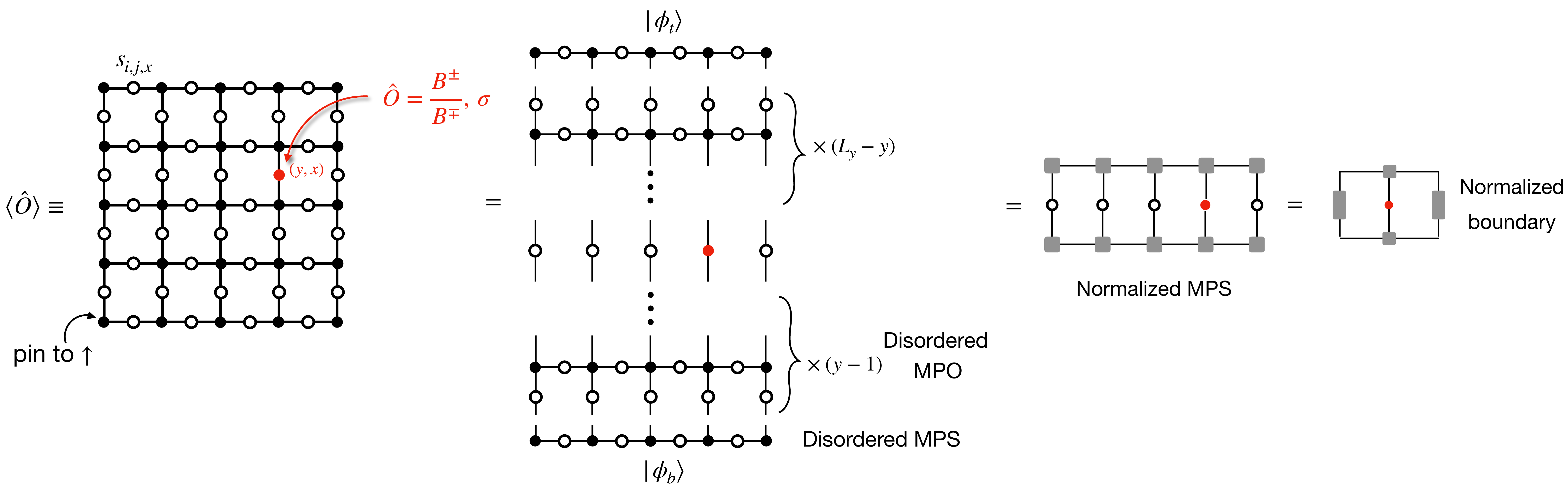} 
   \caption{{\bf Contraction for a finite size disordered TN on Lieb square lattice using MPS techniques.}
   The state space of the effectively evolved MPS is composed of all contributing classical configurations in the same row.}
   \label{fig:contraction}
\end{figure*}

\section{Appendix C: Numerical calculations for 2D Ising protocol}

\subsection{Tensor-network representation of disorder probability function}
For more general perturbation, let us discuss the explicit tensor-network form of the measurement-induced disorder probability function of the 2D Ising protocol, which will be used in the numerical calculation. 
The partition function $p_{\{s\}} = \norm{\bra{\{s\}}\ket{\psi}}^2$ can be cast into the tensor-network formalism:
\begin{equation}
\includegraphics[width=.5\columnwidth]{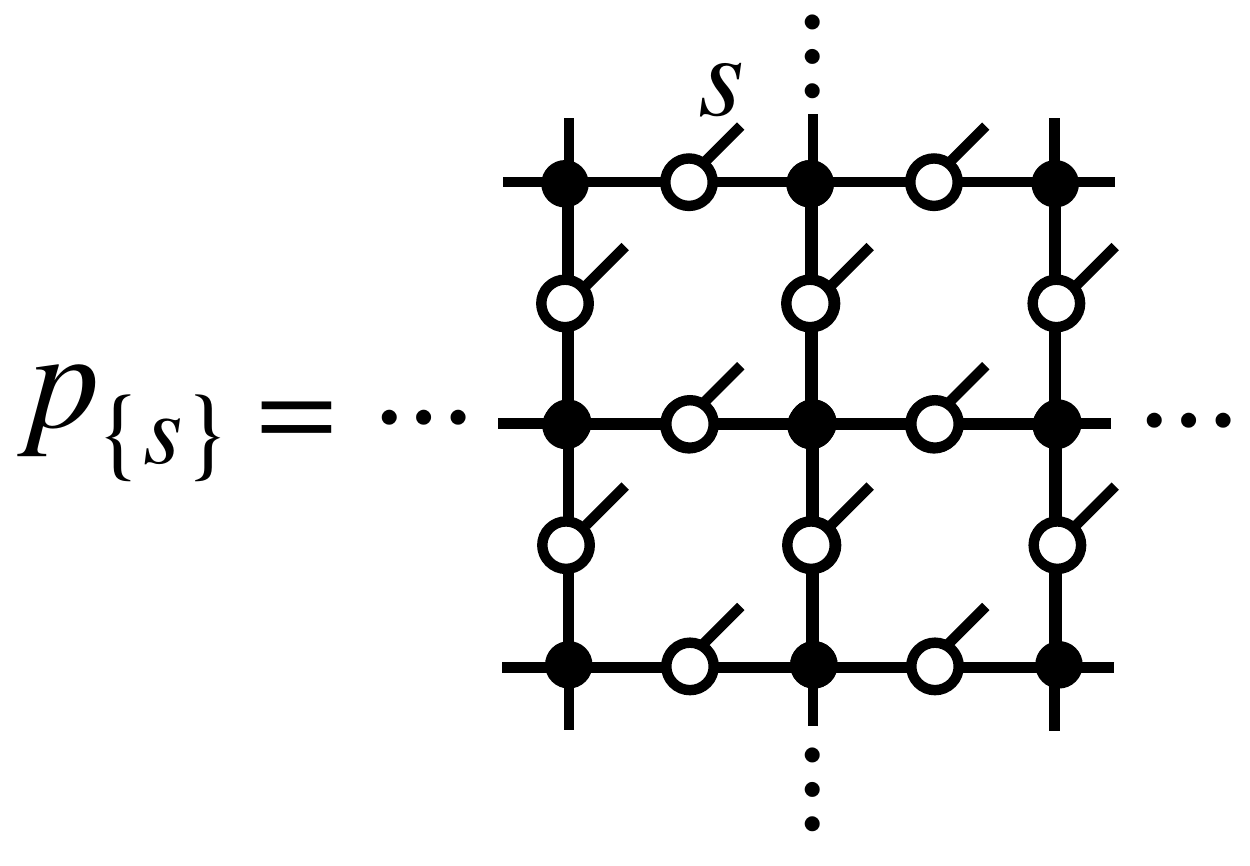},
\label{eq:tensornetwork}
\end{equation}
where the solid dot on each vertex denotes a rank-4 delta type tensor $T_{l,u,r,d} = \delta_{l,u}\delta_{l,r}\delta_{l,d}$, while the hollow circle on bond denotes a bond matrix $B^s$ determined by $s=\pm 1$:
\begin{equation}
\begin{split}
&B^+=\left( \begin{matrix}\cos(t_A+t_B)^2 & \cos(t_A-t_B)^2 \\  \cos(t_A-t_B)^2 & \cos(t_A+t_B)^2 \end{matrix}\right),\\
&B^-=\left( \begin{matrix} \sin(t_A+t_B)^2 &  \sin(t_A-t_B)^2 \\ \sin(t_A-t_B)^2 & \sin(t_A+t_B)^2 \end{matrix}\right).
\end{split}
\label{eq:extBondMtx}
\end{equation}
The matrix elements can be derived from fixing a classical configuration for the site spin and extract the components of the bond acilla spin in $x$ basis. 
Notice that the tensor-network representation for the classical Ising model (in arbitrary dimension) has the same structure with the bond matrix expressing the Ising interaction:
\begin{equation*}
\left( \begin{matrix} e^{-\beta J_l} & e^{\beta J_l} \\ e^{\beta J_l} & e^{-\beta J_l} \end{matrix}\right) \,,
\end{equation*}
where $J_l$ is the bond dependent Ising interaction strength.
One can then check that swapping $t_A\leftrightarrow t_B$ leaves the tensor invariant, and a transformation $t_A\to-t_A$ is equivalent to flipping one site sublattice spins,  while a transformation $t_A\to t_A+\pi/2$ is equivalent to flipping all the bond spins. They all leave the measurement averaged results invariant, and the whole phase diagram in the main text can be deduced, e.g., from the corner region $0\leq t_A\leq t_B \leq \pi/4$. 
Note that due to the global Ising symmetry, $p_{\{s\}} = p_{\{-s\}}$, we can choose to project a single site spin to $\uparrow$ when evaluating the relative probability. 

Moreover, the above tensor-network construction can be generalized to any bipartite lattice in any dimension, keeping the same bond matrix and extending the site delta tensor to include more legs.


\subsection{Hybrid Monte Carlo and tensor network algorithm}
With the TN expression at hand, here we explain in details our hybrid Monte Carlo and tensor network algorithm designed for sampling the postmeasurement 2D projected entangled pair states wave function. 

The numerical algorithm is as follows:
(i) We initialize a (uniform or fully random or random but flux-free ) bond spin configuration $\{s\}$.
(ii) As in the standard Metropolis algorithm, in each Monte Carlo step, we propose a local flip $s_{l}'  = - s_{l}$ at a randomly chosen bond spin, with an acceptance probability min$\{p_{\{s'\}} / p_{\{s\}}, 1\}$.
(iii) The relative probability $p_{\{s'\}} / p_{\{s\}}$ is the ratio between two classical TNs that share the same bond disorder except at the proposed spin-flip bond. Therefore we can contract out the entire finite size TN except the target bond, using the standard density matrix compression algorithm. In this way, the effective \textit{normalized} environment for this bond spin is obtained. The physical observable, either the relative probability, or the magnetization can be inserted to the bond before the final contraction for a number. 
(iv) After each Monte Carlo sweep (one step per bond spin), we measure our bond spin sample, and also use the same TN contraction scheme to measure the magnetization of the central site spin. 
(v) Finally, we collect many independent Markov chains initiating from independent random configurations, discarding the data until equilibration time, and perform binning analysis (implemented with Julia package BinningAnalysis.jl) to perform the statistical average for all the physical observables $[\langle\cdots\rangle]$. 

The detailed TN contraction scheme for the square lattice is as follows:
(i) We start from the bottom boundary row and construct an MPS out of  local rank-3 tensors $|\phi_b\rangle$ as shown in Fig.~\ref{fig:contraction}. 
(ii) Construct the transfer matrix MPOs out of the local rank-4 tensors in the next rows, which are used to update the boundary MPS consecutively, where the virtual bond is truncated with error tolerance e.g. $10^{-10}$. 
(iii) Perform the same MPS evolution for the top boundary in the reverse direction $|\phi_t\rangle$, until we get two MPSs from bottom and top, respectively, which sandwich the target $y$-bond at $(x,y)$ location, see Fig.~\ref{fig:contraction}. 
(iv) Insert physical observables into the target bond and contract it out with the top and bottom MPSs $\langle \psi_t | \hat{O} |\phi_b\rangle$. 

A few remarks:
(i) The state space of the MPS is the possible classical configurations at the boundary. 
(ii) A pinning field at the lattice boundary corner can be applied to the initial boundary MPS. 
(iii) The MPO is generally not a unitary operator, and therefore the evolved boundary MPS could lie in an area law low entangled space\cite{Harrow2022}, such that our algorithm can be applied to very large system sizes. 
(iv) The physical meaning of the norm of the MPS evolved by many layers of MPO is related to the partition function of the 2D classical model, which is expected to decay exponentially with the system size $\propto e^{-L^2}$ for a finite free energy density. 
Therefore, renormalization in each step is indispensable, analogous to generic transfer matrix methods for disorder problem. 
(v) In our problem, when $(t_A,t_B)\to (\pi/4, \pi/4)$, the effective energy landscape for the bond spin consists of extensive number of disconnected valleys, which correspond to different loop configurations and are related to each other by gauge transformation.


\subsection{Numerical results for the Nishimori line}
While the finite-size-scaling for the Nishimori line has been shown in the main text, here we present all the complementary observables for the fixed finite size $L=24$.

\begin{figure}[thb] 
   \centering
   \includegraphics[width=\columnwidth]{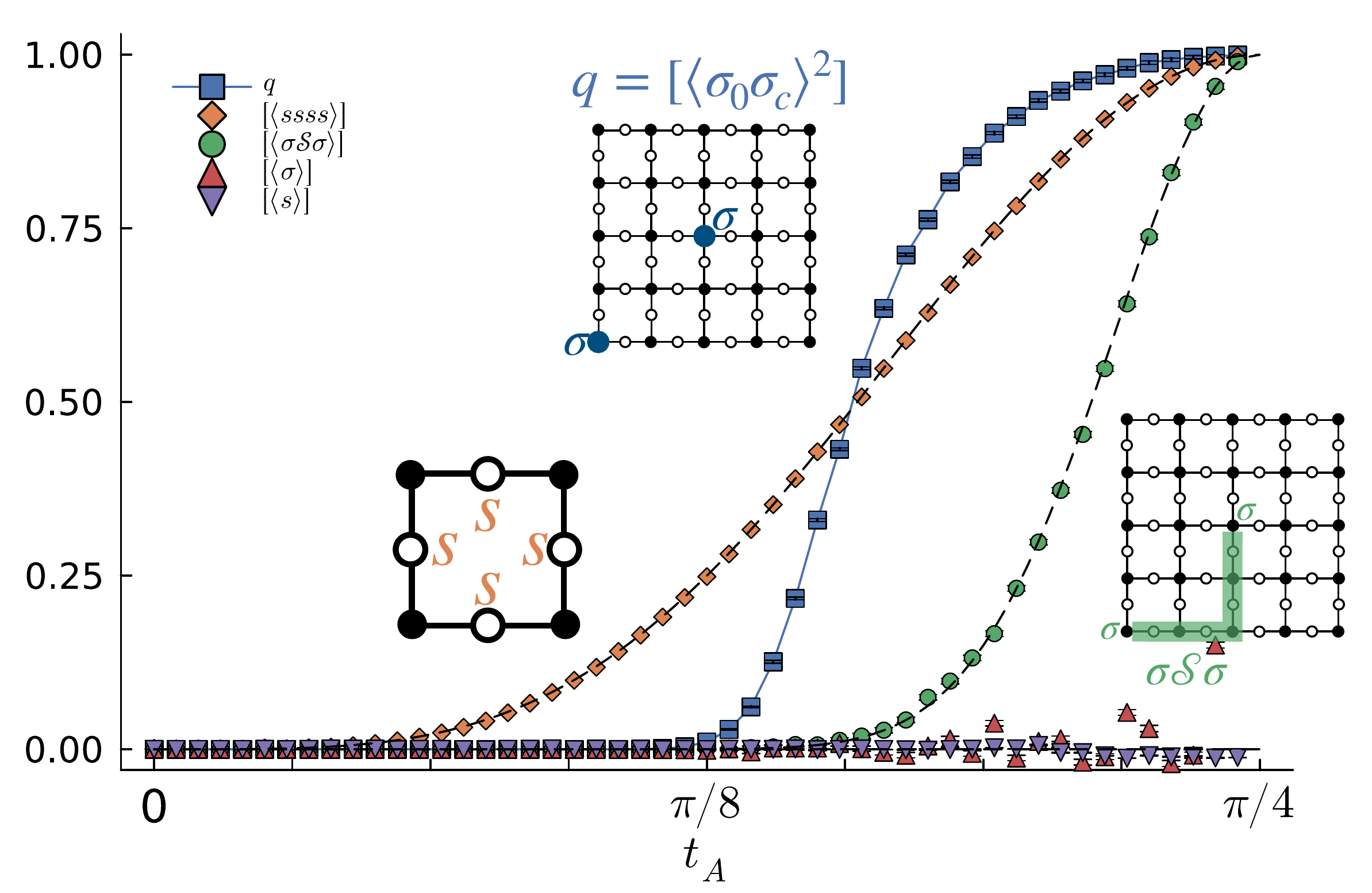} 
   \caption{{\bf Transition from SRE to LRE states after finite evolution time along the Nishimori line} 
    ($t_A, t_B=\pi/4$) in the phase diagram.
   Shown are results from our hybrid Monte-Carlo / tensor-network approach (symbols) for a $24 \times 24$ Lieb lattice with open boundaries,
   compared to analytical predictions (dashed lines).
   The EA order parameter $[\langle \sigma_0 \sigma_c \rangle^2]$ for the site spins signals the onset of LRE / spin glass order. 
   The bond-spin observables $[\langle s\rangle]$, $[\langle ssss\rangle]$ indicate a vanishing average local magnetization and a plaquette Wilson loop
   in agreement with our analytical expectation (dashed lines) for the measurement outcomes obtained in our numerical approach. 
   The orange dashed line indicates the analytical value $\sin(2t_A)^4$ for plaquette Wilson loop, while the green dashed line indicates $\sin(2t_A)^L$ for a Wilson line operator, which is schematically shown in inset. 
    Note that the bottom left corner spin is pinned in our calculations. 
   }
   \label{fig:L24allcorr}
\end{figure}

In Fig.~\ref{fig:L24allcorr} we show numerical results along the Nishimori line ($t_B=\pi/4$) for fixed linear system size $L=24$ and open (free) boundary conditions, a case which can be directly compared to the RBIM. 
First, we find an ancilla bond spin magnetization $[s]\approx 0$, which validates that our Monte Carlo sampling indeed reproduces the vanishing local expectation of ancilla bond spins, while simultaneously observing a finite density of $\pi$-fluxes that frustrate the site spins, with measurements of the gauge invariant plaquette Wilson loop being quantitatively consistent with the analytical derivation from Eq.~\eqref{eq:ssss}, i.e.\ $[ssss] = \sin(2t_A)^4$. 
Second, the gauge invariant open Wilson line, a collective correlation between the site spins and bond spins (illustrated in the inset), 
is also consistent with our analytical expectation $[\langle \sigma s\cdots s\sigma\rangle] = \sin(2t_A)^L$, where $L$ is the length of the line. 
Thirdly, the EA order parameter $[\langle \sigma_0 \sigma_c\rangle^2]$ turns nonzero for an intermediate value $t_A<\pi/4$, while $[\langle \sigma_0 \sigma_c\rangle]\approx 0$, though some visible fluctuations arise for this gauge-sensitive observable for large $t_A$. These numerical fluctuations arise due to the fact that when $(t_A, t_B)\to(\pi/4, \pi/4)$, different gauge-equivalent loop configurations become disconnected deep valleys not connected by local ancilla bond spin flips, 
and therefore our Monte Carlo samples tend to get trapped in some gauge minima. Note that this insufficiency leads to fluctuations  only for the gauge-dependent observables. One can always replicate the Monte Carlo samples and perform a further gauge symmetrization to smoothen out the gauge-dependent observables to zero.


\subsection{Numerical results beyond the Nishimori line}

\begin{figure*}[t!] 
   \centering
   \includegraphics[width=\columnwidth]{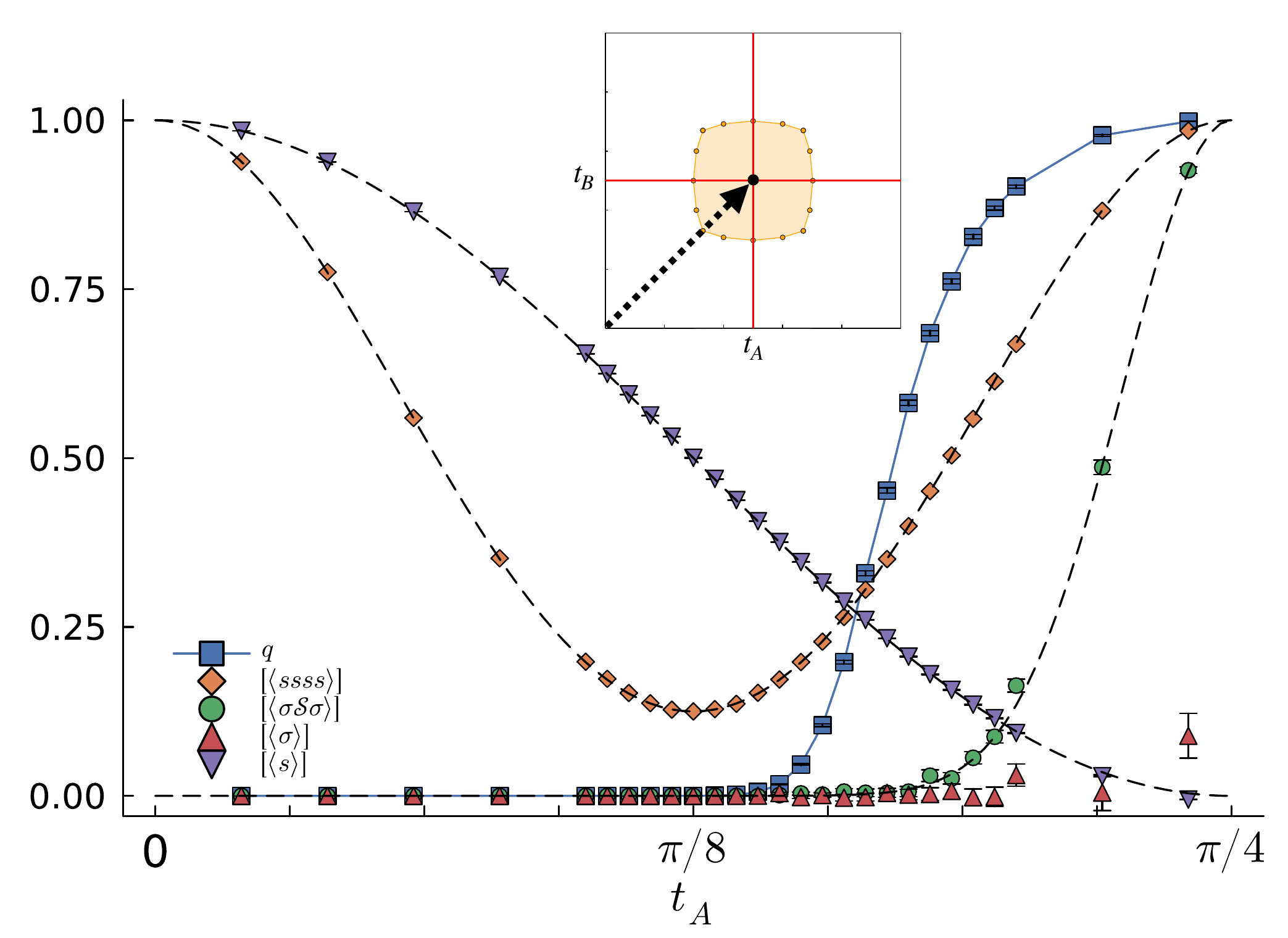} 
   \includegraphics[width=\columnwidth]{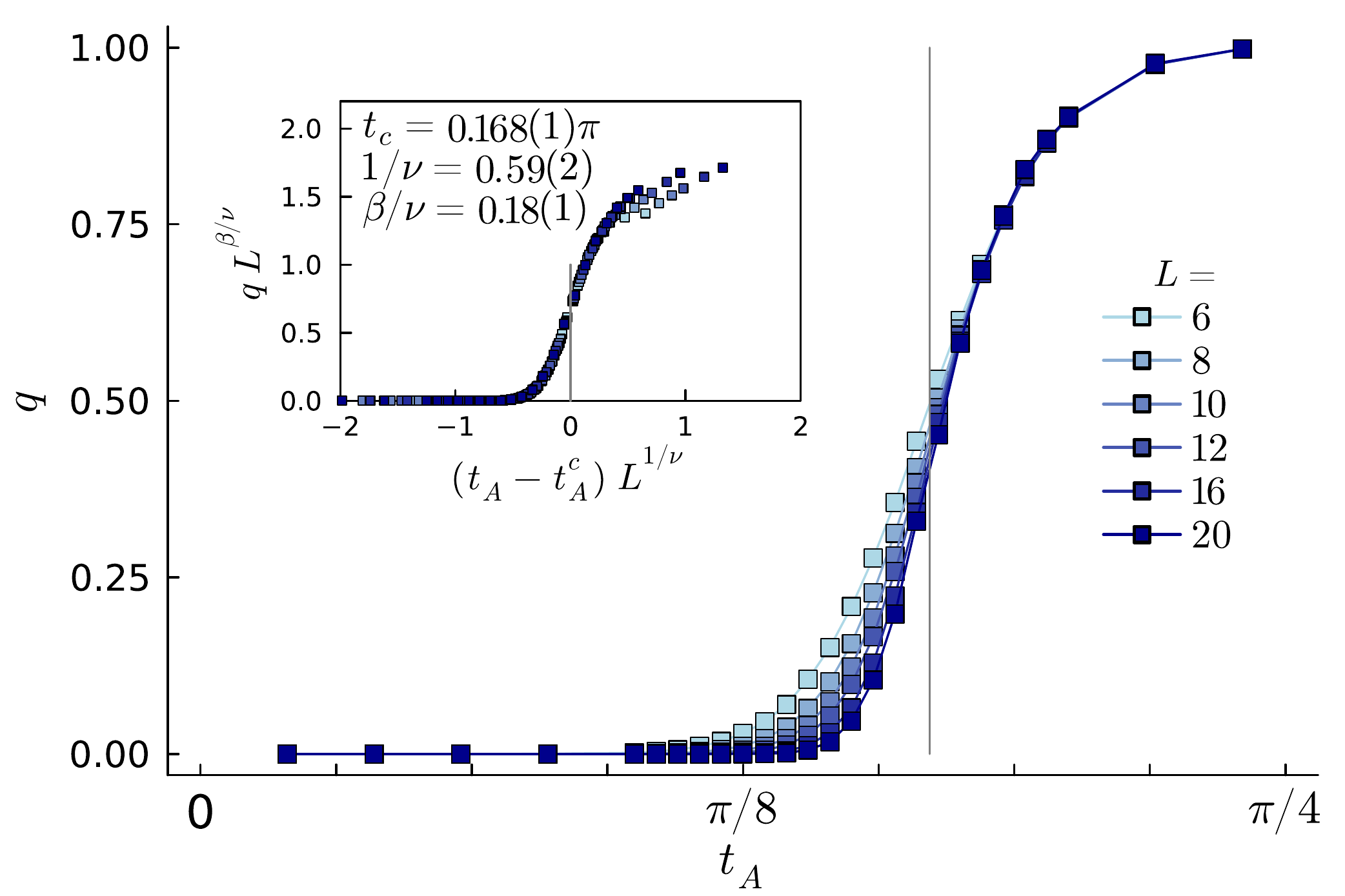} 
   \caption{{\bf Transition from SRE to LRE for maximal coupling imbalance along diagonal cut $\boldsymbol{t_B=t_A}$}. 
   (a) The left panel shows, in analogy to Fig.~\ref{fig:L24allcorr}, several observables for a fixed system size $L=20$, 
   with symbols indicating numerical data and the dashed lines show the analytical expectations. 
   The inset shows the line cut through the 2D phase diagram. 
   (b) The right panel shows finite-size scaling of the EA order parameter, in analogy to Fig.~2 of the main text, 
   including an inset showing a data collapse of the data within the window $0.1\pi \leq t_A \leq 0.2\pi$. 
   The finite size scaling fitting is performed using autoScale.py\cite{FSSpackage}, and the fitting exponents are shown in the figure. 
   }
   \label{fig:tBtA}
\end{figure*}

\begin{figure*}[t!] 
   \centering
   \includegraphics[width=\columnwidth]{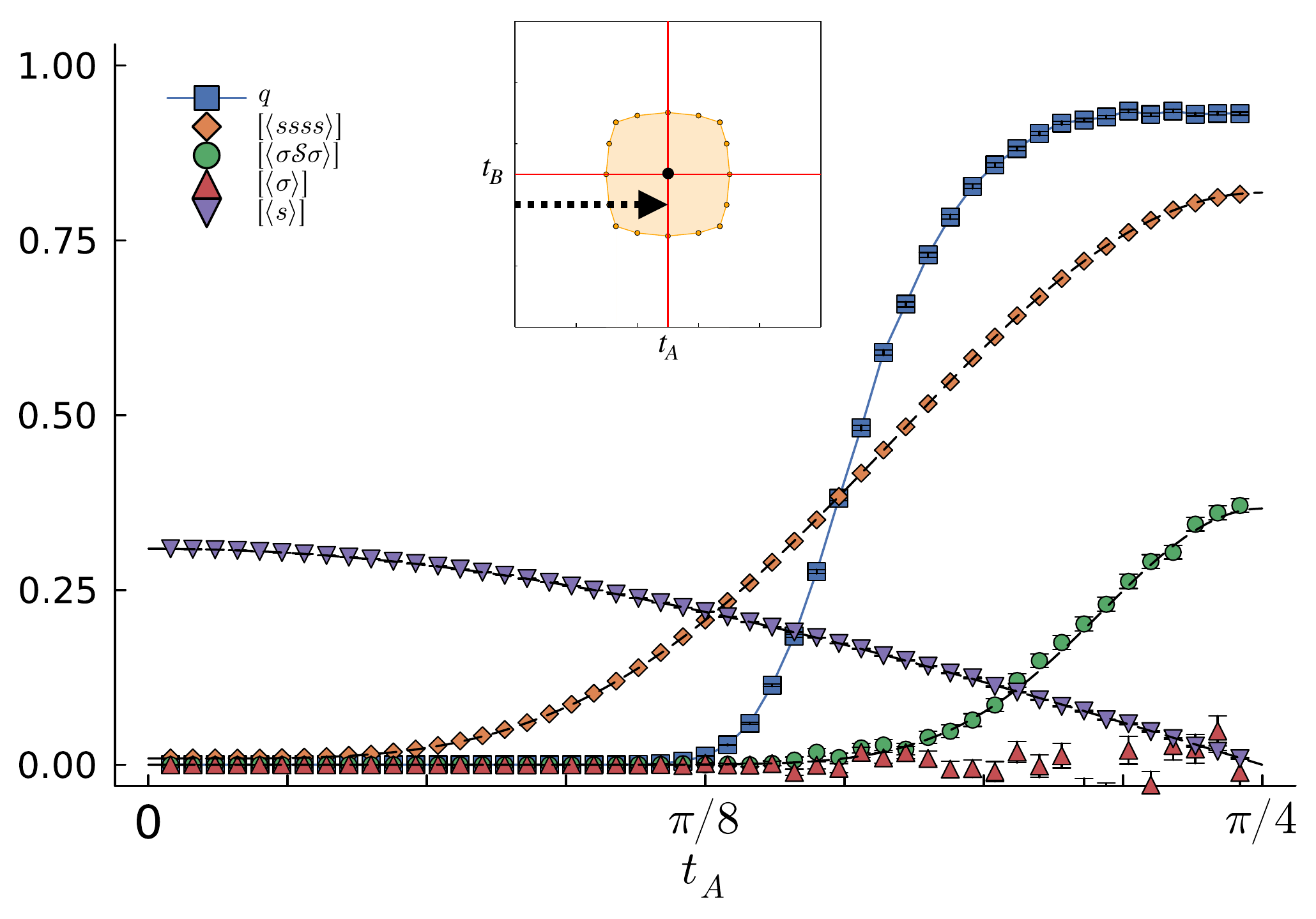} 
   \includegraphics[width=\columnwidth]{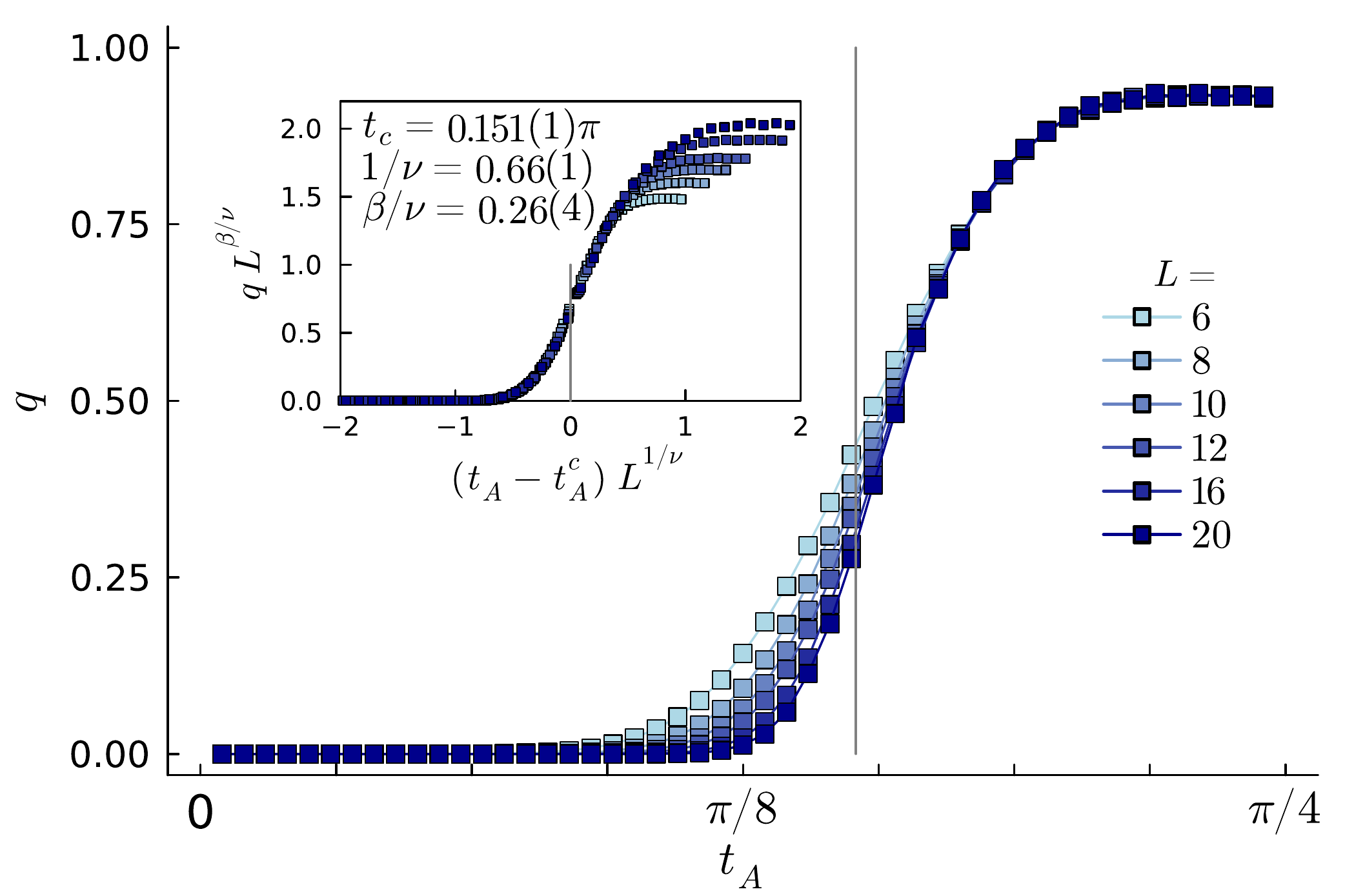} 
   \caption{{\bf Transition from SRE to LRE for imbalanced couplings along horizontal cut for $\boldsymbol{t_B=\pi/5}$}.
   Same as Fig.~\ref{fig:tBtA} but for the indicated horizontal cut.
   }
   \label{fig:tB.2pi}
\end{figure*}

Going beyond the analytically tractable, gauge-symmetric Nishimori line we show explicit numerical results from our hybrid TN/Monte Carlo approach for the two alternate couplings discussed in the main text, i.e.\ for $t_B=t_A$ with maximal coupling imbalance which corresponds to the diagonal line in the phase diagram of Fig.~1(c) in the main text, as well as for a case in between with $t_B=\pi/5$. 
Our numerical results for these two settings are presented in Figs.~\ref{fig:tBtA} and \ref{fig:tB.2pi}, respectively.
The results mirror the plots showing results for the gauge-symmetric case, $t_B=\pi/4$, shown in Figs.~\ref{fig:L24allcorr} and Fig.~2 of the main text.
Note that for the horizontal cut in Fig.~\ref{fig:tB.2pi}, which does not include the origin, the EA order parameter saturates slightly below 1 (as it would in the vicinity of the origin); similar deviations from their maximal values are also seen for other observables.

The finite-size scaling analysis, shown in the right panels of Figs.~\ref{fig:tBtA} and \ref{fig:tB.2pi}, provides estimates for the 
critical point and exponents as summarized in Table~\ref{tab:tc} below.

\begin{table}[h]
\caption{{\bf Phase boundary points} for square lattice obtained from numerical calculations. }
\begin{center}
\begin{tabular}{c|ccc}
\toprule
$t_B$ & $t_A^c$ & $\nu$ & $\beta$ \\
\hline
 $\pi/4$ & $0.149\pi$ & 1.4 & 0.36 \\
 $\pi/5$ & $0.151\pi$ & 1.5 & 0.39 \\
$t_A$   & $0.168\pi$ & 1.7 & 0.31 \\
\bottomrule
\end{tabular}
\end{center}
\label{tab:tc}
\end{table}%
%

\begin{figure*}[htb] 
   \centering
   \includegraphics[width=\columnwidth]{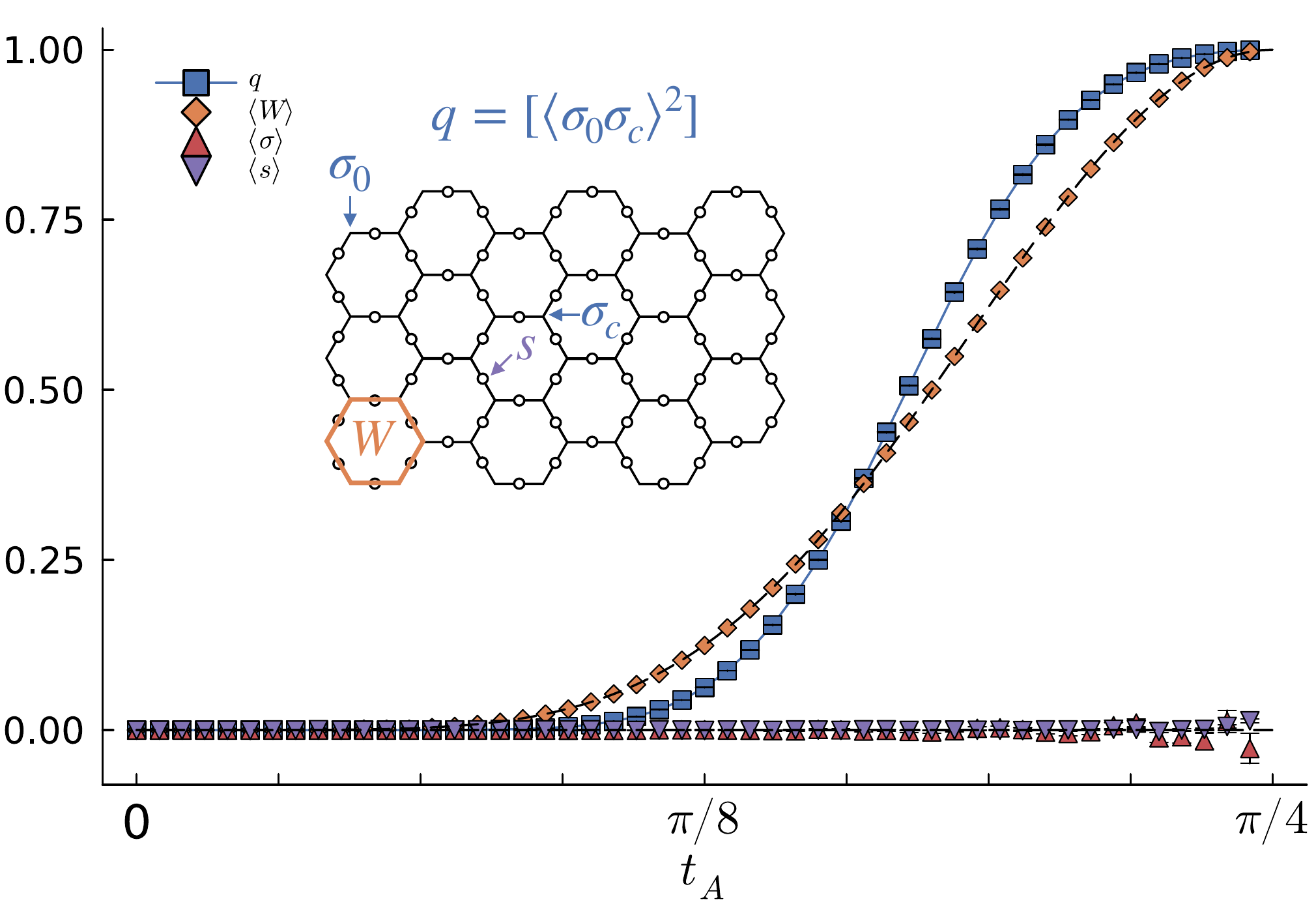} 
   \includegraphics[width=\columnwidth]{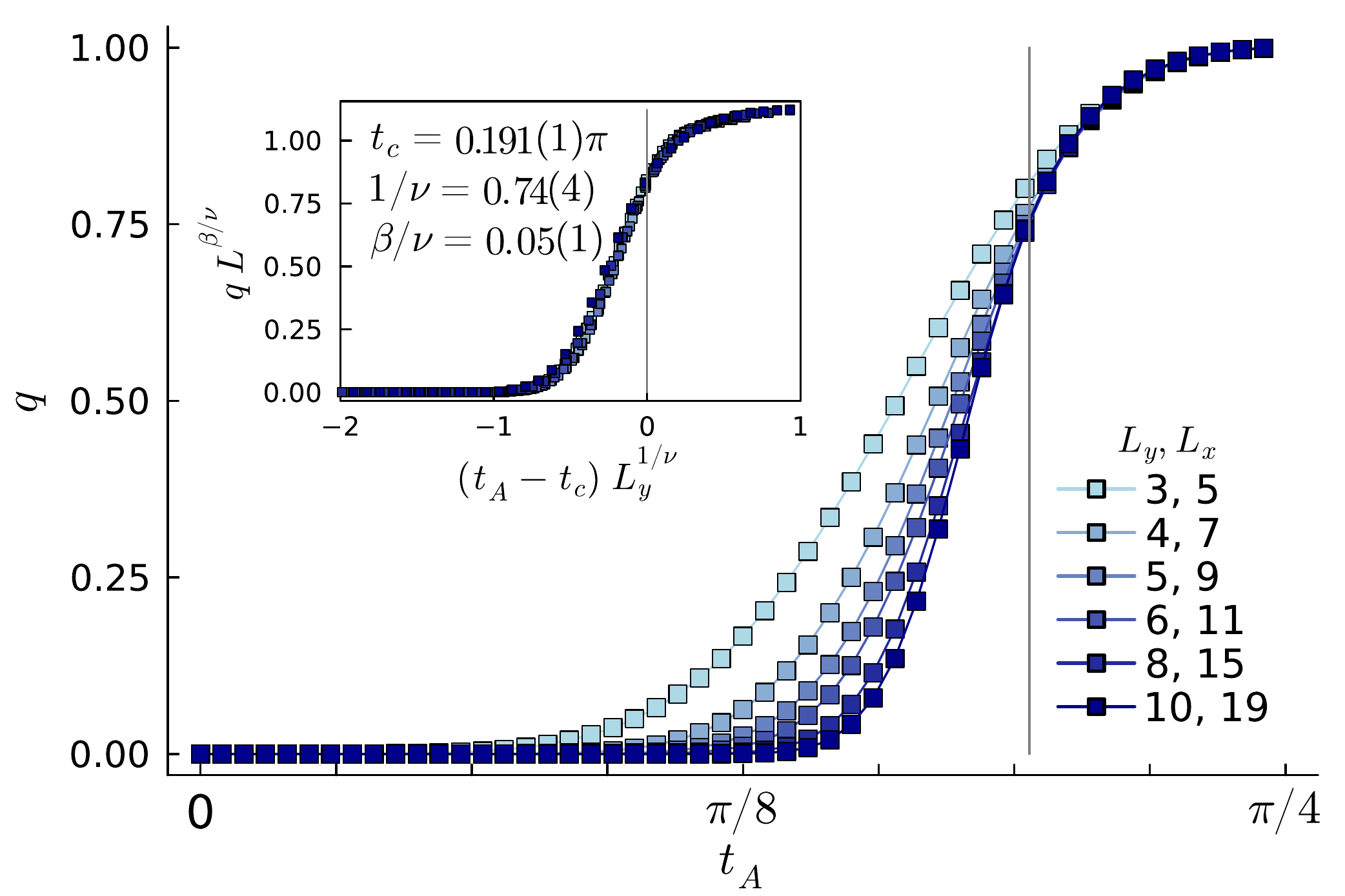} 
   \caption{{\bf Transition from SRE to LRE for the gauge-symmetric model ($\boldsymbol{t_B=\pi/4}$) on the heavy-hexagon lattice}.
   (a) The left panel shows, in analogy to Fig.~\ref{fig:L24allcorr} of the main text, several observables for a fixed $4\times7$ system size, 
   with symbols indicating numerical data and the dashed lines show the analytical expectations. 
   (b) The right panel shows finite-size scaling of the EA order parameter, in analogy to Fig.~2 of the main text, 
   including an inset showing a data collapse of the data within the window $0.1\pi \leq t_A \leq 0.22\pi$.
}
   \label{fig:symhexPhaseDiagram}
\end{figure*}

\begin{figure*}[htb] 
   \centering
   \includegraphics[width=\columnwidth]{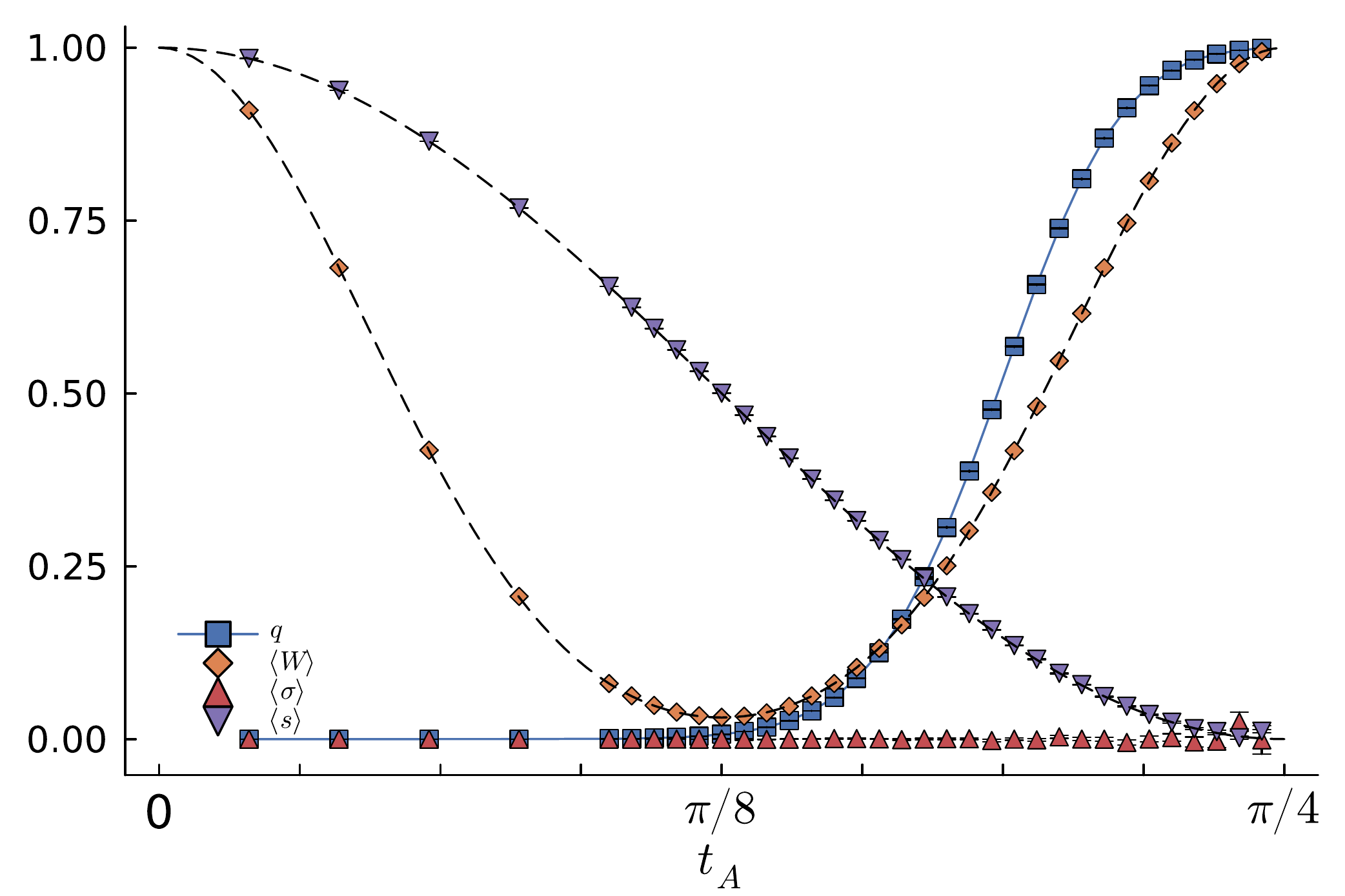} 
   \includegraphics[width=\columnwidth]{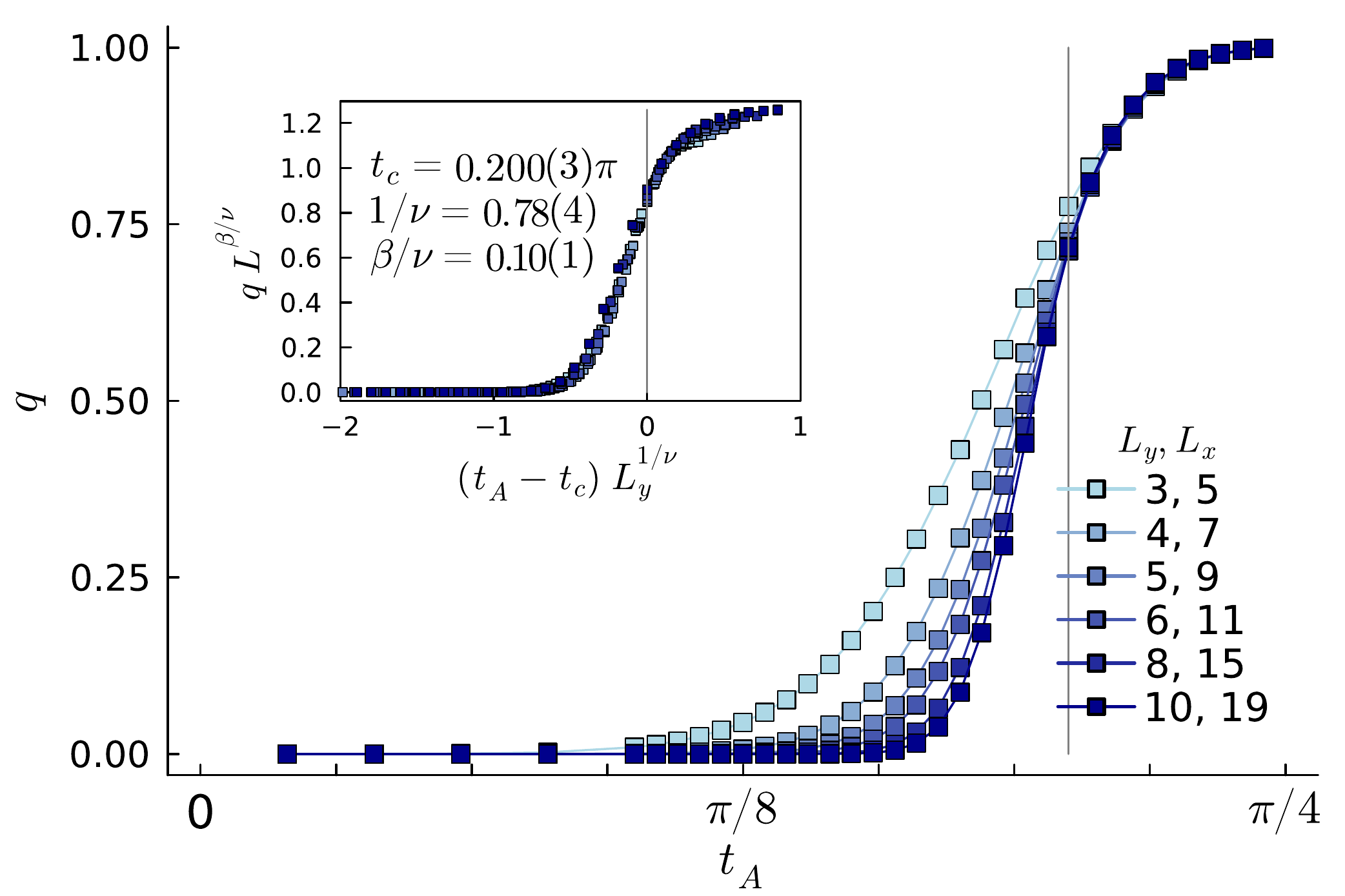} 
   \caption{{\bf Transition from SRE to LRE for the maximally imbalanced model ($\boldsymbol{t_B=t_A}$) on the heavy-hexagon lattice}.
   Same as Fig.~\ref{fig:symhexPhaseDiagram} but for the indicated diagonal cut.
  }
   \label{fig:hexPhaseDiagram}
\end{figure*}

\subsubsection{Technical details}
On a technical level, our simulations here are done for a square Lieb lattice with  open (free) boundary conditions 
and equal linear size $L_x=L_y=L$ with total number of spins $N = 3L^2+4L+1$ for
$L=6,8,\ldots,20$, i.e. the largest system has $N=20\times20\times3+81=1,281$ spins. 

The TN contraction has been performed with MPS evolution under row transfer matrix, where the MPS in each step is truncated with a $10^{-10}$ cutoff in the density matrix eigenvalues as in standard DMRG calculation (details explained in previous algorithm section). 
For the Monte Carlo sampling, we have simulated 11 independent Markov chains initiated from a random-bond but flux-free configurations (bonds at all rows and the first column are random), up to 10,000 sweeps (flips per bond) for $L\leq 10$, 5,000 sweeps for $L=12$, 3,000 sweeps for $L=16$, and 1,000 sweeps for $L=20$. 
For every Markov chain, the data during the first one tenth of the sweeps are discarded, while the rest are treated as equilibrium ensemble 
and analyzed using a binning analysis for statistical averages.  \\[1cm]


\section{Appendix D: Numerical results for IBM's heavy-hexagon geometry}

In the current incarnation of its transmon-based quantum computing devices that IBM deploys in its cloud access program\cite{IBMQuantumCloud},
the underlying lattice geometry of the transmon qubits is a heavy-hexagon lattice -- a lattice geometry that can be conceptualized as a honeycomb lattice whose bonds have been decorated with an additional site, i.e.\ a honeycomb variant of the square Lieb lattice discussed in the main text.

In order to provide quantitative guidance, we have redone some of our principal calculations for this heavy-hexagon lattice geometry. 
Figs.~\ref{fig:symhexPhaseDiagram} and \ref{fig:hexPhaseDiagram} provide information for the transition from SRE to LRE states 
for the gauge-symmetry and maximally imbalanced model, i.e.\ for the horizontal/vertical and diagonal cuts of our principal phase diagram in Fig.~1(c) in the main text. In comparison to the square Lieb lattice, we see a slight shift of the critical point to higher values, indicating a somewhat smaller extend of the LRE phase for this lattice geometry. Otherwise our results remain basically unaltered as expected by universality arguments.


\subsubsection{Technical details}
On a technical level, our simulations here are done for heavy-hexagon lattices with open (free) boundary conditions 
and spatial dimensions $L_y \times (L_x=2L_y-1)$ with $L_y=3,4,5,6,8,10$. 
The total number of bonds are $3L_xL_y-L_x-L_y-2$ and the total number of sites are $2L_yL_x-2$, and thus the total number of spins are $N=5L_xL_y-L_x-L_y-4$. 
The largest system size thus has $N=5\times 19\times 10 - 19-10-4=917$ spins. 
In particular, $L_y=4, L_x=7$ corresponds to $N=125$, which are the IBM 127 qubits quantum computing system\cite{IBMQuantumCloud} when two dangling qubits at top left and bottom right are pinned. 

The TN contraction has been performed with an MPS cutoff of $10^{-10}$. 
For the Monte Carlo sampling, we have run 23 independent Markov chains with 10,000 sweeps for $L_y=3,4,5,6$ and 6,000 sweeps for $L_y=8$ and 4,000 sweeps for $L_y=10$, where the first one tenth of the sweeps are discarded for equilibration ensemble average.


\section{Appendix E: Glassy topological order: 3D toric code on the Nishimori line}
In this section we discuss in details the three-body gate protocol to realize the 3D glassy $Z_2$ topological order. A brief discussion for the two-body gate protocol will be appended in the end. 

\subsubsection{3-body gate protocol}

\begin{figure}[b] 
   \centering
   \includegraphics[width=\columnwidth]{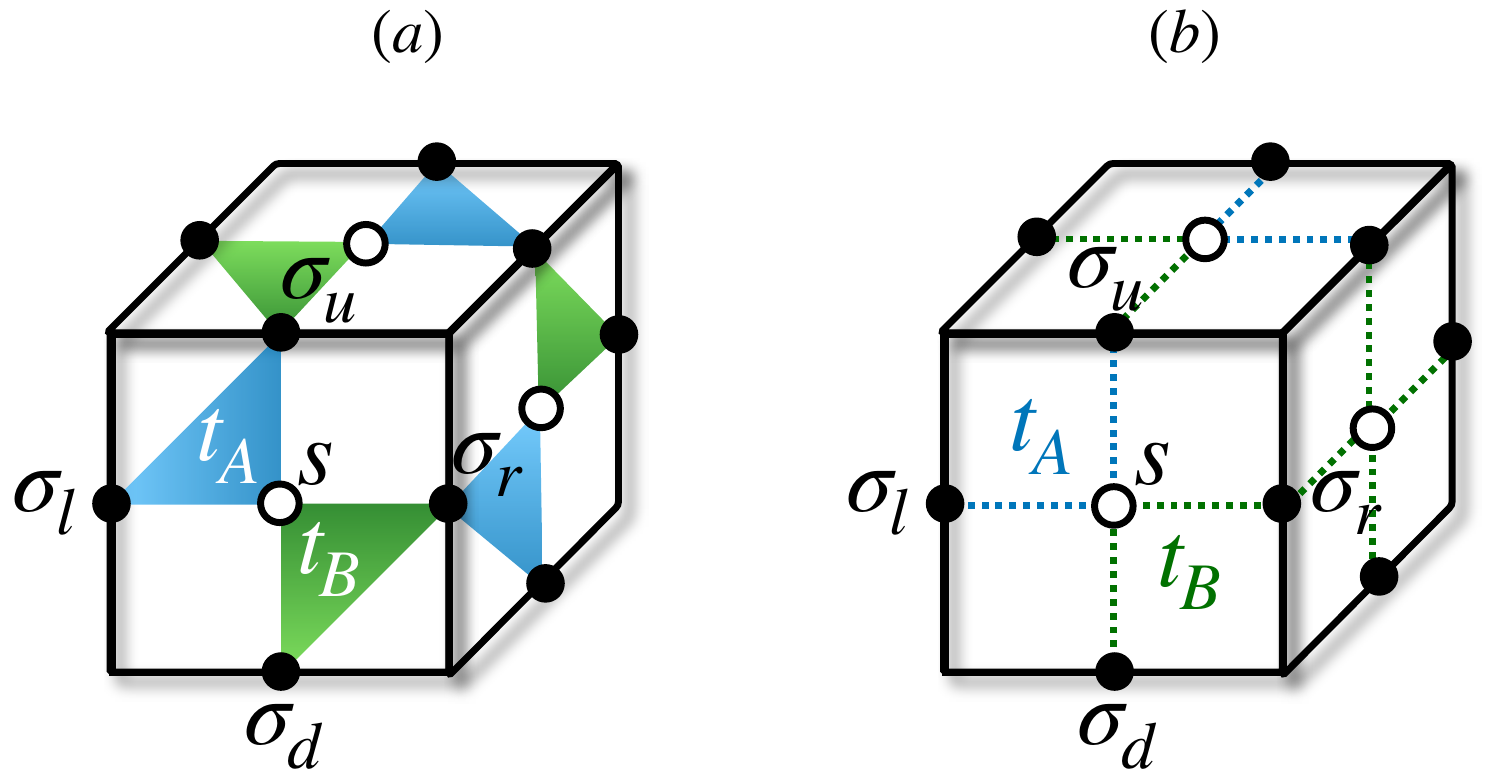} 
   \caption{
   {\bf Protocols for preparing glassy toric code in 3D cubic lattice}. 
   Target spins are placed on bond centers, while ancillas are placed on plaquette centers, which are measured in $x$ direction after the unitary entangling operation. 
   (a) A 3-body gate protocol: the four spins surrounding a given plaquette are grouped into two pairs, which are entangled with the ancilla by ZZZ gate evolution separately.  
   (b) A 2-body gate protocol: the four spins surrounding a given plaquette are entangled with the ancilla by ZZ gate evolution separately. 
   }
   \label{fig:3dprotocol}
\end{figure}

Consider a 3D cubic lattice, with target physical spins on the bond centers and ancillas at the plaquette centers. For each plaquette we label the four spins by $l,u,r,d$ in the meaning of left, up, right, and down. We can take Eq.~\eqref{eq:twotermprotocol} with $\mathcal{O}_A=\sigma_l^z \sigma_u^z$ while $\mathcal{O}_B=\sigma_r^z \sigma_d^z$, as in Fig.~\ref{fig:3dprotocol}a. 
After measuring ancilla spins in $x$ basis, the state is
\begin{equation}
\bra{\{s_q\}}\ket{\psi} \propto e^{-\frac{\beta}{2} \sum_q (J_{s_q} W_q + h s_q)}\ket{+}\, ,
\end{equation}
which satisfies $\prod_{l\in \mbox{\varhexstar}}\sigma_l^x=1$ indicating the electric charge is frozen. 
At the strong measurement fixed point $t_A=t_B=\pi/4$, $\beta=+\infty$, $W_q = \prod_{l\in \partial q}\sigma_l^z=-s_q$, which corresponds to a topological toric code eigen-state with a static magnetic flux tube defect penetrating those plaquettes satisfying $s_q=1$. 
Away from the fixed point, two physical ingredients start to play their roles: 
(i) the fluctuation of magnetic flux; 
(ii) the disorder with nonzero density of magnetic monopoles (in the cubic centers).
To understand the phase transition, the model can be mapped to the classical 3D plaquette Ising gauge model\cite{Wegner71duality} with disordered plaquette interaction:
\begin{equation}
p_{\{s_q\}} 
=
\norm{\bra{\{s_q\}}\ket{\psi}}^2 
= 
\sum_{\{\sigma\}} e^{-\beta \sum_q \left(J_{s_q} W_q + h s_q\right)},
\label{eq:classicalgauge}
\end{equation}
where $q$ labels the plaquette, $J_\pm$ and $h$ are the same as defined before. The disorder probability is the same as the partition function. Here the disorder correlation is nonzero if and only if the ancillas on the plaquette centers form a {\it closed surface}. Namely, the single ancilla measurement average becomes $[s] = \cos(2t_A)\cos(2t_B)$, while the six ancilla spins surrounding a cube exhibit:
\begin{equation}
\begin{split}
&[\prod_{q\in \partial\mbox{\mancube}} s_q]
=
\bra{\psi} \prod_{q\in \partial\mbox{\mancube}} s_q^x \ket{\psi} \\
=& 
\cos^6(2t_A)\cos^6(2t_B) + \sin^6(2t_A)\sin^6(2t_B)\, ,
\end{split}
\end{equation}
which determines the disorder monopole density. 

Analogous to the Ising protocol, the line $t_B=\pi/4$ is mapped to the standard random plaquette Ising gauge model (RPGM)\cite{Preskill2002, Preskill2003, Matsui2004} with gauge symmetric disorder ensemble: same probability for any two disorder plaquette configurations that share common monopole distribution, related by a modified gauge symmetry $\{\tau_l=\pm 1\}$:
\begin{equation}
\sigma_l' = \sigma_l\tau_l,\quad
s_q' = s_q \prod_{l\in \partial q} \tau_l\, .
\end{equation}
As a sanity check, along the Nishimori line the monopole density can be deduced from the uncorrelated flux disorder:
\begin{equation*}
[\prod_{q\in \partial\mbox{\mancube}} s_q] 
= \prod_{q\in \partial\mbox{\mancube}} [ s_q']'
=\sin^6(2t_A) \,.
\end{equation*}
It was found that the monopole disorder does not destroy the topological ordered phase immediately, which spans a finite phase region beyond the strong measurement limit, until a critical point which was analytically proved to be {\it finite}\cite{Preskill2003} and numerically found to be $p_{s'=1}^c\approx 0.033$\cite{Matsui2004} i.e. $t_A^c\approx 0.192\pi$. 
When $t_A > t_A^c$, the magnetic flux fluctuation is prohibited corresponding to the deconfined topological ordered phase. 
Analogous to the spin glass, the linear Wilson loop always vanishes under symmetric disorder average, while its second moment, an EA analog of Wilson loop becomes perimeter-law scaling in the ordered phase.
\begin{equation}
[\langle \prod_{q\in A} W_q\rangle] = 0,
\quad
[\langle \prod_{q\in A} W_q\rangle^2] \propto e^{-|\partial A| / \xi},
\end{equation}
where $A$ is an arbitrary surface, $|\partial A|$ denotes its perimeter, and $\xi$ is a nonuniversal lengthscale.

\subsubsection{2-body gate protocol}

Since the implementation of a three-body ZZZ evolution might be difficult to realize in experimental settings, we propose an alternate two-body Ising evolution protocol, simply by further decomposing the ZZZ evolution into two Ising evolutions, as shown in Fig.~\ref{fig:3dprotocol}b. 
Then the postmeasurement effective non-unitary operator becomes
\begin{widetext}
\begin{equation}
\begin{split}
M_s=&\bra{s^x=s}_s e^{-i s^z (t_A (\sigma_l^z+ \sigma_u^z) + t_B(\sigma_r^z +\sigma_d^z))}\ket{+}_s\\
=&\cos^2 t_A \cos^2 t_B \times\\
&
\begin{cases}
1 + \tan^2 t_A \tan^2 t_B W
- (\tan^2 t_A \sigma_l^z\sigma_u^z + \tan^2 t_B \sigma_r^z\sigma_d^z + \tan t_A \tan t_B (\sigma_l^z+\sigma_u^z)(\sigma_r^z + \sigma_d^z)), & s=+1\\
-i \left(
\tan t_A (\sigma_l^z + \sigma_u^z) + \tan t_B (\sigma_r^z + \sigma_d^z) 
- \tan t_A \tan^2 t_B  W (\sigma_l^z + \sigma_u^z) -\tan^2 t_A \tan t_B W(\sigma_r^z + \sigma_d^z))
\right)
, & s=-1\\
\end{cases}\, .
\end{split}
\end{equation}
\end{widetext}
The fact that multiple terms are connected with the ancilla separately leads to multiple polynomial terms for the non-unitary operator, more complicated than the two-term protocol we discussed above. 
Nevertheless, at the strong measurement fixed point $t_A=t_B=\pi/4$, $\tan t_A=\tan t_B=1$,  the above equation simplifies into
\begin{equation}
\begin{split}
M_s(t=\frac{\pi}{4})
=&\frac{1}{4}
\begin{cases}
2(1+W) (-1)^{\delta_{\sigma_l^z + \sigma_u^z + \sigma_r^z + \sigma_d^z, \pm 4}}
, & s=+1\\
-i (1- W)(\sigma_l^z + \sigma_u^z + \sigma_r^z + \sigma_d^z)
, & s=-1\\
\end{cases}\, ,
\end{split}
\end{equation}
which projects out the same toric code eigenstate up to a prefactor
\begin{equation}
\prod_qM_{s_q}(t=\frac{\pi}{4}) \ket{+} \propto \prod_q \frac{1\pm W_q}{2}\ket{+}.
\end{equation} 
Namely, the charge is frozen because $\prod_{l\in \mbox{\varhexstar}} \sigma_l^x = 1$;
the measurement outcomes determine the static magnetic flux configuration, either expelled or occupied, because $W_q=s_q$.
Away from the strong measurement limit, the diagonal correlation is fully described by the effective classical model defined by Boltzmann weight $M^\dag M$, which would include not only the plaquette interaction $W$ but also the nearest-neighbor and next-nearest-neighbor two-body interactions. On the physical level, these two-body Ising interactions fluctuate the electric charge pairs. 
A detailed study of the phase transition of such model is beyond the scope of our current work. Nevertheless, in the perturbative regime, the topological order established at strong disorder limit could be robust against such short-ranged interactions.


\section{Appendix F: Absence of stable LRE in 1D}

For 1D lattice, there is no loop, and the probability function can be analytically factorized in the domain wall basis $\mu_{i}\equiv\sigma_i\sigma_{i+1}$:
\begin{equation*}
\begin{split}
p_{\{s\}}=\norm{ \bra{\{s\}}\ket{\psi}}^2 
&\propto  
\sum_{\{\sigma\}}e^{-\beta \sum_{i} (J_{s_{i}} \sigma_i \sigma_{i+1} + h s_{i})}\\
&\propto
\prod_i \sum_{\mu_{i}=\pm 1}e^{-\beta (J_{s_{i}} \mu_{i} + h s_{i})},
\end{split}
\end{equation*}
from which one can derive the \textit{uncorrelated} normalized probability of a single bond spin:
\begin{equation}
p_{s=1} = \frac{1+\cos(2t_A)\cos(2t_B)}{2},
\label{eq:probUncorr}
\end{equation}
In the following we analytically derive the EA order parameter for finite system size. Consider a 1D chain with $L$ number of bonds in open boundary condition, thus $\sigma_{L/2} = \prod_{i\leq L/2} \mu_{i }$, and there is a one-to-one correspondence between the site spin and domain wall configurations: $\sum_{\{\sigma\}}\cdots =\sum_{\{\mu\}}\cdots $. In a random bond configuration $\{s_i\}$, the magnetization of the site spin at the central site $i=L/2$ can be derived by integrating out the domain walls :
\begin{equation}
\begin{split}
\langle\sigma_0\sigma_{L/2}\rangle_{\{s\}}
&= \frac{1}{Z}\sum_{\{\sigma\}}e^{-\beta \sum_{i} J_{s_{i}} \sigma_i \sigma_{i+1} }\sigma_0\sigma_{L/2}\\
&= \frac{\prod_{i\leq L/2} \sum_{\mu_i=\pm1}e^{-\beta J_{s_{i}} \mu_i }\mu_{i}} {\prod_{i\leq L/2} \sum_{\mu_i=\pm1}e^{-\beta J_{s_{i}} \mu_i }}\\
&= \prod_{i\leq L/2} \tanh(-\beta J_{s_i}). 
\end{split}
\end{equation}
Recall that $\tanh(\beta J_\pm) = \sin(2t_A)\sin(2t_B)/(\pm 1+\cos(2t_A)\cos(2t_B))$, and when $t_B=\pi/4$, this reduces to $\pm \sin(2t_A)$. 
Then we can combine this with the factorized probability function to obtain the measurement averaged EA order parameter:
\begin{equation}
\begin{split}
q
&=\sum_{\{s\}} p_{\{s\}} \langle\sigma_0\sigma_{L/2}\rangle_{\{s\}}^2\\
&=\left( \sum_{s=\pm 1} p_{s} \tanh(\beta J_{s})^2\right)^{L/2}\\
&=\left(\frac{\sin(2t_A)^2 \sin(2t_B)^2}{1-\cos(2 t_A)^2 \cos (2t_B)^2}\right)^{L/2}.
\end{split}
\end{equation}
Check that when $t_B=\pi/4$, we have 
\begin{equation}
	q=\sin(2t_A)^L \,,
\end{equation}
which determines the correlation length $\xi = -\frac{1}{\ln \sin(2t_A)}$ that diverges only when approaching $t_A=\pi/4$. 
And when $t_B=t_A=t$, we have 
\begin{equation}
	q = \left(\frac{\sin(2t)^4 }{1-\cos(2 t)^4}\right)^{L/2} \,,
	\label{eq:1d-scaling-form}
\end{equation}
where the correlation length again diverges only when approaching $t=\pi/4$. 

\subsubsection{Numerical confirmation}

\begin{figure}[t!] 
   \centering
   \includegraphics[width=\columnwidth]{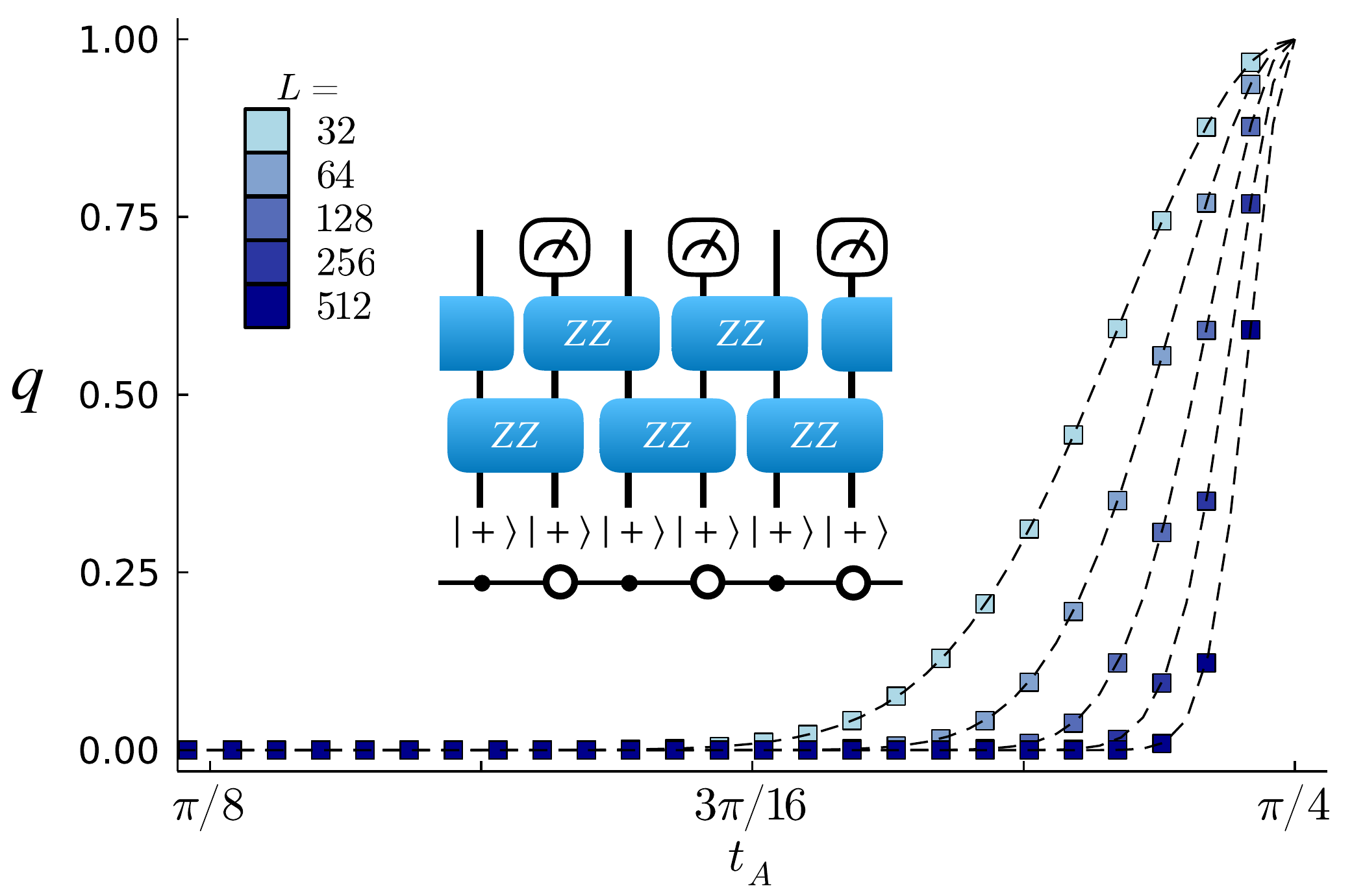} 
   \caption{{\bf 1D numerical results for $t_B=t_A$} and varying finite system sizes. 
   The inset schematically illustrates  the circuit of Ising (ZZ) evolution organized in a brickwall pattern, where every other qubits are measured in $s^x$-basis afterwards. The postmeasurement circuit, folded with its conjugate counterpart, can be compressed into a chain of random 2-by-2 transfer matrices along the space dimension, schematically shown beneath the circuit. By contracting out the transfer matrices one obtains a sample-dependent magnetization. 
   The square markers in the main plot are averaged over 1000 random, uncorrelated samples of the bond spins. 
   The dashed lines denote the analytical scaling form of Eq.~\eqref{eq:1d-scaling-form}.
   }
   \label{fig:1d}
\end{figure}

As a sanity check, we perform the same numerical calculations, where the sampling is reduced to uncorrelated sampling, and the contraction of classical TN reduces to taking product of dimension-2 transfer matrices along the chain. As shown in Fig.~\ref{fig:1d}, the numerical data (denoted by square markers) perfectly agrees with the analytically derived finite-size scaling form (denoted by the dashed line). 


\section{Appendix G: Preparing frustration-free transitions}

\subsubsection{1D SRE phase transition}

While the 1D case does not give rise to a stable LRE phase, we can still prepare a transition between two extended SRE phases. In particular, let us consider the following wave function which is a phase transition between 1D paramagnet ($-1 < \lambda < 1$) and a 1D SPT phase protected by the $\mathbb Z_2^T$ symmetry $\prod_n \sigma^x_n K$ ($|\lambda|>1$)\cite{Jones21}:
\begin{equation}
\ket{ \psi(\lambda) } = \exp \bigg( - \frac{1}{2} \textrm{arctanh}(\lambda) \sum_n \sigma^z_n \sigma^z_{n+1} \bigg) \ket{+_x}^{\otimes N}. \label{eq:SPTtransition}
\end{equation}
As one approaches the transition points at $\lambda = \pm 1$, the correlation length blows up as $\xi \sim \frac{1}{|1-\lambda|}$; at $\lambda = \pm 1$ the state is a cat state with $\xi = \infty$. Indeed, these transitions are effectively the zero-temperature transition of the classical Ising chain. Moreover, we point out that $|\psi(\lambda)\rangle$ is the ground state of
\begin{equation}
H = \sum_n \left( (1-a)^2 \sigma^x_n - a^2 \sigma^z_{n-1} \sigma^x_n \sigma^z_{n+1} - 2 a (1-a) \sigma^z_n \sigma^z_{n+1}  \right),
\end{equation}
with $a = \frac{\lambda}{1+\lambda}$, i.e., $\lambda = \frac{a}{1-a}$.

This transition has been implemented before on a quantum computer using a unitary circuit\cite{Smith22}, which requires a depth scaling linearly with system size. Here we point out that Eq.~\eqref{eq:SPTtransition} can be implemented with a finite-depth circuit and single-site measurements. Let us first consider $|\lambda| \leq 1$, such that $\textrm{arctanh}(\lambda)$ is real; we can interpret it as an effective inverse temperature $\beta = \textrm{arctanh}(\lambda) = \frac{1}{2}\ln \left| \frac{1+\lambda}{1-\lambda} \right|$. As we have already discussed, we can implement this imaginary Ising time-evolution using a depth-2 circuit on a chain and measuring ancillas. If $|\lambda|>1$, then
\begin{equation}
\textrm{arctanh}(\lambda) =  \frac{1}{2}\ln \left| \frac{1+\lambda}{1-\lambda} \right| - \textrm{sign}(\lambda) \; \frac{i \pi}{2}. 
\end{equation}
Hence, for $|\lambda|>1$, the complex evolution in Eq.~\eqref{eq:SPTtransition} consists of the imaginary Ising time evolution, as well as a unitary time-evolution coinciding with the cluster SPT entangler. Thus, the state for $|\lambda|>1$ can be prepared with a depth-4 unitary circuit, and a layer of single site measurement. Note that for any value of $\lambda$, we can always correct for the measurement outcomes with a single unitary feedback layer of spin flips, due to the absence of loops in 1D. In conclusion, we can exactly prepare Eq.~\eqref{eq:SPTtransition} with a finite-time protocol (by correcting for any measurement outcome due to the absence of frustration), independent of system size.

\subsubsection{LRE-to-LRE transition in any dimension}

Consider an arbitrary graph. We consider a family of states interpolating between the $XX$ ferromagnet ($0 \leq \lambda <1$) and $YY$ ferromagnet ($\lambda > 1$), which realize distinct LRE phases in the presence of a particular time-reversal $\mathcal T= K$ (i.e., complex conjugation):
\begin{equation}
\ket{ \psi(\lambda) } = \exp \bigg( - \frac{1}{2} \textrm{arctanh}(\lambda) \sum_v \sigma^z_v \bigg) \left( \frac{\ket{+_x}^{\otimes N} + \ket{-_x}^{\otimes N}}{\sqrt{2}} \right) \,. \nonumber
\end{equation}
It can be shown\cite{pivot} that this state is the ground state of the following Hamiltonian with $\alpha = \frac{\lambda^2}{1+\lambda^2}$:
\begin{equation}
    H = -\sum_{\langle v,v'\rangle} \left[ (1-\alpha) X_vX_{v'} +\alpha Y_vY_{v'}\right]  - \sqrt{\alpha(1-\alpha)}\sum_v  z_v Z_v \,, \nonumber
\end{equation}
where $z_v$ is the coordination number of each vertex. There is a continuous $z_\textrm{dyn}=2$ phase transition separating these two distinct LRE phases.

Using Eq. (1) of the main text, one can \textit{deterministically} prepare the wave function $\ket{\psi(\lambda)}$ (i.e., one can correct for any measurement outcome due to the absence of frustration). In fact, the 1D case is Kramers-Wannier dual to the example which we already discussed in Eq.~\eqref{eq:SPTtransition}.

\end{document}